\documentclass[12pt,letterpaper,usenames,dvipsnames]{article}
\usepackage{jheppub}

\allowdisplaybreaks[4]  
\usepackage{url} 
\usepackage{bm} 
\usepackage{dsfont} 
\usepackage{stmaryrd} 
\usepackage{cancel}
\usepackage{xfrac} 
\usepackage{mathtools}

\usepackage{amsthm}

\usepackage{tikz}
\usetikzlibrary{arrows,calc,snakes,shapes,shapes.arrows,
decorations.markings,decorations.pathmorphing}
\newcounter{sarrow}
\newcommand\xrsquigarrow[1]{%
\stepcounter{sarrow}%
\begin{tikzpicture}[decoration={snake,amplitude=.7mm},>=angle 90]
\node (\thesarrow) {\strut#1};
\draw[->,decorate] (\thesarrow.south west) -- (\thesarrow.south east);
\end{tikzpicture}%
}
     \tikzset{>=triangle 90}
     \tikzset{snake it/.style={decorate, decoration=snake}}
     \tikzstyle{gr}=[draw,circle,green!50!black,fill=green!50!black,scale=.6]
     \tikzstyle{Bl}=[draw,circle,blue,scale=.6]
     \tikzstyle{Blf}=[draw,circle,fill=blue,scale=.5]
     \tikzstyle{R}=[draw,circle,fill=red,scale=.6]
     \tikzstyle{bl}=[draw,circle,fill=black,scale=.6]
     \tikzstyle{bbc}=[draw,circle,fill=black,scale=.75]
     \tikzstyle{bbcs}=[draw,circle,fill=black,scale=.2]
     \tikzstyle{rc}=[circle,fill=red,scale=.4]
     \tikzstyle{wc}=[draw,circle,scale=.75]

\usepackage{multirow} 
\usepackage{colortbl} 
\usepackage{arydshln} 

\usepackage{xcolor}

\def\ccb{\cellcolor{blue!07}}
\def\ccg{\cellcolor{black!40!green!10}}

\def\ccr{\cellcolor{red!07}}

\usepackage[final]{showkeys}   

\def\nn{\nonumber}
\def\del{{\partial}}
\def\bar{\overline}

\def\til{\widetilde}
\def\hat{\widehat}
\def\ds{\ddots}
\def\ph#1{\phantom{#1}}
 
\def\vev#1{{\langle{#1}\rangle}} 
\def\bpm{\begin{pmatrix}}
\def\epm{\end{pmatrix}}
\def\bpms{\begin{pmatrix*}} 
\def\epms{\end{pmatrix*}}
\def\bsm{\begin{smallmatrix}}
\def\esm{\end{smallmatrix}}

\def\^{\wedge}
\def\Tr{\mathop{\rm Tr}}

\def\U{{\rm U}}
\def\SU{\mathop{\rm SU}}

\def\SO{{\rm SO}}
\def\SL{{\rm SL}}
\def\GL{{\rm GL}} 
\def\Sp{{\rm Sp}}

\def\CUB{{\rm CUB}}
\def\FCC{{\rm FCC}}
\def\HEX{{\rm HEX}}
\def\C{\mathbb{C}}

\def\N{\mathbb{N}} 
 
\def\R{\mathbb{R}} 
\def\Z{\mathbb{Z}}
\def\ff{{\mathfrak f}}
\def\gf{{\mathfrak g}}
\def\tf{{\mathfrak t}}
\def\uf{{\mathfrak u}}
\def\bd{{\boldsymbol{d}}}
\def\bl{{\boldsymbol{\ell}}}
\def\bL{{\boldsymbol{L}}}
\def\bm{{\boldsymbol{m}}}
\def\cB{{\mathcal B}}
\def\cE{{\mathcal E}}
\def\cH{{\mathcal H}}

\def\cN{{\mathcal N}}
\def\cO{{\mathcal O}}
\def\cS{{\mathcal S}}
\def\cW{{\mathcal W}}
\def\tcW{{\til\cW}}

\def\hq{{\hat q}}

\def\tQ{{\til Q}}
\def\tX{{\til X}}
\def\a{{\alpha}}
\def\ba{{\boldsymbol\a}}
\def\b{{\beta}}
\def\g{{\gamma}}
\def\bg{{\boldsymbol\g}}
\def\G{{\Gamma}}
\def\d{{\delta}}
\def\bde{{\boldsymbol\d}}
\def\D{{\Delta}}
\def\bDe{{\boldsymbol\D}}

\def\bla{{\boldsymbol\l}}
\def\th{{\theta}}
\def\k{{\kappa}}
\def\l{{\lambda}}
\def\bla{{\boldsymbol\l}}
\def\L{{\Lambda}}
\def\bLa{{\boldsymbol\L}}
\def\m{{\mu}}

\def\n{{\nu}}

\def\r{{\rho}}

\def\t{{\tau}}

\def\vf{{\varphi}}

\def\w{{\omega}}
\def\bw{{\boldsymbol\w}}
\def\Om{{\Omega}}

\title{Flavor symmetries and the topology of special K\"ahler structures at rank 1}
\author[a]{Philip C. Argyres,}
\author[b]{Matteo Lotito}
\affiliation[a]{Physics Department, University of Cincinnati, PO Box 210011, Cincinnati OH 45221}
\affiliation[b]{Amherst Center for Fundamental Interactions, Physics Department, University of Massachusetts, Amherst MA 01003}
\emailAdd{philip.argyres@gmail.com}
\emailAdd{mlotito@umass.edu}

\abstract{
We propose a method for determining the flavor charge lattice of the continuous flavor symmetry of rank-1 4d $\cN=2$ superconformal field theories (SCFTs) and IR free gauge theories from topological invariants of the special K\"ahler structure of the mass-deformed Coulomb branches (CBs) of the theories.  The method is based on the middle homology of the total space of the elliptic fibration over the CB, and is a generalization of the F-theory string web description of flavor charge lattices.  The resulting lattices, which we call ``string web lattices'', contain not only information about the flavor symmetry of the SCFT but also additional information encoded in the  lattice metric derived from the middle homology intersection form.  This additional information clearly reflects the low energy electric and magnetic charges of BPS states on the CB, but there are other properties of the string web lattice metric which we have not been able to understand in terms of properties of the BPS spectrum.  We compute the string web lattices of all rank-1 SCFTs and IR free gauge theories.  We find agreement with results obtained by other methods, and find in a few cases that the string web lattice gives additional information on the flavor symmetry.}
 
\begin{document}

\begin{flushright}  ACFI-T18-20\end{flushright}

\maketitle


\section{Introduction and summary}

Four-dimensional $\cN{=}2$ supersymmetric quantum field theories are remarkable for the diversity of techniques which can be employed to calculate exactly many of their properties at strong coupling.  One such property is the flavor symmetry of a non-lagrangian\footnote{One not continuously connected to a free theory by an exactly marginal deformation.} superconformal field theory (SCFT).  Information about the flavor symmetry can be gleaned from the isometries of the Higgs branch of vacua, from the spectrum of BPS states on the Higgs and Coulomb branches, and from the analytic geometry of the mass deformations of the Coulomb branch (CB) geometry.   These methods often give only partial information about the flavor symmetry, or are not readily calculable in general settings.

Our main goal is to introduce and explore a new general technique for calculating some flavor symmetry information for $\cN{=}2$ SCFTs from the topology of their mass-deformed CBs.  This comes from generalizing a picture of BPS states on the CB as string webs stretched between 7-branes in an F-theory realization of certain $\cN{=}2$ theories \cite{Sen:1996vd, Banks:1996nj, Aharony:1996xr, Sen:1996sk, Gaberdiel:1997ud, Bergman:1997yw, Gaberdiel:1998mv, Mikhailov:1998bx}.  What we calculate is a lattice of flavor charges of gauge-neutral states in the theory, which we call the (neutral) ``string web lattice''.\footnote{There is a straight forward generalization of our construction to non-neutral states --- i.e., those which carry charges under the low energy $\U(1)$ gauge fields on the CB.  We limit ourselves here to gauge-neutral states only for simplicity.}  The shape of this lattice (i.e., the lattice metric ignoring overall normalizations of orthogonal sublattices) carries some information about the flavor symmetry since it is a weight lattice of the flavor algebra.   In the case, extensively explored in the F-theory context in \cite{Gaberdiel:1997ud, Gaberdiel:1998mv, DeWolfe:1998zf, DeWolfe:1998eu, Hauer:2000xy}, where the flavor symmetry is a simple Lie algebra, the neutral string web lattice is the flavor root lattice in a particular normalization.  But in the general, non-simple, case, the string web lattice metric carries other information as well, encoded in the relative normalizations and deviations from orthogonality of the sublattices corresponding to simple or $\U(1)$ flavor symmetry factors.

The bulk of this paper is devoted to defining and calculating the string web lattice for rank-1 $\cN=2$ SCFTs and IR free field theories, and to uncovering the information encoded in the string web lattice by comparing it to the flavor and BPS spectrum data for these theories found by other techniques.  A systematic classification and construction of planar Coulomb branch (CB) geometries \cite{Argyres:2015ffa, Argyres:2015gha, Argyres:2016xua, Argyres:2016xmc} produced a list of all possible CB geometries of rank-1 4d $\cN=2$ QFTs, reproduced in table \ref{tablist} below.
This classification used a method that required determining the Seiberg-Witten (SW) curves and one-form for each allowed deformation, imposing a precise factorization of the discriminant, and performing a complicated and lengthy computation of the one-form using an ansatz introduced in \cite{Minahan:1996fg, Minahan:1996cj}.  From the detailed construction of the SW curves and one-forms the maximal flavor symmetry algebras of the QFTs corresponding to these geometries were determined, and are also shown\footnote{We use Dynkin notation for the simple Lie algebras, where $A_n = \SU(n+1)$, $B_n=\SO(2n+1)$, $C_n=\Sp(2n)$, and $D_n=\SO(2n)$, as well as the exceptional $E_{6,7,8}$, $F_4$ and $G_2$ algebras.  Also, we use a notation where $U_1=\U(1)$.  Finally, it is useful to keep in mind the following low-rank algebra equivalences: $A_1 = B_1 = C_1$, $D_1 = U_1$, $B_2 = C_2$, $D_2 = A_1 \oplus A_1$, and $D_3 = A_3$.} in table \ref{tablist}.  

As we will describe shortly, in addition to being computationally difficult, the identification of these flavor symmetries using this method suffers from unavoidable ambiguities.  This sort of uncertainty about deducing the flavor symmetry from the CB geometry is not only an issue with the direct approach to constructing SW one-forms \cite{Seiberg:1994aj, Minahan:1996fg, Minahan:1996cj, Noguchi:1999xq, Argyres:2015gha} but also is present in $\cS$-class \cite{Gaiotto:2009we, Gaiotto:2009hg} and string \cite{Xie:2015rpa} constructions of CB geometries, and in BPS quiver approaches \cite{Shapere:1999xr, Cecotti:2010fi, Cecotti:2011rv, Alim:2011ae, Alim:2011kw} to flavor symmetries.  A less ambiguous method of determining the flavor symmetry comes from the degeneracies of the spectrum of the Higgs branch operators which form representations of the flavor algebra \cite{Beem:2013sza, Beem:2014rza}.  In particular, the $\SU(2)_R$ isospin-1 operators necessarily fill out the flavor adjoint representation.  Unfortunately this latter method is presently limited to theories which have $\cS$-class realizations.

\begin{table}[h]
\centering
\small
$\begin{array}{|c|c|c|c|c|}
\hline
\text{\#}&\text{UV sing.}&\text{deformation}&\text{maximal flavor}&\text{string web}\\
&\text{type}&\text{pattern}&\text{symmetry}&\text{lattice shape}\\
\hline
1. & &\ccb\{{I_1}^{10}\} & E_8 & {\rm E}_8  \\
2. & &\{{I_1}^6,I_4\} & C_5 & \FCC_5 \\
3. &  &\ccr \{{I_1}^2,{I_4}^2\} &\ccr  C_2 & \CUB_2 \\
4. &  &\{{I_1}^4,I^*_0\} & F_4 & {\rm F}_4\\
5. &  &\{{I_1}^3,I^*_1\} &\ccr  C_3 \ \text{or}\ B_3
& \FCC_3 \\
6. &  &\{I_3,I^*_1\} & A_1 & \CUB_1\\
7. &  &\{{I_1}^2,I^*_2\} &  C_2 & \CUB_2 \\
8. &  &\{I_1,I^*_3\} & A_1 & \CUB_1 \\
9. &  &\{I_2,IV^*\} & A_1 & \CUB_1\\
10. &  &\{{I_1}^{2},IV^*\} & G_2 & \HEX_2\\  
11. & \multirow{-11}{*}{$II^*$}   
&\{I_1,III^*\} & A_1 & \CUB_1 \\
\hline 
12. &  & \ccb \{{I_1}^9\} & E_7 & {\rm E}_7 \\ 
13. &  &\{{I_1}^5,I_4\} &\ A_1\oplus C_3\ \,
& \CUB_1 \oplus \FCC_3\\
14. &  &\{{I_1}^3,I^*_0\} & B_3 & \CUB_3 \\
15. &  &\{{I_1}^2,I^*_1\} &\ccr C_2 & \CUB_1 \oplus \CUB_1\\
16. &  &\{I_2,I^*_1\} & A_1 & \CUB_1\\
17. &  &\{I_1,I^*_2\} & A_1 & \CUB_1\\ 
18. &  \multirow{-7}{*}{$III^*$}  
&\{I_1,IV^*\} &A_1  & \CUB_1\\
\hline
19. & &\ccb \{{I_1}^8\} & E_6 & {\rm E}_6 \\
20. & &\{{I_1}^4,I_4\} & C_2 \oplus U_1 &
\ccg \CUB_2 \niplus \CUB_1 \\ 
21. &  &\{{I_1}^2,I^*_0\} & G_2 & \HEX_2 \\ 
22. &  \multirow{-4}{*}{$IV^*$} 
&\{I_1,I^*_1\} & U_1 & \ccg \CUB_1\\
\hline
23. & &\ccb\{{I_1}^6\} & D_4 & \FCC_4\\
24. & &\{{I_2}^3\} & C_1 & \CUB_1\\
25. &  \multirow{-3}{*}{$I^*_0$} 
&\{{I_1}^2,I_4\} & C_1 & \CUB_1\\
\hline
26. & IV &\ccb\{{I_1}^4\} & A_2 & \HEX_2\\
\hline
27. & III &\ccb \{{I_1}^3\} & A_1 & \CUB_1\\
\hline
28. & II &\ccb \{{I_1}^2\} & \emptyset & \emptyset\\
\hline
\hline
& &\ccb \{{I_1}^n\} & A_{n-1} \oplus U_1 & 
\HEX_{n-1} \\
\multirow{-2}{*}{IRF $U(1)$}& 
\multirow{-2}{*}{$I_{n>0}$} 
& \text{many more (\S\ref{sec6.1})} 
& \oplus_i\, (A_{n_i} \oplus U_{1_i} )
& \ccg \oplus_i\, \HEX_{n_i} \niplus_i \CUB_{1_i} \\[1mm]
\hline
& &\ccb \{{I_1}^{6+n}\} & D_{n+4} & \FCC_{n+4}\\
\multirow{-2}{*}{IRF $SU(2)$}& 
\multirow{-2}{*}{\ $I^*_{n>0}$\ } 
&\ \text{many more (\S\ref{sec6.2})}\ 
& {\oplus} C_{m_i} {\oplus} 
B_{n_j} {\oplus} D_{p_k}
& \ccg {\oplus} \CUB_{m_i} {\oplus} 
(\CUB {\niplus} \til\FCC)_{n_j,p_k}
\\[1mm]
\hline
\end{array}$
\caption{Allowed CB deformations and assignment of the maximal flavor symmetry for the corresponding rank-1 SCFT.  The notation and the meaning of the shaded entries are explained in the text.}
\label{tablist}
\end{table}

All this prompts us to look for a different construction of the spectrum of gauge-neutral states from a given CB geometry.  One place such a construction has appeared is in the F-theory realization \cite{Sen:1996vd, Banks:1996nj, Aharony:1996xr, Sen:1996sk, Gaberdiel:1997ud, Bergman:1997yw, Gaberdiel:1998mv, Mikhailov:1998bx, DeWolfe:1998zf, DeWolfe:1998eu, Hauer:2000xy} of the effective theory on the CB of certain $\cN=2$ QFTs as the worldvolume theory of a D3-brane probing a background IIB string geometry provided by a set of parallel $(p,q)$-7branes.  Here electrically and magnetically neutral BPS states are realized as string webs stretching between the 7-branes, and so the lattice, $\L_F$, of possible flavor charges is simply given by the lattice of allowed boundary conditions on the $(p,q)$-7branes for such ``neutral'' string webs.  Furthermore, a natural metric on $\L_F$ is induced from a symmetric pairing related to the intersection form of string webs.  As explored extensively in \cite{Gaberdiel:1997ud, Gaberdiel:1998mv, DeWolfe:1998zf, DeWolfe:1998eu, Hauer:2000xy}, the identification of this metric with the Killing metric determined by the flavor Lie algebra gives consistent results.

In this paper we abstract this string web construction of the flavor charge lattice, $\L_F$, away from F-theory to a purely field theoretic setting.  In particular, we show that the flavor charge lattice can be deduced just from the topology of the special K\"ahler (SK) structure of the CB, at least in the rank 1 case.  So its computation has a much simpler combinatoric character as compared to the algebraically complicated determination of the SW curve and one-form data.  In particular, string webs appear as a certain subspace of the middle homology lattice $H_2(X,\Z)$ where $X \xrightarrow{\pi}$ CB is the total space of the elliptic fibration over the CB, and the metric on this lattice is given by the intersection pairing of the middle homology cycles.

Once abstracted away from the F-theory setting, this string web construction can be generalized to apply to arbitrary rank-1 $\cN=2$ QFTs, shown in table \ref{tablist}, and not just the subset that have F-theory realizations (which are the blue rows in table \ref{tablist}).  We will argue that this generalization is essentially unique.  Unlike in the cases described by F-theory, the total space, $X$, of the CB is now singular in general, and this generalization requires a careful treatment of the middle homology of $X$.  Our prescription for defining the flavor charge lattice is to set
\begin{align}\label{flavlatt}
\L_F = H_2(\tX,\Z) / \text{ker} Z,
\end{align}
where 
\begin{align}\label{Xtil}
\tX := X \setminus \{\text{singular fibers}\},
\end{align}
$H_2$ denotes the compact homology, and $Z$ is the central charge map $Z: H_2(\tX,\Z) \to \C$ given by
\begin{align}\label{Zmap}
Z(\a) = \int_\a \Om ,
\end{align}
where $\Om$ is the holomorphic symplectic form on $\tX$ \cite{Donagi:1994, Seiberg:1994aj, Donagi:1995cf}.  The intersection pairing on $H_2(\tX,\Z)$ restricts to a unique pairing on $\L_F$ because of the fiber bundle structure of $\tX$ and the property that the fibers are lagrangian cycles with respect to $\Om$.

Because, as we have emphasized above, the metric on $\L_F$ carries information beyond its identification as a flavor weight lattice, we will call $\L_F$ computed by the prescription \eqref{flavlatt} a ``string web lattice''.  Also, we outline the generalization of this string web lattice prescription to higher-rank CBs in section \ref{sec7}, at the end of the paper.

\paragraph{Results.}

The above prescription determines a flavor charge lattice with metric given by the intersection pairing on the middle homology cycles.  For simple factors of the flavor algebra it is natural to associate this lattice with the flavor root lattice with its metric inherited from the Killing form for that simple Lie algebra.  Indeed, as mentioned above, this identification was checked already in the blue-shaded cases in table \ref{tablist} \cite{Gaberdiel:1997ud, Gaberdiel:1998mv, DeWolfe:1998zf, DeWolfe:1998eu, Hauer:2000xy}.  However, we find that that is not the case in general:  the green-shaded string web lattices in table \ref{tablist} are cases where the string web lattice is a flavor charge lattice which is strictly larger than the flavor root lattice.

One situation where this always occurs is when the flavor symmetry has $\uf(1) := U_1$ flavor factors.  The flavor root lattice of a $U_1$ factor is the trivial (rank-$0$) lattice since the adjoint irrep of $U_1$ is the trivial irrep.   But in field theories with vacua in a Coulomb phase a global flavor $U_1$ can mix in the IR with the (global parts of) the unconfined electric and magnetic $\U(1)$ gauge groups.  In this case even states which carry no electric or magnetic charges --- e.g., the neutral string webs we are generalizing here --- can carry $U_1$ flavor charges.    Again, this is easy to verify for IR free $\U(1)$ gauge theories, and we do that in section \ref{sec6.1}.  The result is that $U_1$ flavor factors correspond to rank-1 sublattices of the flavor charge lattice, but these sublattices need not be orthogonal to those of the simple factors.

The shape of the string web lattices that we find are recorded in the right-most column of table \ref{tablist}.  The flavor root lattices of simple Lie algebras have the forms 
\begin{align}\label{introlatts}
\begin{array}{c|cccccc}
\ \text{Lie algebra}\  & A_n & B_n &\ C_n\ \text{or}\ D_n\ \  
&\ E_n\ &\ F_4\ & G_2 \\
\hline
\ \text{root lattice}\ &\ \HEX_n\ &\ \CUB_n\ & \FCC_n
&\ {\rm E}_n &\ {\rm F}_4 &\ \HEX_2\ 
\end{array} 
\end{align}
where the subscript denotes the rank of the lattice.  They are defined in appendix \ref{appA}, but briefly: $\CUB_n$, $\FCC_n$, and $\HEX_n$ are $n$-dimensional generalizations of cubic, face-centered cubic, and hexagonal lattices, respectively, while ${\rm F}_4$ and ${\rm E}_{6,7,8}$ are certain exceptional lattices.  There are some low-rank equivalences between these lattices, recorded in appendix \ref{appA}.
Also, in table \ref{tablist} orthogonal sums of these lattices are denoted as a direct sum $\oplus$, while non-orthogonal combinations with the rank-1 $\CUB_1$ lattices of $\U_1$ flavor factors are denoted as semi-direct sums, $\niplus$.  

The metric computed from the intersection form has a natural normalization, unlike the Killing form on a simple Lie algebra.  We compute it explicitly in equations \eqref{webinner2} and \eqref{webinner3}.  For a semi-simple flavor algebra, the flavor charge lattice decomposes as the orthogonal sum of weight lattices for each simple factor, but their relative normalizations are not determined by the Killing form.  However these relative normalizations \emph{are} determined by our prescription, thus supplying additional data beyond their identification as a flavor weight lattice.  The relative normalizations of the various lattice factors are not recorded in table \ref{tablist}, but are worked out in sections \ref{sec5} and \ref{sec6}; see in particular tables \ref{typeIn}, \ref{tab4}, and \ref{tab5}.  This relative normalization data reflects the relative low energy $\U(1)$ electric and magnetic charges of the (tower of) fields in the field theory which are charged under the different flavor symmetry factors.  This can be seen in the case of IR free gauge theories, and is explained and illustrated in some detail in section \ref{sec6} below.  In particular, in section \ref{sec6.3} we try to formulate a conjecture for how some properties of the BPS spectrum are reflected in the flavor lattice metric.  However, we have not been able to successfully associate all properties of the string web lattice metric to properties of the BPS spectrum.

Determining the flavor charge lattice (as a lattice with metric) is still not enough information to uniquely determine the flavor Lie algebra.  Even in the cases where the flavor charge lattice is the root lattice, the Lie algebra is not determined, as is apparent from \eqref{introlatts}.  The Lie algebra is determined by its \emph{root system}, which is simply the finite subset of the root lattice which are the roots of the Lie algebra.  In the case of a simply-laced Lie algebra, the root system is the set of shortest non-zero vectors in the root lattice, but a given lattice may also support a non-simply-laced root system.  If knowledge of the root lattice is supplemented by knowledge of the flavor Weyl group, then the flavor algebra is determined uniquely.  Unfortunately, we have not been able to determine the Weyl group action from the SK structure topological data in any simple way;  see \cite{Hauer:2000xy} for a discussion of this problem in the F-theory context.  

Nevertheless, as table \ref{tablist} shows, there is still a rich class of examples of rank-1 $\cN=2$ QFTs with associated flavor symmetries against which we can compare the flavor lattices as computed using our generalized string web prescription \eqref{flavlatt} and \eqref{Zmap}.   We find agreement with the maximal flavor assignments in table \ref{tablist}.  Furthermore, for theories \# 5 and 15 in the list the flavor lattice approach of this paper gives additional information:
\begin{itemize}
\item For theory \#5, the flavor lattice is identified as an $\FCC_3$ lattice (see appendix \ref{appA} for its definition) which is consistent with a $C_3$ and not a $B_3$ flavor algebra.  This was a case where the SW curve and one-form construction is not able to differentiate between the $B_3$ or $C_3$ flavor algebra assignment.
\item For theory \#15, the flavor lattice is a rectangular lattice --- the orthogonal sum of two rank-1 lattices with generating vectors of different lengths --- so not the $\CUB_2$ lattice expected from the maximal symmetry assignment $C_2$.  This is not a contradiction because it was shown in \cite{Argyres:2015gha, Argyres:2016xua} that the maximal $C_2$ flavor assignment of theory \#15 is not self-consistent (see below).  This is an example of how the extra lattice normalization data from our homological approach can extract flavor symmetry information that was not apparent from the SW curve construction of the CB geometry.
\end{itemize}
Indeed, the flavor algebra assignments deduced from the SW curve and 1-form and given in table \ref{tablist} are uncertain for a few reasons, discussed at length in \cite{Argyres:2015ffa, Argyres:2015gha, Argyres:2016xua, Argyres:2016xmc}.  The main point of uncertainty is that all the flavor symmetries shown in table \ref{tablist} were deduced by assuming that the discrete symmetry of the family of mass-deformed SW curves is in fact the Weyl symmetry of the flavor Lie algebra.  But it is possible that the flavor Lie algebra is smaller, and some of the discrete symmetry group of the curve is simply a discrete global symmetry of the theory.  
Indeed, for theories \# 3, 5, and 15 (shaded red in the table), the maximal flavor assignment shown is not internally consistent with the structure of the RG flows turned on by the mass deformations as deduced from their SW curves.
Furthermore, for theory \#3 it was found that there is no sub-maximal flavor assignment which is consistent with the RG flows of the curve, so there does not exist any SCFT corresponding to this geometry.

In addition to SCFT CB geometries, we have also calculated the string web lattices for all rank-1 IR free gauge theories, shown in the bottom two lines of table \ref{tablist}.  This computation is more difficult than for the SCFT CB geometries for a few reasons: (1) there are an infinite number of distinct mass deformations of these geometries, (2) the topology of their CB SK structures has not previously been worked out, and (3) we can encounter multiple singularities on the CB coming from a hypermultiplet in a single irreducible representation of the gauge group.  Section \ref{sec6} is devoted to surmounting these difficulties.  In particular, we are able to determine the topology of the CB SK structure for all mass deformations by fairly elementary arguments using the fact that these are weakly coupled field theories.  As mentioned above, these examples have been useful for diagnosing the ``extra'' information contained in the string web lattice metric.

(We have also worked out the SW curves and one-forms for these IR free theories, even though we do not need them for the homological calculation of their string web lattices.  But since they have not appeared elsewhere in the literature, we present them in appendices \ref{appU1} and \ref{appSU2}.)

\paragraph{Relation to the BPS spectrum.}

Another possible way of determining the flavor symmetry from its charge lattice is to determine the subset of the flavor charge lattice whose elements are realized as BPS states at various places on the CB, since these will necessarily form Weyl orbits of the flavor symmetry.  

But the determination of the BPS spectrum is also a complicated problem which seems to depend on more than just the SK structure topological data.  Indeed, in the F-theory string web picture, the BPS states are associated to string webs of minimal length \cite{Sen:1996sk, Bergman:1997yw} in the background IIB string geometry.  Upon T-dualizing to an M-theory picture, BPS states correspond to M2 branes stretched along a holomorphic curve in a background complex geometry \cite{Mikhailov:1998bx}.  

The IIB background metric data is reflected in the K\"ahler metric on the CB and the M-theory background complex structure data is reflected in the complex structure of the total space, $X$, of the CB.  Since neither of these structures are used directly\footnote{The string web lattice definition rests on the fiber structure, $\pi:X\to$ CB, of the total space, and the compatibility of the symplectic structure, $\Om$, on $X$ with that fiber structure.  The CB K\"ahler form is $\w(u) = \int_{X_u} \Om \wedge \bar\Om$, and $\pi$ and $\Om$ are holomorphic with respect to the complex structure of $X$.} in our homological definition of the string web lattice, it is unclear to us how much or what information about the BPS spectrum is captured by the string web lattice.

\paragraph{Relation to the rational geometry of $X$.}

Relatedly, Caorsi and Cecotti \cite{Caorsi:2018ahl} have proposed a method for determining the flavor algebras of rank-1 SCFTs from the rational geometry of the total space of the CB.  \cite{Caorsi:2018ahl} works on a smooth compactification $\cE$ of the total space $\tX$ of \eqref{Xtil}, which is rationally equivalent to $X$, and essentially computes $\L_F$ as a certain subspace of $H_2(\cE,\Z)$ defined through natural projections on the rational sections of $\cE$.  

While this prescription is undoubtedly closely related to our proposal \eqref{flavlatt} --- we both get the same results for rank-1 SCFTs with semi-simple maximal flavor symmetries by looking at intersections of middle-dimension cycles on the total space of the CB --- the precise connection between the two is not clear to us.  In particular, the compact middle homology of $\tX$ modulo $\text{ker} Z$ which we construct does not contain any sections of the elliptic fibration; indeed, it does not have representatives which are holomorphic curves in $\tX$.  This reflects the fact that the construction of \cite{Caorsi:2018ahl} uses a compact smooth version of the total space where the singular fibers are replaced by certain resolutions, whereas in our construction we use a non-compact space where these fibers are entirely absent.  This difference perhaps reflects different philosophies:  on our part, since the singular fibers are places where the effective description of the CB breaks down, we wish to derive as much information as we can about the flavor physics without making any assumptions about what happens at the singularities.  

Relatedly, here we do not compactify the metric infinities of the CB.  In the language of \cite{Caorsi:2018ahl} we remove their (dual) fiber ``at infinity".   This has the practical implication that we can also examine CB geometries which not only describe (relevantly deformed) SCFTs, but also those which describe IR free field theories.  These are theories whose CB geometry is only unambiguous in the vicinity of the origin (and for small-enough deformation parameters) because of the ambiguity in the choice of  their UV completions, should those even exist.

By not using specific models of the geometry at the singularities and infinity, our proposal can also be generalized to cases where the CB does not have ``planar" topology.  As discussed in \cite{Argyres:2017tmj} such cases might occur if there are rank-1 SCFTs with non-freely-generated CB chiral rings.  Caorsi and Cecotti's approach might also be generalizable to these cases were they to lift their assumption that the holomorphic symplectic form is non-degenerate at the singular fibers.  We will not discuss this generalization here.

Finally, note that \cite{Caorsi:2018ahl} also make a proposal for how to identify the root system (not just the root lattice) of rank-1 SCFTs in their approach.  As mentioned above, in our approach we did not find any natural way to identify the root system.  It would be interesting to understand if and how the rational section data that Caorsi and Cecotti use is translated into the topological data we use.

\paragraph{Organization of the rest of the paper.}

In the next section we motivate our prescription \eqref{flavlatt} for the flavor charge lattice by reviewing the F-theory string web picture, and abstracting it away from string theory by reformulating it just in field theory terms.  We also review the topological invariants of the SK structure of rank-1 CBs.
Section \ref{sec3} then shows how to compute the middle homology of \eqref{flavlatt} in terms of the topological data of the SK structure.  In section \ref{sec4} we then show how to compute the middle homology intersection form, and then in the next two sections we compute the resulting lattices for all rank-1 SCFTs and for the rank-1 IR free field theories, and compare the results we find to the answers known from other techniques.  We end in section \ref{sec7} with a short discussion on the generalization of the string web approach to higher ranks.
The appendices review root systems and root lattices of semisimple Lie algebras, and describe the construction of Seiberg-Witten curves and one forms for IR free $\U(1)$ and $\SU(2)$ gauge theories, which have not appeared before in the literature.


\section{Flavor charge lattice from the topology of the SK structure of the CB}\label{sec2}

\subsection{Review of string webs in F-theory}\label{Fthry}

In the F-theory realization \cite{Sen:1996vd, Banks:1996nj} of the CB effective actions of certain 4d $\cN=2$ field theories as the worldvolume action of D3-brane probes of parallel $(p,q)$-7branes, BPS states are realized \cite{Aharony:1996xr, Sen:1996sk, Gaberdiel:1997ud, Bergman:1997yw} as $(p,q)$-string webs ending on the 7-branes and 3-branes.  

The 2d plane --- which we will describe as the complex $u$-plane --- transverse to the worldvolume of parallel 7-branes is the CB of a rank-1 4d $\cN=2$ QFT.  The low energy $\U(1)$ $\cN=2$ electrodynamics at the vacuum at a given point $u\in$ CB is given by the worldvolume theory of a probe D3-brane placed at that point.\footnote{Certain higher-rank theories can be realized by placing more than one probe D3-brane in this 7-brane background geometry.  In this paper we only consider the rank-1 case.}  The complex axion-dilaton field, $\t(u)$, is determined by its response to the 7-brane configuration and is the low energy $\U(1)$ gauge coupling.

String webs consist of segments of $(p,q)$-strings --- bound states of $q$ fundamental strings and $p$ D1-branes in IIB string theory which exist for $\gcd(p,q)=1$ --- which end on the D3-brane or on a $(p,q)$-7brane, or at junctions with other $(p',q')$-string segments satisfying charge neutrality at each junction.   Minimal-length configurations of these string webs are the BPS states of the theory at the vacuum given by the position, $u$, of the probe 3-brane.  In the low energy $\U(1)$ gauge theory on the D3 brane, the $(p,q)$ charge of the string segment ending on the D3 brane is the (magnetic, electric) charge of the BPS state.  In \cite{Argyres:2001pv} it was shown how to derive this string web picture of BPS states from the low energy effective theory on the CB without reference to any string construction, showing that it can be applied to general $\cN=2$ QFTs and not just those special ones found in F-theory constructions.   

The string web picture of BPS states can be lifted to M theory by a series of dualities \cite{Mikhailov:1998bx} where the BPS string web appears instead as an M2-brane whose 2d spatial world volume is a Riemann surface with boundary on an M5-brane (the lift of the D3-brane probe) and wrapping nontrivial cycles in the bulk background geometry (the lift of the $(p,q)$-7brane configuration).  The symmetric intersection form of the M2-brane Riemann surfaces induce a symmetric pairing of string webs in the F-theory picture.

The set of string webs thought of topologically --- i.e., up to homotopies of the string segments and splitting and joining of parts of webs where it is allowed by charge conservation --- forms a lattice labeled by the sets of charges of all their endpoints.  Thus topologically each string web is labeled by the numbers, $n_i$, of (oriented) string segments it has ending on the $i$th 7-brane and the net $(p,q)$ charge of the segments ending on the D3-brane.  Since the intersection form is a topological invariant, it defines a (perhaps indefinite) metric on this lattice.

\begin{figure}[htbp]
\centering
\begin{tikzpicture}[decoration={markings,
mark=at position .75 with {\arrow{>}}}]
\begin{scope}[scale=1.5]
\fill[color=black!05] (7.5,0) rectangle (15.5,4);
\node[R] (br6) at (8.5,1.0) {};
\node at (8.5,0.75) {$(0,1)$};
\draw[decorate,decoration=snake,red!50] (br6) -- (8.5,0);
\node[R] (br7) at (10,1.0) {};
\node at (10,0.75) {$(0,1)$};
\draw[decorate,decoration=snake,red!50] (br7) -- (10,0);
\node[R] (br8) at (11.5,1.0) {};
\node at (11.5,0.75) {$(0,1)$};
\draw[decorate,decoration=snake,red!50] (br8) -- (11.5,0);
\node[R] (br9) at (13,1.0) {};
\node at (13,0.75) {$(-1,2)$};
\draw[decorate,decoration=snake,red!50] (br9) -- (13,0);
\node[R] (br10) at (14.5,1.0) {};
\node at (14.5,0.75) {$(1,1)$};
\draw[decorate,decoration=snake,red!50] (br10) -- (14.5,0);
\node[bl] (jnc1) at (10,2.5) {};
\node[bl] (jnc2) at (13.75,2.5) {};
\draw[ultra thick,blue,postaction={decorate}] (br6) .. controls (8.5,1.75) and (9,2.5) ..(jnc1);
\draw[ultra thick,blue,postaction={decorate}] (br7) -- (jnc1);
\draw[ultra thick,blue,postaction={decorate}] (br8) .. controls (11.5,1.75) and (11,2.5) ..(jnc1);
\draw[ultra thick,blue,postaction={decorate}] (jnc1) .. controls (10,3.5) and (13.75,3.5) ..(jnc2);
\draw[ultra thick,blue,postaction={decorate}] (br9) .. controls (13,1.75) and (13,2) .. (jnc2);
\draw[ultra thick,blue,postaction={decorate}] (br10) .. controls (14.5,1.75) and (14.5,2) .. (jnc2);
\node[blue] at (8.25,2.0) {$-(0,1)$};
\node[blue] at (10.4,1.5) {$(0,1)$};
\node[blue] at (11.6,2.25) {$2\cdot(0,1)$};
\node[blue] at (12,3) {$(0,3)$};
\node[blue] at (12.45,1.5) {$-(-1,2)$};
\node[blue] at (15.0,1.5) {$-(1,1)$};
\end{scope}
\end{tikzpicture}
\caption{A neutral string web (in blue) ending on five 7branes (red dots).  Each 7brane is labeled by its $(p,q)$ charges, satisfying $\gcd(p,q)=1$.  An integer number, $n$, of $(p,q)$ strings can end on a $(p,q)$-7brane, and so carries $n\cdot(p,q)$ string charge away from the 7brane.  String segments can meet at junctions as long as the total string charge is conserved.  The pictured web is ``neutral" because it has no segments ending on probe D3 branes.\label{web}}
\end{figure}
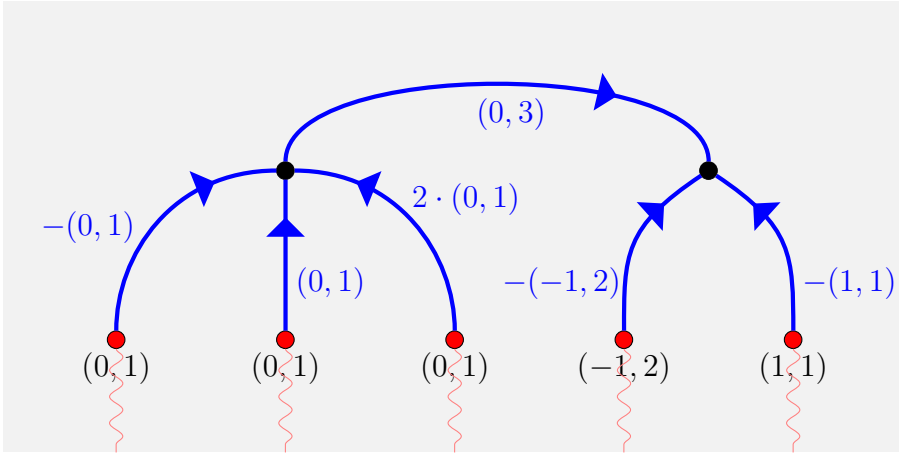

The positions of the 7branes in the $u$-plane are governed by the masses (or other dimensionful scales) of the 4d $\cN=2$ QFT.  A 4d SCFT corresponding to zero masses is thus given by a configuration where all the 7branes are coincident.  If the 4d CFT has flavor symmetry $F$, then upon turning on generic masses, it is broken to $\U(1)^f$ with $f=\text{rank}(F)$, since the masses transform in the adjoint of $F$.  The conserved charges, $n_i$, of these $\U(1)$ global symmetry groups at generic masses are called the ``quark number" charges.   Thus, in terms of the low energy $\U(1)$ gauge theory on the CB, the string web charge lattice is the lattice of possible (magnetic, electric) charges for the $\U(1)$ factor (D3-brane) and the ``quark number" charges.  It thus forms a lattice of rank $2+f$.

The flavor symmetry of the 4d CFT appears as the gauge symmetry of the 8d YM theory on the worldvolume of the stack of coincident 7branes.  Separating the 7branes corresponds to Higgsing this gauge group by giving a vev to a complex adjoint scalar field.  In particular, the massive W-bosons of this Higgsing are the string webs which end only on the 7branes.  Since there are no string segments ending on D3 branes for such webs, this subset of string webs are those with $(p,q)=(0,0)$ $\U(1)$ gauge charges, and so form a rank-$f$ sublattice of ``neutral" string webs.  Thus, the neutral string web charge lattice is identified with the flavor charge lattice of the 4d flavor symmetry algebra.

A series of papers \cite{Gaberdiel:1997ud, Gaberdiel:1998mv, DeWolfe:1998zf, DeWolfe:1998eu, Hauer:2000xy} developed this picture of the lattice of string webs.  In particular, they assumed that the metric on the flavor charge lattice coming from the string web intersection pairing is the Killing metric on the weight lattice determined by the flavor Lie algebra.  They then compute in cases to show that this string web lattice with metric given by the intersection pairing gives the correct flavor root lattices corresponding to the then-known A-D-E series of rank-1 $\cN=2$ SCFTs.

In the rest of this section we will rephrase this string theory construction solely in terms of the low energy $\U(1)$ gauge theory on the CB.  In particular, we will see that it has a simple expression in terms of the topology of the special K\"ahler (SK) structure on the CB.   We will then show how in this field theory language it can be easily generalized to CBs with SK structure of topological types not present in the string construction.  But first we provide a quick reminder of the main topological ingredients characterizing the SK structure of a rank-1 CB.  (See \cite{Freed:1997dp} for a more general discussion of various definitions of SK geometry and their interrelations.)

\subsection{Review of rank-1 SK geometry}\label{rk1SK}

We describe a rank-1 CB as the complex $u$-plane with a K\"ahler metric given by the kinetic terms of the complex scalar of the low energy $\U(1)$ vector multiplet.  This metric will be singular at points where additional $\U(1)$-charged modes become massless.  It has additional structure, called a special K\"ahler (SK) structure, reflecting the way the lattice of $\U(1)$ gauge charges varies over the CB.

Denote by $\bl$ the row vector of the $\U(1)$ magnetic and electric charges of a particle,
\begin{align}\label{zdef}
\bl := (p\ q).
\end{align}
The Dirac-Schwinger-Zwanziger charge inner product of two dyons is
\begin{align}\label{dszform}
\vev{\bl_1,\bl_2} := q_1 p_2- p_1 q_2.
\end{align}
The freedom to choose different definitions of what we call electric and magnetic charges is the freedom to choose different bases of the charge lattice which preserve the inner product.  This is the group of electric-magnetic (EM) duality transformations, $\Sp(2,\Z)\simeq \SL(2,\Z)$.  It is generated by two elements, 
\begin{align}\label{STdef}
S := \bpm 0&-1\\ 1&\ph{-}0 \epm, 
\qquad\text{and}\qquad 
T := \bpm 1&1\\ 0&1 \epm,
\end{align}
which satisfy $S^2=(ST)^3=-I$.  Note that the Dirac product can be written as $\vev{\bl_1,\bl_2} := \bl_1 S \bl_2^T$.  An $\SL(2,\Z)$-invariant of a single dyon charge, $\bl$, is its EM duality invariant charge, $Q_\bl$, defined by
\begin{align}\label{Qdef}
Q_\bl^2 := \gcd(p^2,q^2).
\end{align}

The low energy $\U(1)$ gauge coupling, $\t(u)$, is complex analytic in $u$ and is allowed to be discontinuous by EM duality transformations which act on $\t$ by fractional linear transformations, $\t \to (A\t+B)(C\t+D)^{-1}$, for $M_\g := \left(\bsm  A&B\\ C&D \esm\right) \in\SL(2,\Z)$.  Here $M_\g$ is the EM duality monodromy of a closed path $\g$ in the CB.  By continuity, if $\g$ is contractible then $M_\g=I$ since $\SL(2,\Z)$ is discrete, so non-trivial monodromies are associated to paths encircling singularities on the CB.  More generally, the monodromies give a representation in $\SL(2,\Z)$ of the fundamental group of the CB minus its singularities.

$\t(u)$ can be realized geometrically as the complex structure modulus of a complex torus, $X_u$.  Realize $X_u$ as the complex plane modulo lattice translations, $X_u = \C/\L_{\t(u)}$, where $\L_{\t(u)}= \{ n+m\t(u) , n,m\in\Z\}$.  Then the SK structure of the CB geometry is encoded in the complex geometry of the total space, $X$, of the torus fibration, $X \xrightarrow{\pi}$ CB, with fibers $X_u = \pi^{-1}(u)$.

The magnetic and electric $\U(1)$ charge numbers $\bl=(p\ q)$ parameterize the homology class of a cycle $[\l(\bl)]$ on the fiber $X_u$.  More precisely, the 1-homology classes of $X_u$ are parameterized by a pair of integers $\bla=(c\ d)\in\Z^2$ as $[\l]=c[\a]+d[\b]$, where we take $\a$ and $\b$ to be a canonical basis of 1-cycles on the torus, i.e., one with intersection form
\begin{align}\label{abbasis}
\a\cdot\a = \b\cdot\b = 0, \qquad \b\cdot\a = -\a\cdot\b = 1.
\end{align}  
The integers $\bla=(c\ d)$ are related to the magnetic and electric $\U(1)$ gauge charges by $\bl=(p\ q)= \sqrt{P}\, (c\ d)$ where $P\in\N$ is a positive integer characterizing a \emph{polarization} of $X_u$.  More physically, we call $\sqrt P$ the \emph{charge quantization unit}, relative, of course, to some normalization conventions; see appendix A and section 4 of \cite{Argyres:2015ffa} for a more detailed discussion of the normalization of EM charges.  We will also review this in a bit more detail in section \ref{sec6.2} below where it will be important.

Most of the rank-1 SCFTs theories listed in table \ref{tablist} --- all those except numbers 6, 9, 16, and 24 --- have $P=1$, so the distinction between the homology classes and the charges is irrelevant for them.  But this is not true in every case, and, indeed, infinitely many of the IR free field theories in the last rows of table \ref{tablist} have $P\neq1$.  Because of this, we have introduced a notation emphasizing the difference between EM charges and fiber homology classes, recapitulated here:
\begin{align}\label{zmdef}
\text{EM charge vectors:}& &
\bl &= (p\ q) \in \sqrt P \, \Z^2 , \quad P\in\N,
\nonumber\\ 
\text{Fiber homology classes:}& &
\bla &= (c\ d) \in \Z^2 \simeq H_1(X_u,\Z) ,\\
\text{related by}& &
\bl &= \sqrt P \, \bla .\nonumber
\end{align}
The homology class of a 1-cycle, $\l$, on a fiber associated to a state of EM charge $\bl$ is
\begin{align}\label{}
[\l] = c [\a] + d[\b] = \bla \bpm [\a] \\ [\b] \epm.
\end{align}
Then the Dirac charge inner product \eqref{dszform} is encoded in terms of the intersection form of 1-cycles on $X_u$ by 
\begin{align}\label{Dcip}
\vev{\bl_1,\bl_2} &= P\, \l_1\cdot\l_2, &
&\text{where} &
\l_1\cdot\l_2 &= \bla_1 S \bla_2^T.
\end{align}
Note also that the invariant charge \eqref{Qdef} of a state with charge $\bl$ is
\begin{align}\label{QtoQ}
Q_\bl &= \sqrt P \, \gcd(\bla), &
&\text{where} &
\gcd(\bla) &:= \gcd(c,d) .
\end{align}

In this paper, since we are interested in computing homological invariants of the SK geometry of the CB, we will be using fiber homology classes, $\bla\in H_1(X_u,\Z)$, exclusively, and will only relate them to physical charge vectors, $\bl$, when we want to clarify the physical interpretation of a given result. 

There is one other geometric object on $X$ that is needed to specify its SK structure.  This is the natural holomorphic symplectic form on $X$, i.e., the closed, non-degenerate holomorphic (2,0) form on $X$ given by \cite{Donagi:1994, Donagi:1995cf}\footnote{In the case of SK geometries of rank greater than one there are further conditions that $\Om$ must satisfy.  In terms of SW theory, $\Om$ is closely related to the exterior derivative of the SW one-form \cite{Seiberg:1994aj}.}
\begin{align}\label{DW2form}
\Om := du \wedge dz .
\end{align}
The periods of $\Om$ compute the central charge of the $\cN=2$ supersymmetry algebra in the low energy theory on the CB.  In particular, if $\a$ is a 2-chain in $X$ whose boundary is a 1-cycle $\g(\bl)$ on $X_u$, then
\begin{align}\label{ZOm}
Z(u) = \int_\a \Om ,
\qquad \del\a \subset X_u,
\end{align}
where $Z(u)$ is the central charge in the $u$ vacuum and in the $\bl$ charge sector.  Since $\Om$ is closed and since its restriction to any fiber $X_u$ vanishes, $Z(u)$ is a linear map $H_1(X_u) \oplus H_2(X) \to \C$,
\begin{align}\label{rank1CC}
Z(u) = p\, a_D(u) + q\, a(u) + \bw(\bm).
\end{align}
$a(u)$ and $a_D(u)$ defined in this way are the special coordinate and dual special coordinate on the CB, $\bm \in \C^f$ are the complex linear mass parameters, and $\bw$ is a vector in the lattice $\L_F$ of global $\U(1)^f$ ``quark number" charges --- i.e., the flavor weights.  We use a notation in which weights and masses are dual: $\bw\in\L_F$ and $\bm\in\L_F^*\otimes_\Z \C \simeq \C^f$, and $\bw(\bm) := \w^i m_i$ is the dual pairing.

\subsection{Proposal for identifying the flavor charge lattice}

The linear mass parameters $\bm$ transform in the adjoint representation of the global flavor symmetry Lie algebra, $F$, of the massless theory (the SCFT or the IR free theory).  Thus $f=\text{rank}(F)$, and the lattice of the quark number charges, $\L_F$, is identified with a lattice of weights, $\bw$, of $F$.  So if $\ff$ is the Cartan subalgebra of $F$, then $\bw\in\ff^*$ and $\bm\in\ff_\C$.

Whenever $\ba(\bm)=0$ for $\ba$ a root of $F$ there should be a degeneracy in the BPS spectrum since on these subspaces of the mass parameter space the flavor symmetry is not completely broken to abelian factors.  So we expect that all $\bw$ in the root lattice of $F$ should occur in the central charge \eqref{rank1CC}.  Conversely, if there were some $\bw$ not in the root lattice of $F$, then there will be additional degeneracies in the BPS mass spectrum at every point on the CB for masses satisfying $\bw(\bm)=0$ which are not due to an enhanced symmetry.  Discounting the existence of such ``accidental'' degeneracies which persist for all values of $u\in$ CB, one concludes that the lattice, $\L_F$, of quark number charges (flavor charges) should be the root lattice of $F$:  $\L_F = \L_{F\text{-root}}$.  Indeed, in many of the cases we check later in this paper we find this to be the case.  But we will also see that there are cases where $\L_F \supsetneq \L_{F\text{-root}}$.  In particular, this always seems to be the case when there are $\U(1)$ flavor symmetry factors; also in section \ref{sec6} we will see many examples of lagrangian theories with flavor charge lattices which include flavor weights not in the root lattice.


Since from the expressions \eqref{ZOm} and \eqref{rank1CC} for the central charges we have that the flavor weights $\bw$ are classes in $H_2(X,\Z)$, we thus propose to locate the flavor charge lattice as a subgroup $\L_F \subset H_2(X,\Z)$.  

Since $X$ is not compact we have to specify the homology theory used to compute $H_2$.  Physically, cycles whose central charges diverge correspond to infinitely massive states which are decoupled from the theory.  So we are only interesed in cycles $\g\in H_2(X,\Z)$ for which $\int_\g \Om <\infty$.  This is equivalent to keeping only homology cycles with compact support.

Not all classes in $H_2(X,\Z)$ are in $\L_F$ since there may be some elements of the middle homology which are annihilated by the central charge map \eqref{ZOm} on 2-cycles.  Thus we would like to identify the flavor charge lattice with the quotient
\begin{align}\label{LFH2}
\L_F \overset{?}{=} H_2(X,\Z)/\ker Z.
\end{align}

The flavor Lie algebra $F$ does not enter directly in the SK structure of the CB.  This is not surprising physically:  the flavor symmetry has no action on the CB, and can only be reconstructed from its action on the dependence of the mass parameters which explicitly break the flavor symmetry to the $\U(1)^f$ quark number symmetry.  In particular, the Lie algebra Killing forms on each simple factor of $F$, which determine up to overall normalizations metrics on the corresponding flavor charge sublattices of $\L_F$, is not an obvious ingredient in the SK structure.  But the intersection form on $H_2(X,\Z)$ gives a natural symmetric bilinear form, so from \eqref{LFH2} it is natural to guess that this gives the charge lattice Killing metric.  Indeed, this is effectively what was assumed (or conjectured and checked) in the F-theory construction described in the last subsection.   Our aim is to generalize and check this conjecture in the case of general $X$.  To check the self-consistency of the proposal requires that one checks that modding out by $\ker Z$ respects the intersection form on $H_2(X,\Z)$, and that the resulting lattice bilinear form has definite signature.  One then must check whether the resulting lattice coincides with the physically known flavor charge lattice for each $X$.

But, unlike the F-theory cases where under a generic deformation the total space $X$ is a manifold, the general rank-1 CB geometry $X$ has complex singularities.  Even though compact homology is still defined for these spaces, Stokes' theorem can break down, and so the appropriateness of modding out by $\ker Z$ in \eqref{LFH2} must be re-examined.  Also, from a physical perspective, the description of the low energy effective action as a theory of massless $\U(1)$ vector multiplets breaks down at the singularities of the CB.  Instead of trying to resolve or otherwise interpret any singularities of the total space that occur over the metrically singular points of the CB base, we will pursue a strategy of trying to extract the flavor physics from the structure of $X$ away from the singular fibers.  Thus we propose to replace $X$ with the (non-compact) smooth manifold $\tX$ given by removing all singular fibers from $X$.  $\tX$ is thus a smooth elliptic fibration over the CB minus its metrically singular points with regular fiber everywhere.  

In the next subsections we will show that the most straight forward identification,
\begin{align}\label{LFH3}
\L_F = H_2(\tX,\Z)/\ker Z,
\end{align}
with $H_2$ computed with compact support makes sense physically.  The intuitive reason one expects that this prescription should work is that since the holomorphic 2-form $\Om$ does not diverge  as one approaches the singular fibers\footnote{The reason $\Om$ is regular at the singular fibers follows from a combination of unitarity of the underlying $\cN=2$ QFT and from the regularity of the complex structure of the CB base at its metric singularities (i.e., at the singular fibers).  See \cite{Argyres:2017tmj} for a detailed discussion of this point.} and also vanishes when restricted to any fiber, the details of any cycle near a singular fiber should be killed by the $Z$-map.  Thus after modding out by the kernel of $Z$, it should make no difference whether we use the singular total space $X$, or restrict to only its smooth fibers, $\tX$.  Indeed, we will see explicitly that \eqref{LFH3} defines a lattice (no torsion appears) of the correct rank, and we will show how to compute the intersection form induced on $\L_F$ purely topologically.   

\begin{table}[t]
\centering
$\begin{array}{|c|c|c|}
\hline
\multicolumn{3}{|c|}{\text{\bf Possible singular fibers of $X \xrightarrow{\pi}$ CB}}\\
\hline\hline
\ \text{Kodaira type}\ \ & M &\ \text{order of}\ M\ \ \\
\hline
II^*   &ST            &6 \\
III^*  &S               &4 \\
IV^*  &(ST)^2      &3 \\
I_0^* &-I              &2 \\
IV      &(ST)^{-2} &3 \\
III      &S^{-1}     &4 \\
II       &(ST)^{-1} &6 \\
I_0     & I              &1 \\
\hline\hline
I^*_n\ \ (n>0) & {-T^n} & \infty \\
I_n\ \ (n>0)     & {T^n}   & \infty \\
\hline
\end{array}$
\caption{Singular fibers of the elliptic fibration over the CB describing its SK structure, labeled by their Kodaira type, a representative $M\in\SL(2,\Z)$ of the conjugacy class of the monodromy around the singular fiber, and the order of $M$.  The $I_0$ row describes the regular fiber and corresponds to a free $\U(1)$ vector multiplet.  The last two rows give infinite series of singularities describing IR free theories.\label{kodaira}}
\end{table}

\subsection{Topological invariants of rank-1 SK structures}

The torus fibration over the CB has singular fibers over a set of points $\{u_i, \ i=1,\ldots,A\}$ in the CB, and the fibers have non-trivial $\SL(2,\Z)$ monodromies around these points.   In fact, the set of possible degenerations of the fibers and their associated monodromies have been classified by Kodaira \cite{Kodaira:1964, Kodaira:1966} in the context of the classification of elliptic surfaces.  Indeed, $X$ is an elliptic fibration with a non-vanishing canonical class, $\Om$, so meets Kodaira's requirement for the resolutions of its singular fibers to give an elliptic surface.  Kodaira's classification\footnote{As pointed out in \cite{Caorsi:2018ahl}, the non-appearance of multiple fibers --- the ${}_mI_n$ in Kodaira's classification --- follows from the existence of a chosen section of the elliptic fibration.} is shown in table \ref{kodaira}.  

\begin{figure}[tbp]
\centering
\begin{tikzpicture}[decoration={markings,
mark=at position .5 with {\arrow{>}}}]
\begin{scope}[scale=1.5]
\clip (9,0) rectangle (15,4);
\fill[color=black!05] (9,0) rectangle (15,4);
\node[Bl] (or1) at (12,3) {};
\node[R] (br1) at (10.5,2) {};
\node[red] at (10.5,2.35) {$u_1$};
\node[R] (br2) at (12,1.5) {};
\node[red] at (12,1.85) {$u_2$};
\node[R] (br3) at (13.5,2) {};
\node[red] at (13.5,2.35) {$u_3$};
\draw[decorate,decoration=snake,red!50] (br1) -- (10,0);
\draw[decorate,decoration=snake,red!50] (br2) -- (12,0);
\draw[decorate,decoration=snake,red!50] (br3) -- (14,0);
\draw[thick,blue,postaction={decorate}] (or1) .. controls (8,3) and (10,-1) .. (or1);
\node[blue] at (10,3) {$M_1$};
\draw[thick,blue,postaction={decorate}] (or1) .. controls (10.5,0) and (13.5,0) .. (or1);
\node[blue] at (11.3,1) {$M_2$};
\draw[thick,blue,postaction={decorate}] (or1) .. controls (14,-1) and (16,3) .. (or1);
\node[blue] at (14,3) {$M_3$};
\end{scope}
\end{tikzpicture}
\caption{Singularities and their monodromies on the $u$-plane.  The solid points with coordinates $u_i$ are the singularities shown with a choice of ``branch cuts" emanating from them.    The $M_i\in SL(2,\Z)$ are EM duality monodromies associated to the closed paths looping around these singularities starting from a conventional base point given by the open circle.\label{CBtopol}}
\end{figure}
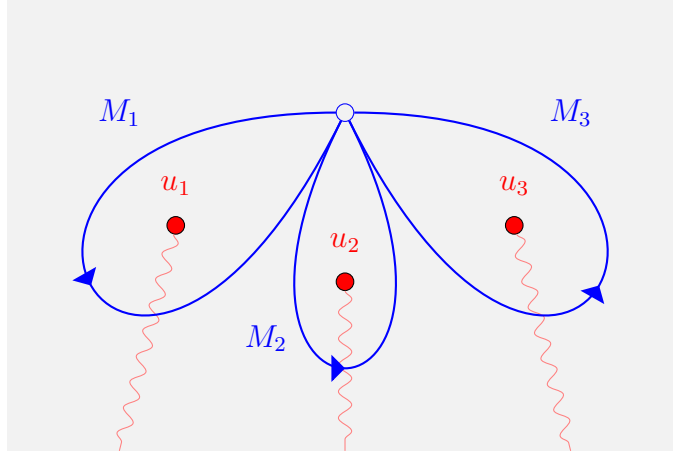

Each singular fiber $X_{u_i}$ has a corresponding EM duality monodromy $M_i\in\SL(2,\Z)$.  More explicitly, choose a canonical basis $[\a],[\b]\in H_1(X_0,\Z)$ of 1-cycles of the fiber over a chosen base point $0\in$ CB, satisfying \eqref{abbasis}.  This basis is then extended by continuity to the whole CB minus the conventional set of cuts emanating from the singularities of the CB; see figure \ref{CBtopol}.  Upon dragging this basis along a simple closed curve $\m$ encircling one of the singularities (so crossing just one of the cuts) counterclockwise, it suffers a monodromy
\begin{align}\label{monodact0}
\bpm[\a]\\ [\b]\epm 
\ \xrsquigarrow{\small{$\ \ \m\ \ $}}\ \,
M \bpm[\a]\\ [\b]\epm, \qquad
M \in \SL(2,\Z).
\end{align}
A general fiber homology class is $[\g] = c [\a] + d [\b] = \bg {[\a]\choose[\b]}$, so under the $\m$-monodromy becomes 
\begin{align}\label{monodact}
\g  \overset{\m}{\rightsquigarrow} \g' &:= M\circ \g &
&\text{with}&
\bg' &= \bg M .
\end{align}
Here we have introduced the shorthand $M\circ \g$ to denote the action of a monodromy on a fiber cycle.

If we drag a fiber cycle $\g = \bg {\a\choose\b}$ counterclockwise around a curve $\m_1$ with monodromy $M_1$, then $\g \overset{\m_1}{\rightsquigarrow} \g' = \bg M_1 {\a\choose\b} := \bg' {\a\choose\b}$.  Following this with a second curve, $\m_2$, with monodromy $M_2$ then gives $\g' \overset{\m_2}{\rightsquigarrow} \g'' = \bg' M_2 {\a\choose\b} = \bg M_1 M_2 {\a\choose\b}$.  So, in these conventions we have 
\begin{align}\label{M1M1rule}
M_1 \circ M_2 = M_1 M_2 ,
\end{align}
where $\circ$ means concatenation of base cycles (with first cycle on the left).\footnote{Note that the opposite convention, namely $M_1\circ M_2 = M_2 M_1$, was used in \cite{Argyres:2015ffa, Argyres:2015gha, Argyres:2016xua, Argyres:2016xmc}, even though \eqref{monodact0} was also adopted there, which is an inconsistent choice of conventions.  Nevertheless, the results of those papers are not affected: one should simply reverse the order of the monodromy data for the CB geometries constructed there.}

%
%
%
%
%
%
%
%
%
%

Under a change in choice of canonical 1-cycle basis of the fibers by an element $g\in\SL(2,\Z)$, $M_i$ changes by $M_i \to g M_i g^{-1}$, so only the conjugacy class of $M_i$ in $\SL(2,\Z)$ is an invariant property of the singularity.  Representative monodromies of the Kodaira singularities are listed in table \ref{kodaira}.  The Kodaira type of the singularities are uniquely determined by the conjugacy class of their EM duality monodromy, $[M]$, though not all such conjugacy classes occur.

The set of EM duality monodromies, $\{ M_i,\ i=1,\ldots,A\}$ around each of the singularities on the CB characterizes the topology of the SK structure of the CB.  This set of monodromies can be specified up to an overall $\SL(2,\Z)$ conjugation by picking a base point, an ordering of the singularities, and a set of simple closed paths encircling each singularity in the same sense.  A convenient way of specifying the ordering and paths is by choosing a set of non-intersecting ``branch cuts" emanating from each singularity and going to $u=\infty$ parallel to the negative imaginary $u$-axis.  Then the singularities are ordered according to increasing Re$(u)$-values of the cuts at Im$(u)\to-\infty$, and a basis of (homotopy classes of) closed paths on the punctured $u$-plane, $\{\m_i\}$, are defined by demanding that $\m_i$ crosses only the $i$th branch cut just once counterclockwise; see figure \ref{CBtopol}.  

Denote the $\SL(2,\Z)$ monodromy around $\m_i$ by $M_i$ for $i=1,\ldots, A$.  These monodromies are specified up to a common $\SL(2,\Z)$ conjugation, reflecting the freedom to choose an arbitrary EM duality basis at the base point.  Furthermore, the ordering of the singularities is arbitrary, and can be changed by moving cuts across neighboring singularities (therefore passing them through neighboring cuts).  Upon dragging the monodromy paths, one finds that neighboring monodromies can get interchanged or conjugated by each other, giving an action of the braid group on $n$ strands on $\SL(2,\Z)$ matrices.  This braid group action is described in more detail in \cite{DeWolfe:1998zf, DeWolfe:1998eu, Argyres:2015gha}; we will not need it here.  Thus it is really only the set of $\{M_i\}$ monodromies up to an overall $\SL(2,\Z)$ conjugation and braid equivalences which is a topological invariant of a CB geometry.

Since the set of EM duality monodromies, $\{M_i\}$, up to overall $\SL(2,\Z)$ conjugation and braid equivalences is an algebraically complicated topological invariant, it is useful to have some simpler, though less detailed, invariant of the SK structure.  One is simply the unordered list, $\{ K_1 , \ldots, K_A\}$ of Kodaira types $K_i$ of the singular fibers $X_{u_i}$.  This is equivalent to an unordered list of $\SL(2,\Z)$ conjugacy classes of monodromies around each singularity.  We call this list the deformation pattern of the CB geometry \cite{Argyres:2015ffa, Argyres:2015gha}, since if we have an ``initial" singular fiber $X_0$ of Kodaira type $K_0$, then upon (mass) deformation the deformation pattern shows how it splits into a collection of singular fibers $X_{u_i}$ of Kodaira types $K_i$.  The possible deformation patterns for all rank-1 SCFTs were classified in \cite{Argyres:2015ffa, Argyres:2015gha} and are shown in table \ref{tablist}.  It turns out that the specification of an initial singularity type and its deformation pattern completely determines the SK geometry of the family of mass-deformed CBs. 


\paragraph{Summary.}

So far we have proposed that the flavor charge lattice is given by middle-homology cycles of the total space $\tX$ of the smooth fibers over the CB.  More precisely, we proposed that $\L_F = H_2(\tX,\Z)/\ker Z$, i.e., it is the lattice of middle homology cycles which are not annihilated by the central charge map \eqref{ZOm}, and a metric on this lattice is given by the intersection form on the middle homology.  We have also seen how the topology of the SK structure on a rank-1 CB is characterized by the ordered set of EM duality monodromies, $\{M_1,\ldots,M_A\}$, around the singularities (modulo equivalences having to do with choice of a basis of cycles on the CB and modulo an overall $\SL(2,\Z)$ conjugation).  We will now show how to compute $H_2(\tX,\Z)/\ker Z$ and its intersection form from this topological data, and in doing so we will also show that we precisely recover the F-theory neutral string web picture.

\section{String webs from the topology of the SK structure}\label{sec3}

\subsection{Middle homology cycles as string webs}

We now show that the part of the compact middle homology of $\tX$ which is not annihilated by the central charge map is determined by the EM duality monodromy data on the CB.

Since $\tX$ is a fiber bundle, we can understand its homology in terms of a basis of cycles which can be deformed to be in the CB base (i.e., is homologous to a cycle which is contained in a section of $\tX$), in a fiber, $X_u$, or is extended in both the base and fiber directions.  A cycle homologous to one extended only in the base or in a fiber is annihilated by the $Z$ map \eqref{ZOm} since $\Om$ has one leg in the base and one in the fiber.
 
So to understand $H_2(\tX)/\ker Z$ we need focus only on cycles which extend in both the base and fiber directions.  Locally in the base these are homotopic to segments on the base times a cycle in the fiber (continuously dragged along the segment).  Since $H_1(X_u) = \Z^2$, each such segment is labeled by a pair of integers $\bla=(c\ d)$ describing the 1-homology cycle in the fiber, $[\l]=\bla ([\a]\ [\b])^T = c[\a]+d[\b]$, in terms of a canonical basis, $\{\a,\b\}$, of $H_1(X_u)$.  As $u$ varies along the segment, $\bla$ remains constant, i.e., the fiber cycle is continuously dragged using the natural (Gauss-Manin) connection on the local system of $H_1(X_u)$ over the CB.  Thus, locally on the CB these 2-chains in $\tX$ look like $(p,q)$-string segments, and we will simply call them $\bla$-segments.

An orientation on a $\bla$-segment is determined by a choice of orientation on the line segment in the CB base along with an orientation of the cycle in the fiber.  Changing $\bla \to -\bla$ reverses the orientation of the fiber cycle.  Thus the orientation of the $\bla$-segment is unchanged upon simultaneously reversing the orientation of the line segment in the CB and the sign of $\bla$.

Note that just like $(p,q)$-strings, multiple $\bla_i$-segments, $i=1,\ldots,k$ can join at a point $u\in$ CB to give a locally closed 2-chain as long as $\sum_i \bla_i =0$.  In this way segments can be joined together to form webs of segments, much as in figure \ref{web}.

\begin{figure}[tbp]
\centering
\begin{tikzpicture}[decoration={markings,
mark=at position .5 with {\arrow{>}}}]
\begin{scope}[scale=0.5]
\node at (4.5,-0.5) {(a)};
\fill[color=black!10] (0,0) -- (2,4) -- (10,4) -- (8,0) -- cycle;
\node[R] (or) at (5,2) {};
\node[Blf] (or1) at (5,6) {};
\draw[very thick,blue] (or) -- (9,2);
\draw[thick,red,dashed] (or) -- (5,8);
\draw[very thick,blue,fill=blue!25] (or1) arc (180:90:0.5) 
-- (9,6.5) arc (90:270:0.2 and 0.5)
-- (5.5,5.5) arc (270:180:0.5);
\node[Blf] (or1) at (5,6) {};
\draw[very thick,blue,fill=blue!15] (9,6) ellipse (0.2 and 0.5);
\draw[thick,dashed] (9,2) -- (9,5.5);
\end{scope}
\begin{scope}[scale=0.5,xshift=10cm]
\node at (4.5,-0.5) {(b)};
\fill[color=black!10] (0,0) -- (2,4) -- (10,4) -- (8,0) -- cycle;
\node[R] (or) at (5,2) {};
\draw[very thick,blue] (or) ellipse (2 and 1);
\draw[thick,red,dashed] (or) -- (5,4.5);
\draw[very thick,blue,fill=blue!25] (7.5,6) arc (0:180:2.5 and 1.5) arc (180:335:0.5) arc (135:45:2.2) arc (210:360:0.5);
\draw[very thick,blue,fill=blue!25] (7.5,6) arc (360:180:2.5 and 1.5) arc (180:45:0.5) arc (225:315:2.35) arc (135:0:0.5);
\draw[very thick,blue!50] (3,6) circle (0.5);
\draw[very thick,blue!50] (7,6) circle (0.5);
\draw[thick,dashed] (7,2) -- (7,5.5);
\draw[thick,dashed] (3,2) -- (3,5.5);
\draw[thick,red,dashed] (5,5.75) -- (5,8);
\end{scope}
\begin{scope}[scale=0.5,xshift=20cm]
\node at (4.5,-0.5) {(c)};
\fill[color=black!10] (0,0) -- (2,4) -- (10,4) -- (8,0) -- cycle;
\node[R] (or) at (5,2) {};
\draw[very thick,blue] (or) ellipse (2 and 1);
\draw[thick,red,dashed] (or) -- (5,4.5);
\draw[very thick,blue,fill=blue!25] (7.5,6) arc (0:180:2.5 and 1.5) arc (180:335:0.5) arc (135:45:2.2) arc (210:360:0.5);
\draw[very thick,blue,fill=blue!25] (7.5,6) arc (360:180:2.5 and 1.5) arc (180:45:0.5) arc (225:315:2.35) arc (135:0:0.5);
\draw[very thick,blue!50] (3,6) circle (0.5);
\draw[very thick,blue!50] (7,6) circle (0.5);
\draw[thick,dashed] (7,2) -- (7,5.5);
\draw[thick,dashed] (3,2) -- (3,5.5);
\draw[thick,red,dashed] (5,5.75) -- (5,8);
\draw[very thick,blue,fill=blue!25, draw opacity=1.0] (7,6) arc (180:90:0.3 and 0.5) 
-- (9,6.5) arc (90:270:0.2 and 0.5)
-- (7.3,5.5) arc (270:180:0.3 and 0.5);
\draw[very thick,blue,fill=blue!15] (9,6) ellipse (0.2 and 0.5);
\draw[very thick,blue] (7,2) -- (9,2);
\draw[thick,dashed] (9,2) -- (9,5.5);
\end{scope}
\end{tikzpicture}
\caption{Ways for $\bla$-segments to close: (a) by pinching off, (b) by circling back on itself, or (c) by circling back and joining another $\bla$-segment.}\label{worms}
\end{figure}
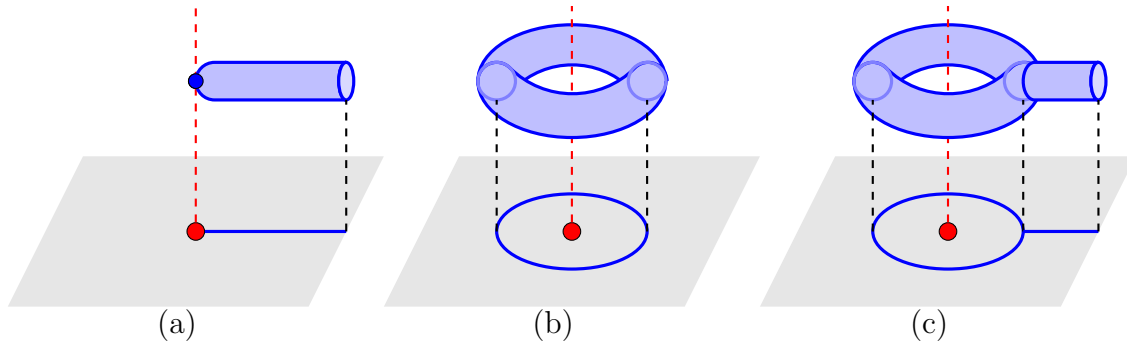

$\bla$-segments are not closed 2-chains: if the segment has boundary points $u_0$ and $u_1$ in the CB, then the $\bla$-segment has boundary $\l\times\{u_0\} - \l\times\{u_1\}$.   There are basically two ways to ``close off" an end of a $\bla$-segment at $u_0\in$ CB to make it into a cycle.  One is to put $u_0$ at a point in the CB where the $\l$ cycle pinches off.  This is illustrated in figure \ref{worms} (a), and can only happen at one of the metric singularities on the CB where the fiber degenerates.  But cycles of $X_u$ only pinch off when the fiber is singular, and the singular fibers are removed from $\tX$, as discussed above.  So ends like in figure \ref{worms} (a) do not contribute to the compact middle homology of $\tX$.  

The other way to close off an end of a $\bla$-segment is for two ends of the segment (or web of segments) to coincide, with the two segments carrying opposite $\bla$ charges.  Examples are shown in figure \ref{worms} (b) and (c).  If the cycle in the CB in figures \ref{worms} (b) or (c) can be deformed to a point without encountering a metric singularity, then the cycles are homologous to zero.  Thus the only way of making homologically non-trivial 2-cycles from $\bla$-segments involves segments which  forms a closed loop around at least one singularity on the CB.     

If a 2-cycle, $\a$, has a component whose segments only loop around a single singularity, $u=u_i\in$ CB, as in figure \ref{worms} (b), then the cycle can be shrunk homologously to an arbitrarily small neighborhood of the singularity.  Because the holomorphic 2-form $\Om$ is regular on $X$, including at the singular fibers, we have $\lim_{\a\to X_{u_i}}\int_\a \Om = 0$.  But since $\Om$ is closed, its period depends only on the homology class of $\a$, so we learn that $\int_\a \Om=0$, i.e., $\a\in\ker Z$.

Thus the only middle homology classes of $\tX$ that may not be in the kernel of the central charge map are those with representatives given by a web of $\bla$-segments which ends on two or more singularities on the CB as in figure \ref{worms} (c).  Note that if the web of $\bla$-segments loops around more than one CB singularity, it can always be deformed into a set of simple loops each enclosing a single singularity.  (The fundamental group of an $A$-punctured plane is freely generated on $A$ generators which can be taken to be the homotopy classes of simple loops around each puncture.)  Thus the most general such homology class can be deformed into a tree of $\bla$-segments with ends given by small loops around each CB singularity.  Furthermore, by the argument of the last paragraph, each small loop can be shrunk to be arbitrarily close to its singular fiber without affecting the value of the central charge (its $\Om$ period).  

This characterizes all the elements of $H_2(\tX,\Z)/\ker Z$ as tree-like webs of oriented intervals on the CB ending at the singularities $u_i\in$ CB with a 1-cycle with in a given fiber homology class, $\bla_i\in H_1(X_u,\Z)$, suspended above each such interval.  At an intersection of $A$ intervals in the base labeled by $\{i_1,\ldots,i_A\}$ (say all oriented into the intersection), the condition that the suspended 1-cycles join to form a closed 2-cycle is the condition that
\begin{align}\label{neutrality}
\sum_{a=1}^A \bla_{i_a}=0 .
\end{align}
We will call this condition ``charge neutrality" of the web.  This is precisely the string web picture shown in figure \ref{web}.   The set of such webs forms a lattice with elements labeled by charges $\bla_i$ of their intervals ending on CB singularities, subject to one overall charge neutrality condition \eqref{neutrality}.

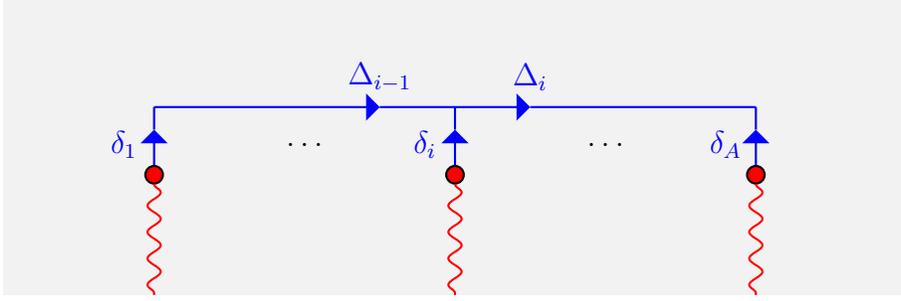
\begin{figure}[htbp]\centering
\begin{tikzpicture}[thick]
\fill[color=black!05] (-6,-2) rectangle (6,2);
\node[R] (M1) at (-4,-0.4) {};
\draw[red,snake it] (M1) -- (-4,-2);
\draw (-2,0) node {$\ldots$};
\node[R] (Mi) at (0,-0.4) {};
\draw[red,snake it] (Mi) -- (0,-2);
\draw (2,0) node {$\ldots$};
\node[R] (Mn) at (4,-0.4) {};
\draw[red,snake it] (Mn) -- (4,-2);
\draw[blue,->] (-4,.5) -- (-1,.5);
\draw[blue,->] (-1,.5) -- (1,.5);
\draw[blue] (1,.5) -- (4,.5);
\draw[blue,->] (M1) -- (-4,.2);
\draw[blue] (-4,.2) -- (-4,.5);
\draw[blue,->] (Mi) -- (0,.2);
\draw[blue] (0,.2) -- (0,.5);
\draw[blue,->] (Mn) -- (4,.2);
\draw[blue] (4,.2) -- (4,.5);
\node[blue] at (-1,.9) {$\D_{i-1}$};
\node[blue] at (1,.9) {$\D_{i}$};
\node[blue] at (-4.4,0) {$\d_1$};
\node[blue] at (-.4,0) {$\d_i$};
\node[blue] at (3.6,0) {$\d_A$};
\end{tikzpicture}
\caption{A CB with $A$ singularities (red dots) and their associated cuts (red wavy lines).  Any class in $H_2(\tX,\Z)/\ker Z$ can be represented as a tree-like ``string web" shown in blue with ends on the singularities and each edge carrying a class of $H_1(X_0,\Z)$.}
\label{web2}
\end{figure}

Thus any class in $H_2(\tX,\Z)/\ker Z$ can be represented as in figure \ref{web2}.  Here the 2-cycle in $\tX$ has been deformed into the blue ``web"  on the cut CB with a class of $H_1(X_0,\Z)$ suspended over each edge of the web.   Thus the edge emanating from the $i$th singularity on the CB carries a fiber 1-cycle $\d_i$ with homology class $\bde_i \in H_1(X_0,\Z)$, and
\begin{align}\label{Deltadef}
\bDe_i = \sum_{k=1}^{i} \bde_k 
\end{align}
by the charge neutrality of junctions of $\bla$-segments, i.e., the requirement that fiber classes suspended over the segments join together to form a 2-cycle in the total space.   The charge neutrality condition \eqref{neutrality} is thus
\begin{align}\label{neutrality2}
0= \bDe_A := \sum_{k=1}^A \bde_k.
\end{align}

Here $H_1(X_0,\Z)$ are the homology classes of 1-cycles of the fiber $X_0$ over a chosen base point $0\in$ CB.  $H_1(X_u,\Z)$ for any $u$ in the CB minus the cuts is then identified with that $H_1(X_u,\Z)$ by continuity (i.e., by transporting the $H_1(X_0,\Z)$ linear system via the Gauss-Manin connection).  

\subsection{Boundary charges at singular fibers}
\label{sec3.2}

Thus to compute the lattice $\L_F = H_2(\tX,\Z)/\ker Z$ we need only determine the possible fiber homology classes, $\bde\in H_1(X_u,\Z) \simeq \Z^2$, of a $\bla$-segment ending at a singular fiber.  We will call this set of allowed charges for a segment ending at a given singularity the ``boundary charges" for that singular fiber.  We will now show that the set of boundary charges is fixed by the EM duality monodromy, $M\in\SL(2,\Z)$, associated to that CB singularity relative to a choice of base point and a basis of monodromy cycles as illustrated in figure \ref{CBtopol}.

\begin{figure}[htbp]\centering
\begin{tikzpicture}[thick]
\fill[color=black!05] (-6,-2) rectangle (-1,2);
\fill[color=black!05] (0.5,-2) rectangle (6.5,2);
\node[R] (Ml) at (-3.5,-0.4) {};
\draw[red,snake it] (Ml) -- (-3.5,-2);
\draw (3.5,0) node {$M_i$};
\node[R] (Mr) at (3.5,-0.4) {};
\draw[red,snake it] (Mr) -- (3.5,-2);
\draw[blue,->] (-6,1) -- (-4.5,1);
\draw[blue,->] (-4.5,1) -- (-2.2,1);
\draw[blue] (-2.2,1) -- (-1,1);
\draw[blue] (4.7,1) -- (6.5,1);
\draw[blue,->] (2.5,1) -- (4.7,1);
\draw[blue,->] (0.5,1) -- (2.5,1);
\draw[blue,->] (Ml) -- (-3.5,.4);
\draw[blue] (-3.5,.4) -- (-3.5,1);
\draw[blue,->] (3.5,.2) -- (3.5,.7);
\draw[blue] (3.5,.7) -- (3.5,1);
\draw[blue,->] (3.5,.2) -- (2.8,.2) -- (2.8,-.5);
\draw[blue,->] (2.8,-.5) -- (2.8,-1) -- (4.2,-1) -- (4.2,-.3);
\draw[blue] (4.2,-.3) -- (4.2,.2) -- (3.5,.2);
\node[blue] at (-4.5,1.4) {$\D_{i-1}$};
\node[blue] at (-2.5,1.4) {$\D_{i}$};
\node[blue] at (2.5,1.4) {$\D_{i-1}$};
\node[blue] at (4.5,1.4) {$\D_{i}$};
\node[blue] at (-3.9,.2) {$\d_i$};
\node[blue] at (3.1,.5) {$\d_i$};
\node[blue] at (2.5,-.5) {$\g_i$};
\node[blue,font=\small] at (5.4,-.5) {$\g_i'=M_i\circ\g_i$};
\node at (-0.25,0) {\Large{$\simeq$}};
\end{tikzpicture}
\caption{A convenient homology equivalence in $H_2(\tX,\Z)/\text{ker} Z$.}
\label{prong}
\end{figure}
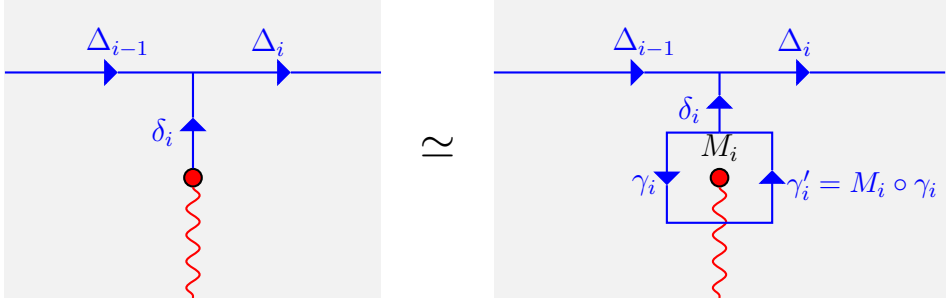

As argued above, in the compact middle homology of the non-singular but non-compact total space, $\tX$, the only way a $\bla$-segment can end at a singularity is by a configuration shown in figure \ref{worms} (c).  This is illustrated in figure \ref{prong}.

Here $M_i$ is the $\SL(2,\Z)$ monodromy associated with the counterclockwise cycle around the $i$th singularity labeled by the $\bg_i$ and $\bg_i'$ classes.  Thus, if $\bg_i\in H_1(X_0,\Z)$ is the fiber homology class associated to the left side of the cycle, then $\bg_i'$ is the image of $\bg_i$ under the $M_i$.  The argument of the previous subsection shows that the homology class in $H_2(\tX,\Z)/\ker Z$ of the string web depends only on $\bde_i = \bg_i'-\bg_i$.  There are infinitely many different $\bg_i$ and $\bg_i'$ which give the same $\bde_i$, but they all differ by classes in $H_2(\tX,\Z)$ which can be shrunk to the singular fiber and so are annihilated by the central charge map.

The fiber homology, $\bde$, that can be carried away from a CB singularity with monodromy $M$ is then (see figure \ref{prong})
\begin{align}\label{deltamonod}
\bde = \bg' - \bg = \bg (M-I) 
\end{align}
for any $\bg \in \Z^2$.  Thus the lattice of charges of $\bla$-segments ending at a CB singularity with monodromy $M$ is the image of $M-I$ acting on the $\Z^2$ EM charge lattice.  Thus, the conclusion is that the fiber homology class $\bde_i$ at the $i$th singularity satisfies
\begin{align}\label{prong2}
\bde_i \in {\rm Im}(M_i-I) \quad \text{acting on the lattice of fiber 1-cycles}.
\end{align}
This is our desired characterization of the boundary charges associated to a given singularity.

Note that allowing multiple windings around a singularity does not enlarge the set of boundary charges.  This follows since any multiple winding is homologous to a sum of single windings.  Algebraically, the boundary charges from an $s$-fold winding is in the image of $M^s-I$, but $M^s-I = (M^{s-1} + \cdots + I) (M - I)$ so its image is in the image of $M-I$.

Since $M-I$ is an integral linear map, its image is a sublattice of the total (rank-2) lattice of EM charges.  If $M$ does not have an eigenvalue $+1$ then $\det(M-I)\neq0$, and the rank of the image sublattice is also 2.  Since $M\in \SL(2,\Z)$, if one of its eigenvalues is $+1$, then both are.  In that case the rank of the image sublattice is smaller than 2.  In case $M=I$ (the trivial monodromy) then the image sublattice has rank 0; i.e., the boundary charge is zero in this case, reflecting the fact that string webs have non-trivial ends only at singularities.  In other cases where both eigenvalues of $M$ are $+1$ but $M$ is a non-trivial Jordan block, the rank of the image lattice is 1.  

We will now compute explicitly the image lattices for the monodromies appearing in the deformation patterns listed in table \ref{tablist} in the introduction.  Inspection of that list shows that only singularities of Kodaira types $I_n$, $I_n^*$, $III^*$, and $IV^*$ enter into the deformation patterns of CFTs or of IR free theories.  (As explained at length in \cite{Argyres:2015ffa}, this follows from a combination of the safely irrelevant conjecture on $\cN=2$ renormalization group flows and the conservation of the order of vanishing of the discriminant under relevant deformations.)  We examine each of these four classes of singularities in turn.

\subsubsection{$I_n$ singularities}

From table \ref{kodaira}, a singularity of $I_n$ type is associated to a monodromy in the conjugacy class of $T^n= \left(\bsm 1 & n\\ 0&1\esm\right)\in\SL(2,\Z)$.  The image of
$T^n-I$ is then simply
\begin{align}\label{}
\bla (T^n-I) = \bpm c & d\epm \bpm 0 & n \\ 0 & 0 \epm
= \bpm 0 & nc \epm
\qquad \text{for all $(c\ d)\in\Z^2$.}
\end{align}
It thus spans a rank-1 sublattice of the lattice of EM charges, namely the sublattice generated by electric charge $n$, i.e., by the charge $\bLa_n :=(0\ n)$.  Thus the fiber 1-cycles that can be carried away from a singularity with $T^n$ monodromy are $\bde$ with $\bde = c \bLa_n$ for any $c\in\Z$.

As is apparent from table \ref{tablist}, theories exist with deformation patterns containing multiple $I_n$-type singularities.  So even if an EM duality basis were chosen so that one of the $I_n$ monodromies is $T^n$, the other $I_{n_i}$ monodromies will in general be $T^{n_i}$ conjugated by some $g_i\in\SL(2,\Z)$.

So consider an $I_n$ singularity with monodromy $T^n$ conjugated by some $g := \left(\bsm A&B\\C&D\esm\right)\in\SL(2,\Z)$,
\begin{align}\label{Tnmonod}
M(I_n,g) = g^{-1} T^n g
= \bpm 1 + n CD & n D^2 \\ -n C^2 & 1 - n CD \epm .
\end{align}
Note that $\det(g)=1$ implies that $\gcd(C,D)=1$.  Then $\text{Im}[M(I_n,g)-I] = (\text{Im}[T^n-I])g$ which is therefore the sublattice of $\Z^2$ generated by $(0\ n)g = n(C\ D)$.  Denote this rank-1 lattice by
\begin{align}\label{InH1classes}
\G(I_n,g) = \{ \bla = c(nC\ nD), c\in\Z \}.
\end{align}
Then the possible 1-cycles $\d(I_n,g)$ that a segment ending at the $I_n$ singularity can carry are
\begin{align}\label{InBC}
\bde(I_n,g) \in \G(I_n,g).
\end{align}
Thus the set of pairs $(C,D)\in\Z^2$ with $\gcd(C,D)=1$  
modulo their overall sign, $(C,D) \sim (-C,-D)$, parameterizes the possible inequivalent lattices of boundary charges for an $I_n$ singularity.  

This gives a picture of the flavor lattice for theories with only $I_{n_i}$ singularities as string webs with string charges for segments ending on an $I_{n_i}$ singularity labeled by an integer multiple of a pair of integers $(C_i,D_i)$ with $\gcd(C_i,D_i)=n_i$.  This reproduces the string web picture found from the F-theory construction reviewed in section \ref{Fthry} and illustrated in figure \ref{web}, as long as all $n_i=1$.  Indeed, the $(p,q)$-7branes of IIB string theory have $\gcd(p,q)=1$ and axio-dilaton monodromy conjugate to $T^1$.   

Thus for SCFTs with deformation pattern including $I_n$ singularities with $n>1$, we have generalized the F-theory picture simply by allowing $(p,q)$-7branes with $\gcd(p,q)=n>1$.  These would seemingly correspond in IIB string theory to bound states of $n$ parallel same-charge 7branes; but no such IIB brane configurations are known.  So such SCFTs seemingly cannot be realized in string theory via the F-theory construction.  (Many of these theories can be realized in M-theory or class-$\cS$ constructions \cite{Argyres:2016xua}.)

From the field theory perspective, the interpretation of the $I_n$ singularities is very simple \cite{Argyres:2015ffa}:  an $I_n$ type singularity in the deformation pattern corresponds to an IR free $\U(1)$ theory with a single massless hypermultiplet with charge vector $\bL$ with invariant charge $Q=\sqrt{n}$.   Thus in some EM duality basis, its EM charge vector is 
\begin{align}\label{InEMcharges}
\bL_n := \sqrt n (C\ D)
\qquad \text{with}\qquad \gcd(C,D)=1,
\end{align}
for some integers $(C,D)$.  The monodromy \eqref{Tnmonod} then follows from the one-loop beta function of the $\U(1)$ theory with a hypermultiplet of charge $\bL$.

The factor of $\sqrt n$ difference between the EM charges \eqref{InEMcharges} and the fiber $H_1$ classes \eqref{InH1classes} is closely related, but not identical, to the factor of $\sqrt P$ from the polarization of the Dirac pairing entering in the EM charges, reviewed in section \ref{rk1SK}.  In case $n$ is square-free, then an $I_n$ deformation pattern singularity is only consistent with a non-principally polarized Dirac pairing with $P=n$.  Thus, for example, a deformation pattern with an $I_2$ singularity (as in lines 9, 16, and 24 of table \ref{tablist}) is only consistent with a theory with $P=2$.  But if $n$ is not square-free, then multiple different choices of polarization are consistent.  For instance, a deformation pattern with an $I_4$ singularity (as in lines 2, 3, 13, 20, and 25 of table \ref{tablist}) is consistent with either $P=4$ or $P=1$ since $\sqrt P$ is an integer.   Note, however, that the value of $P$ is fixed for the whole theory, so all $I_n$ singularities in the deformation pattern must be consistent with the choice of $P$.  (This is just a reflection of the consistency of the low energy theory with the Dirac quantization condition, as discussed in \cite{Argyres:2015ffa}.)  For example, since all the $I_4$ examples in table \ref{tablist} also have $I_1$ singularities, the only consistent polarization for them is the principal one, $P=1$.

\subsubsection{$I_n^*$ singularities}

An $I_n^*$ singularity has an associated monodromy in the conjugacy class of 
\begin{align}\label{In*monod}
M(I_n^*) = -T^n= \left(\bsm -1 & -n\\ 
\ph{-}0 & -1\esm\right) .
\end{align}
Since every deformation pattern in table \ref{tablist} contains at most a single $I_n^*$ singularity, we can always choose an EM duality basis so that the $I_n^*$ monodromy is in a particular conjugacy class, which we will choose to be that shown in \eqref{In*monod}.

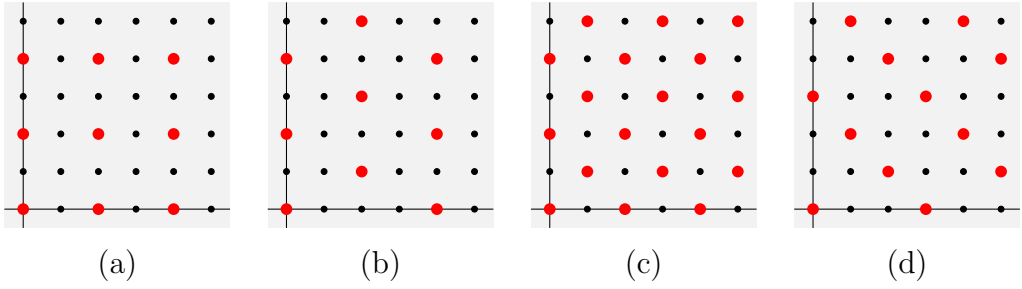
\begin{figure}[htbp]\centering
\begin{tikzpicture}
\begin{scope}[scale=0.5,xshift=0cm]
\node at (2.5,-1.5) {(a)};
\fill[color=black!05] (-.5,-.5) rectangle (5.5,5.5);
\draw[black] (-.5,0) -- (5.5,0);
\draw[black] (0,-.5) -- (0,5.5);
\node[rc] at (0,0) {}; \node[bbcs] at (1,0) {}; 
\node[rc] at (2,0) {}; \node[bbcs] at (3,0) {}; 
\node[rc] at (4,0) {}; \node[bbcs] at (5,0) {};
\node[bbcs] at (0,1) {}; \node[bbcs] at (1,1) {}; 
\node[bbcs] at (2,1) {}; \node[bbcs] at (3,1) {}; 
\node[bbcs] at (4,1) {}; \node[bbcs] at (5,1) {};
\node[rc] at (0,2) {}; \node[bbcs] at (1,2) {}; 
\node[rc] at (2,2) {}; \node[bbcs] at (3,2) {}; 
\node[rc] at (4,2) {}; \node[bbcs] at (5,2) {};
\node[bbcs] at (0,3) {}; \node[bbcs] at (1,3) {}; 
\node[bbcs] at (2,3) {}; \node[bbcs] at (3,3) {}; 
\node[bbcs] at (4,3) {}; \node[bbcs] at (5,3) {};
\node[rc] at (0,4) {}; \node[bbcs] at (1,4) {}; 
\node[rc] at (2,4) {}; \node[bbcs] at (3,4) {}; 
\node[rc] at (4,4) {}; \node[bbcs] at (5,4) {};
\node[bbcs] at (0,5) {}; \node[bbcs] at (1,5) {}; 
\node[bbcs] at (2,5) {}; \node[bbcs] at (3,5) {}; 
\node[bbcs] at (4,5) {}; \node[bbcs] at (5,5) {};
\end{scope}
\begin{scope}[scale=0.5,xshift=7cm]
\node at (2.5,-1.5) {(b)};
\fill[color=black!05] (-.5,-.5) rectangle (5.5,5.5);
\draw[black] (-.5,0) -- (5.5,0);
\draw[black] (0,-.5) -- (0,5.5);
\node[rc] at (0,0) {}; \node[bbcs] at (1,0) {}; 
\node[bbcs] at (2,0) {}; \node[bbcs] at (3,0) {}; 
\node[rc] at (4,0) {}; \node[bbcs] at (5,0) {};
\node[bbcs] at (0,1) {}; \node[bbcs] at (1,1) {}; 
\node[rc] at (2,1) {}; \node[bbcs] at (3,1) {}; 
\node[bbcs] at (4,1) {}; \node[bbcs] at (5,1) {};
\node[rc] at (0,2) {}; \node[bbcs] at (1,2) {}; 
\node[bbcs] at (2,2) {}; \node[bbcs] at (3,2) {}; 
\node[rc] at (4,2) {}; \node[bbcs] at (5,2) {};
\node[bbcs] at (0,3) {}; \node[bbcs] at (1,3) {}; 
\node[rc] at (2,3) {}; \node[bbcs] at (3,3) {}; 
\node[bbcs] at (4,3) {}; \node[bbcs] at (5,3) {};
\node[rc] at (0,4) {}; \node[bbcs] at (1,4) {}; 
\node[bbcs] at (2,4) {}; \node[bbcs] at (3,4) {}; 
\node[rc] at (4,4) {}; \node[bbcs] at (5,4) {};
\node[bbcs] at (0,5) {}; \node[bbcs] at (1,5) {}; 
\node[rc] at (2,5) {}; \node[bbcs] at (3,5) {}; 
\node[bbcs] at (4,5) {}; \node[bbcs] at (5,5) {};
\end{scope}
\begin{scope}[scale=0.5,xshift=14cm]
\node at (2.5,-1.5) {(c)};
\fill[color=black!05] (-.5,-.5) rectangle (5.5,5.5);
\draw[black] (-.5,0) -- (5.5,0);
\draw[black] (0,-.5) -- (0,5.5);
\node[rc] at (0,0) {}; \node[bbcs] at (1,0) {}; 
\node[rc] at (2,0) {}; \node[bbcs] at (3,0) {}; 
\node[rc] at (4,0) {}; \node[bbcs] at (5,0) {};
\node[bbcs] at (0,1) {}; \node[rc] at (1,1) {}; 
\node[bbcs] at (2,1) {}; \node[rc] at (3,1) {}; 
\node[bbcs] at (4,1) {}; \node[rc] at (5,1) {};
\node[rc] at (0,2) {}; \node[bbcs] at (1,2) {}; 
\node[rc] at (2,2) {}; \node[bbcs] at (3,2) {}; 
\node[rc] at (4,2) {}; \node[bbcs] at (5,2) {};
\node[bbcs] at (0,3) {}; \node[rc] at (1,3) {}; 
\node[bbcs] at (2,3) {}; \node[rc] at (3,3) {}; 
\node[bbcs] at (4,3) {}; \node[rc] at (5,3) {};
\node[rc] at (0,4) {}; \node[bbcs] at (1,4) {}; 
\node[rc] at (2,4) {}; \node[bbcs] at (3,4) {}; 
\node[rc] at (4,4) {}; \node[bbcs] at (5,4) {};
\node[bbcs] at (0,5) {}; \node[rc] at (1,5) {}; 
\node[bbcs] at (2,5) {}; \node[rc] at (3,5) {}; 
\node[bbcs] at (4,5) {}; \node[rc] at (5,5) {};
\end{scope}
\begin{scope}[scale=0.5,xshift=21cm]
\node at (2.5,-1.5) {(d)};
\fill[color=black!05] (-.5,-.5) rectangle (5.5,5.5);
\draw[black] (-.5,0) -- (5.5,0);
\draw[black] (0,-.5) -- (0,5.5);
\node[rc] at (0,0) {}; \node[bbcs] at (1,0) {}; 
\node[bbcs] at (2,0) {}; \node[rc] at (3,0) {}; 
\node[bbcs] at (4,0) {}; \node[bbcs] at (5,0) {};
\node[bbcs] at (0,1) {}; \node[bbcs] at (1,1) {}; 
\node[rc] at (2,1) {}; \node[bbcs] at (3,1) {}; 
\node[bbcs] at (4,1) {}; \node[rc] at (5,1) {};
\node[bbcs] at (0,2) {}; \node[rc] at (1,2) {}; 
\node[bbcs] at (2,2) {}; \node[bbcs] at (3,2) {}; 
\node[rc] at (4,2) {}; \node[bbcs] at (5,2) {};
\node[rc] at (0,3) {}; \node[bbcs] at (1,3) {}; 
\node[bbcs] at (2,3) {}; \node[rc] at (3,3) {}; 
\node[bbcs] at (4,3) {}; \node[bbcs] at (5,3) {};
\node[bbcs] at (0,4) {}; \node[bbcs] at (1,4) {}; 
\node[rc] at (2,4) {}; \node[bbcs] at (3,4) {}; 
\node[bbcs] at (4,4) {}; \node[rc] at (5,4) {};
\node[bbcs] at (0,5) {}; \node[rc] at (1,5) {}; 
\node[bbcs] at (2,5) {}; \node[bbcs] at (3,5) {}; 
\node[rc] at (4,5) {}; \node[bbcs] at (5,5) {};
\end{scope}
\end{tikzpicture}
\caption{Allowed sublattices, $\G(M)\subset\Z^2$, of fiber homology cycles for a $\bla$-segment ending at a singularity with monodromy $M$ for (a) $\G(I_n^*)$, $n$ even, (b) $\G(I_n^*)$, $n$ odd, (c) $\G(III^*)$, and (d) $\G(IV^*)$.}
\label{pronglattices}
\end{figure}

The image of $M(I_n^*)-I$ is then
\begin{align}\label{}
\bla(-T^n-I) = \bpm c & d\epm 
\bpms[r] -2 & -n \\ 0 & -2 \epms
= \bpm -2c & \ -nc{-}2d \epm
\qquad \text{for all $(c\ d)\in\Z^2$.}
\end{align}
This is a rank-2 sublattice of the lattice of EM charges, $\G(I_n^*)$, which it is easy to see is 
\begin{align}\label{In*H1classes}
\G(I_n^*) = 
\begin{cases}
\{ \bla = (2c\ \ 2d) \ \text{for}\ c,d\in\Z \} & 
\text{if $n$ is even,} \\
\{ \bla = (2c\ \ c{+}2d) \ \text{for}\ c,d\in\Z \} & 
\text{if $n$ is odd.} 
\end{cases}
\end{align}
These are the two index-4 sublattices shown in figure \ref{pronglattices} (a) and (b).  Thus the possible 1-cycles $\d(I_n^*)$ that a segment ending at an $I_n^*$ singularity can carry are
\begin{align}\label{In*BC}
\bde(I^*_n) \in \G(I_n^*).
\end{align}

\subsubsection{$III^*$ singularity}

A $III^*$ singularity has an associated monodromy in the conjugacy class of 
\begin{align}\label{III*monod}
M(III^*) = S= \left(\bsm 0 & -1\\ 1 & \ph{-}0\esm\right) .
\end{align}
Since every deformation pattern in table \ref{tablist} contains at most a single $III^*$ singularity, we can always choose an EM duality basis so that the $III^*$ monodromy is in a particular conjugacy class, which we will choose to be that shown in \eqref{III*monod}.

The image of $M(III^*)-I$ is then
\begin{align}\label{}
\bla(S-I) = \bpm c & d\epm 
\bpms[r] -1 & -1 \\ 1 & -1 \epms
= \bpm -c{+}d & \ -c{-}d \epm
\qquad \text{for all $(c\ d)\in\Z^2$.}
\end{align}
This rank-2 sublattice, $\G(III^*)$, can also be expressed as
\begin{align}\label{III*H1classes}
\G(III^*) = 
\{ \bla = (c\ \ 2d{+}c) \ \text{for}\ c,d\in\Z \} .
\end{align}
This is the index-2 sublattice shown in figure \ref{pronglattices} (c).  Thus the possible 1-cycles $\d(III^*)$ that a segment ending at a $III^*$ singularity can carry are
\begin{align}\label{III*BC}
\bde(III^*) \in \G(III^*).
\end{align}

\subsubsection{$IV^*$ singularity}

A $IV^*$ singularity has an associated monodromy in the conjugacy class of 
\begin{align}\label{IV*monod}
M(IV^*) = (ST)^2= \left(\bsm -1 & -1\\ 
\ph{-}1 & \ph{0}0\esm\right) .
\end{align}
Since every deformation pattern in table \ref{tablist} contains at most a single $IV^*$ singularity, we can always choose an EM duality basis so that the $IV^*$ monodromy is in a particular conjugacy class, which we will choose to be that shown in \eqref{IV*monod}.

The image of $M(IV^*)-I$ is then
\begin{align}\label{}
\bla((ST)^2-I) 
= \bpm c & d\epm \bpms[r] -2 & -1 \\ 1 & -1 \epms
= \bpm -2c{+}d & \ -c{-}d \epm
\qquad \text{for all $(c\ d)\in\Z^2$.}
\end{align}
This rank-2 sublattice, $\G(IV^*)$, can also be expressed as
\begin{align}\label{IV*H1classes}
\G(IV^*) = 
\{ \bla = (3c{-}d\ \ d) \ \text{for}\ c,d\in\Z \} .
\end{align}
This is the index-3 sublattice shown in figure \ref{pronglattices} (d).  Thus the possible 1-cycles $\d(IV^*)$ that a segment ending at a $IV^*$ singularity can carry are
\begin{align}\label{IV*BC}
\bde(IV^*) \in \G(IV^*).
\end{align}

\subsubsection{Summary}

We have thus effectively computed $\L_F = H_2(\tX,\Z)/\text{ker}Z$ in terms of topological data of the SK structure of the CB.  In particular, given an ordered set of $\SL(2,\Z)$ monodromies, $(M_1,\ldots,M_A)$, corresponding to $A$ singularities on the CB ordered by a choice of ``cuts" as described above, the general element of $\L_F$ is given by the ordered set $(\bde_1, \ldots, \bde_A)$ where each $\bde_i$ is an element of a rank-1 sublattice of $\Z^2$ given by \eqref{InBC} if $[M_i]=I_n$, or a rank-2 sublattice of $\Z^2$ given by \eqref{In*BC}, \eqref{III*BC}, or \eqref{IV*BC} if $[M_i]=I_n^*$, $III^*$, or $IV^*$, respectively.  
Furthermore, the $\bde_i$ must satisfy the linear charge neutrality condition \eqref{neutrality2} in $\Z^2$.  We will call a general element of this lattice a ``string web" and denote it by the formal sum
\begin{align}\label{webdef}
\cW = \sum_{i=1}^A \bde_i \, S_i, \qquad
\text{with} \quad \bde_i \in \G(M_i)\subset \Z^2 \quad \text{such that} \quad
\sum_{i=1}^A \bde_i =0,
\end{align} 
where $S_i$ is a placeholder symbol denoting the $i$th singularity.  This expression denotes the 2-cycle in $\tX$ depicted in figure \ref{web2}.

\section{Computing the middle homology intersection form}\label{sec4}

\begin{figure}[htbp]\centering
\begin{tikzpicture}[thick]
\fill[color=black!05] (-0.5,-2) rectangle (7.5,4);
\draw (3.5,-0.1) node {$M_i$};
\node[R] (Mr) at (3.5,-0.4) {};
\draw[red,snake it] (Mr) -- (3.5,-2);
\draw[blue] (4.7,2) -- (6,2);
\draw[blue,->] (2.5,2) -- (4.7,2);
\draw[blue,->] (1,2) -- (2.5,2);
\draw[blue] (-.1,2) node {$\bullet$};
\draw[blue] (.25,2) node {$\bullet$};
\draw[blue] (.6,2) node {$\bullet$};
\draw[blue] (6.4,2) node {$\bullet$};
\draw[blue] (6.75,2) node {$\bullet$};
\draw[blue] (7.1,2) node {$\bullet$};
\draw[blue,->] (3.5,.2) -- (3.5,.7);
\draw[blue] (3.5,.7) -- (3.5,2);
\draw[blue,->] (3.5,.2) -- (2.8,.2) -- (2.8,-.5);
\draw[blue,->] (2.8,-.5) -- (2.8,-1) -- (4.2,-1) -- (4.2,-.3);
\draw[blue] (4.2,-.3) -- (4.2,.2) -- (3.5,.2);
\node[blue] at (2.5,2.4) {$\D_{i-1}$};
\node[blue] at (4.5,2.4) {$\D_{i}$};
\node[blue] at (3.1,.6) {$\d_i$};
\draw[green!50!black] (4.7,3) -- (6,3);
\draw[green!50!black,->] (2.5,3) -- (4.7,3);
\draw[green!50!black,->] (1,3) -- (2.5,3);
\draw[green!50!black] (-.1,3) node {$\bullet$};
\draw[green!50!black] (.25,3) node {$\bullet$};
\draw[green!50!black] (.6,3) node {$\bullet$};
\draw[green!50!black] (6.4,3) node {$\bullet$};
\draw[green!50!black] (6.75,3) node {$\bullet$};
\draw[green!50!black] (7.1,3) node {$\bullet$};
\draw[green!50!black,->] (4,1) -- (4,1.5);
\draw[green!50!black] (4,1.5) -- (4,3);
\draw[green!50!black,->] (4,1) -- (2.,1) -- (2.,-.5);
\draw[green!50!black,->] (2.,-.5) -- (2.,-1.5) -- (5,-1.5) -- (5,-.3);
\draw[green!50!black] (5,-.3) -- (5,1) -- (4,1);
\node[green!50!black] at (2.5,3.4) {$\til\D_{i-1}$};
\node[green!50!black] at (4.5,3.4) {$\til\D_{i}$};
\node[green!50!black] at (4.5,1.4) {$\til\d_i$};
\node[green!50!black] at (1.5,-.5) {$\til\g_i$};
\node[green!50!black,font=\small] at (6.3,-.5) {$\til\g_i'=M_i\circ\g_i$};
\node[bl] at (3.5,1) {};
\node[bl] at (4,2) {};
\draw (3.2,1.3) node {$P_i$};
\draw (3.7,2.3) node {$Q_i$};
\end{tikzpicture}
\caption{A convenient configuration for two string webs, $\cW$ (in blue) and $\tcW$ (in green), showing their intersection points, $P_i$ and $Q_i$, on the CB in the vicinity of the $i$th singularity, and with their edges labeled by 1-homology cycles of the fiber.}
\label{webint}
\end{figure}
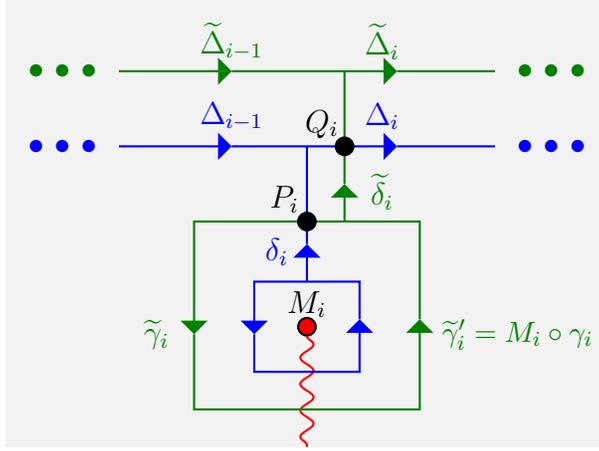

It is now straightforward to compute the intersection form of two middle homology cycles of $\tX$ of the form \eqref{webdef}.   By using the homology equivalence of figure \ref{prong}, any two string webs, $\cW$ and $\tcW$, given as in figure \ref{web2} can be deformed into a configuration, shown in figure \ref{webint}, where they only intersect at two points in the vicinity of each singularity.

Their intersection number is thus the sum over the points $P_i$ and $Q_i$ in figure \ref{webint} of the product of the intersection numbers of the webs in the CB with the intersection numbers of the homology cycles in the fibers at those points.  By inspection of figure \ref{webint} we thus find
\begin{align}\label{webinner}
\cW \cdot \tcW 
= \sum_{i=1}^A 
\left( \d_i \cdot \til\g_i  + \D_i \cdot \til\d_i \right)
=
\sum_{i=1}^A \d_i \cdot \til\g_i 
+ \!\!\!\!\! \sum_{1\le j\le i\le A} \!\!\!\!\! \d_j \cdot \til\d_i ,
\end{align}
where we used \eqref{Deltadef} in the last equality.  We have chosen the overall sign of the intersection form so that the inner product is positive definite on the string web lattice.  (Although the total space, $\til X$, of the CB has a complex structure and therefore a preferred orientation and sign of its middle homology intersection form, our purely topological construction is insensitive to this complex structure.)  

This inner product, coming as it does from an intersection form of 2-cycles, is symmetric.  The expression \eqref{webinner} is not manifestly symmetric, but its symmetric nature can be checked:
\begin{align}\label{symmcheck}
\cW \cdot \tcW - \tcW\cdot\cW &=
\sum_i (\d_i \cdot \til\g_i - \til\d_i\cdot \g_i)
+ \sum_{j\le i} (\d_j \cdot \til\d_i - \til\d_j\cdot\d_i)
\nonumber\\
&= 
\sum_i (\d_i \cdot \til\g_i + \g_i \cdot \til\d_i )
+ \sum_{j\le i} (\d_j \cdot \til\d_i + \d_i \cdot \til\d_j)
\nonumber\\
&= 
\sum_i ( \d_i \cdot \til\g_i + \g_i \cdot \til\d_i 
+ \d_i \cdot \til\d_i )
+ \Bigl( \sum_j\d_j \Bigr)\cdot \Bigl(\sum_i \til\d_i \Bigr)
\\
&= 
\sum_i (\d_i + \g_i)\cdot (\til\d_i + \til\g_i) 
- \sum_i  \g_i \cdot \til\g_i
\nonumber\\
&= 
\sum_i (M_i \circ \g'_i )\cdot (M_i \circ \til\g'_i) 
- \sum_i  \g_i \cdot \til\g_i
=0 .\nonumber
\end{align}
In the second line we used the antisymmetry of the intersection product of 1-cycles on the fibers and in the fourth line the charge neutrality condition \eqref{neutrality2}.  In the last line we used the relation \eqref{deltamonod} between $\d_i$ and $\g_i$ that 
\begin{align}\label{dmonod2}
\d_i = M_i\circ\g_i - \g_i ,
\end{align}
where $M\circ \g$ denotes the action of the monodromy on the 1-cycle $\g$ as in \eqref{monodact}.  Finally, in the last step we used the fact that the intersection product on 1-cycles is $\SL(2,\Z)$-invariant.  It is useful, in the light of \eqref{symmcheck} to write the web inner product \eqref{webinner} in a manifestly symmetric manner $\cW\cdot\tcW = \frac12(\cW\cdot\tcW + \tcW\cdot\cW)$ giving
\begin{align}\label{webinner1}
\cW \cdot \tcW 
&=
\frac12 \sum_{i=1}^A 
\left( \d_i \cdot \til\g_i - \g_i \cdot \til\d_i  \right)
- \frac12 \sum_{i,j=1}^A 
\left( \th_{ij}-\th_{ji} \right)\d_i \cdot \til\d_j ,
\end{align}
where we have defined
\begin{align}\label{thdef}
\th_{ij} := \begin{cases}
1 & \text{if}\ i>j \\
0 & \text{if}\ i\le j
\end{cases} .
\end{align}

Less obvious is the claim that this inner product is well-defined on $H_2(\tX,\Z)/\text{ker} Z$.  The issue is that the elements of $H_2(\tX,\Z)/\text{ker} Z$ are given by the string webs \eqref{webdef} which are defined in terms of the $\d_i$ alone, whereas the inner product in \eqref{webinner1} seems to depend on the $\g_i$, which, recall, were a priori ambiguous cycles associated with 2-chains in $\text{ker} Z$; see figure \ref{prong}.  Nevertheless, \eqref{webinner1} is, in fact, independent of the possible choices of the $\g_i$.  Indeed, two $\g_i$ satisfying \eqref{dmonod2} can differ by a cycle, $\k_i$, in the kernel of $M_i-I$, i.e., satisfying $\k_i= M_i\circ\k_i$.  Then the $\til\k_i$-dependent terms in \eqref{webinner1} vanish:
\begin{align}\label{}
\d_i\cdot\til\k_i = (M_i \circ \g_i - \g_i) \cdot \til\k_i
= (M_i \circ \g_i) \cdot (M_i \circ \til\k_i) 
- \g_i \cdot \til\k_i
=0 .
\end{align}
Thus \eqref{webinner1} can be re-expressed solely in terms of the $\d_i$'s which characterize the string webs.

In the cases where $[M_i] \in \{ I_n^*, III^*, IV^*\}$, this is easy to do since $M_i-I$ has zero kernel so is invertible.   Thus the $\g_i$-dependent terms in \eqref{webinner1} can be written in terms of the $\d_i$'s using $\g_i = (M_i-I)^{-1}\circ \d_i$, for $[M_i] \in \{ I_n^*, III^*, IV^*\}$.  If the image of $M-I$ is an index-$d$ sublattice, i.e.,
\begin{align}\label{}
d = \det(I-M) = 2 - \Tr(M),
\end{align}
then $(M-I)^{-1} = d^{-1}\left( [\Tr(M)-1] I - M \right)$, as follows from the identity $M^2-\Tr(M) M+I = 0$ valid for $M\in\SL(2)$.  Using this, the symmetrized form of the $\d_i\cdot\til\g_i$ terms appearing in \eqref{webinner1} then simplifies to
\begin{align}\label{MmIi1}
\frac12 \left( \d_i \cdot \til\g_i - \g_i \cdot \til\d_i \right)
&= 
\frac1{2d_i} \bde_i (M_i S - S M_i^T) \til\bde_i^T .
\end{align}
Here we are using a notation, introduced earlier, where $\d :=\bde {[\a]\choose [\b]}$,  $\{\a, \b\}$ are a canonical basis of fiber homology 1-cycles satisfying \eqref{abbasis}, $\bde$ is row a vector in $\Z^2$, and $\d\cdot\g  =\bde S \bg^T$.  From the explicit form of the monodromies chosen in \eqref{In*monod}, \eqref{III*monod}, and \eqref{IV*monod}, it follows that
\begin{align}\label{MmIi2}
\frac1{2d} (MS-SM^T) = 
\begin{cases}
-\frac{n}{4} \left( \bsm 1 & 0 \\ 0 & 0\esm \right)
& \text{if}\ M=M(I_n^*) , \\
-\frac{1}{2} \left( \bsm 1 & 0 \\ 0 & 1\esm \right) 
& \text{if}\ M=M(III^*) , \\
-\frac{1}{6} \left( \bsm \ph{-}2 & -1 \\ -1& \ph{-}2\esm \right) 
& \text{if}\ M=M(IV^*) .
\end{cases}
\end{align}

For a singularity with monodromy $M(I_n,g)$ of $I_n$ type, $M-I$ has a 1-dimensional kernel, so is not invertible.  But it is easy to solve \eqref{dmonod2} for $\g$ given an allowed $\d$ for such a singularity given by \eqref{InBC}
\begin{align}\label{Inbde}
\bde = c n (C\ D)
\quad c\in\Z \quad\text{and}\quad \gcd(C,D)=1,
\end{align}
from the explicit expression \eqref{Tnmonod} for $M(I_n,g)$ to find
\begin{align}\label{Inbg}
\bg = c (A\ B)
\quad\text{for any}\quad A,B\in\Z
\quad\text{such that}\quad AD-BC=1.
\end{align} 
Infinitely many integer pairs $(A\ B)$ exist since $\gcd(C,D)=1$, but the argument from 2 paragraphs earlier shows that the $\d\cdot\til\g'$ terms in \eqref{webinner1} are, in fact, independent of the choice of $(A\ B)$.  Indeed, one computes explicitly
\begin{align}\label{Ingterm}
\frac12 \left( \d \cdot \til\g - \g \cdot \til\d \right)
&= 
\frac12 \left( \bde S \til\bg^T - \bg S \til\bde^T \right)
= nc\til c.
\end{align}
Note that $nc$ is an $\SL(2,\Z)$-invariant\footnote{As is clear from \eqref{Tnmonod}, the pair $(C,D)$ characterizing the conjugated $M(I_n,g)$ monodromy is only defined up to overall sign, $(C,D) \sim (-C,-D)$, and thus the sign of $\gcd(\bde)$ is not unambiguously defined.  But the \emph{relative sign} between $\gcd(\bde)$ and $\gcd(\bde')$ for any two fiber 1-cycles carried away from a singularity with given $M(I_n,g)$ monodromy is unambiguous, and is significant.} of $\d$ since
\begin{align}\label{nc=gcd}
nc = \gcd(ncC,ncD) := \gcd(\bde),
\end{align}
the invariant charge introduced in \eqref{QtoQ}.  In this notation \eqref{Ingterm} reads
\begin{align}\label{Ingterm2}
\frac12 \left( \d \cdot \til\g - \g \cdot \til\d \right)
= \frac{1}{n} \gcd(\bde) \gcd(\til\bde) .
\end{align}

Putting \eqref{MmIi1}, \eqref{MmIi2}, and \eqref{Ingterm2} into \eqref{webinner1} gives
\begin{align}\label{webinner2}
\cW\cdot \tcW =
\sum_{i=1}^A G_i(\bde_i,\til\bde_i) 
- \frac12 \sum_{i,j=1}^A (\th_{ij} - \th_{ji}) \, 
\bde_i S \til\bde_j^T ,
\end{align}
with
\begin{align}\label{webinner3}
G_i(\bde_i,\til\bde_i) = \begin{cases}
+ \frac{1}{n} \gcd(\bde_i) \gcd(\til\bde_i) &
\text{if}\  M_i = M(I_n,g) \\
- \frac{n}{4} \bde_i 
\left(\bsm \ph{-} 1& \ph{-}0\\ \ph{-}0& \ph{-}0\esm\right) 
\; \til\bde_i^T &
\text{if}\  M_i = M(I_n^*) \\
- \frac12 \bde_i 
\left(\bsm \ph{-}1& \ph{-}0\\ \ph{-}0& \ph{-}1\esm\right) 
\; \til\bde_i^T &
\text{if}\  M_i = M(III^*) \\
- \frac16 \bde_i 
\left(\bsm \ph{-}2&-1\\-1&\ph{-}2\esm\right) 
\; \til\bde_i^T &
\text{if}\  M_i = M(IV^*) 
\end{cases} ,
\end{align}
Equations \eqref{webinner2} and \eqref{webinner3} are our main result, expressing the inner product on the middle homology cycles in terms of the ``web charges'' $\cW = \sum_i \bde_i S_i$ and the ordered monodromies $M_i=M(S_i)$ around each singularity.

\paragraph{Comparison to the F-theory result.}  

Recall from \eqref{zmdef} that to each homology cycle $\bde\in\Z^2$ there is a corresponding EM charge vector $\bd = \sqrt{P} \, \bde$, where $\sqrt{P}$ for $P\in\N$ is the charge quantization unit for the theory in question.  From \eqref{QtoQ} and \eqref{Dcip}, the EM duality invariant charge, $Q_\bd$, of $\bd$, and the Dirac inner product of two charges satisfy
\begin{align}\label{}
Q_\bd &= \sqrt{P} \, \gcd(\bde) ,&
& \text{and}&
\vev{\bd,\til\bd} = P\, \bde S \til\bde^T .
\end{align}
Rewriting \eqref{webinner2} in terms of EM charge vectors $\bd_i$, and in the case where all the singularities have monodromies of type $M_i = M(I_{n_i},g_i)$, the string web inner product becomes
\begin{align}\label{Fswip}
\cW\cdot \tcW =
\sum_{i=1}^A \frac{1}{P n_i} Q_{\bd_i} Q_{\til\bd_i} 
- \frac{1}{2P} \sum_{i,j=1}^A (\th_{ij} - \th_{ji}) \, 
\vev{\bd_i, \til\bd_j} .
\end{align}
If we specialize to the F-theory case, where $P=n_i=1$ (i.e., all singularities are of $I_1$-type and the unit of charge quantization is therefore necessarily $P=1$), this coincides precisely with the string web inner product found there \cite{Gaberdiel:1997ud, Gaberdiel:1998mv, DeWolfe:1998zf}.


\section{String web lattices of rank-1 SCFTs}\label{sec5}

We now apply the formalism developed in the last two sections to compute the string web lattices and their metrics for all the rank-1 SCFT CB geometries listed in table \ref{tablist}.   

In more detail, our method is as follows.
\begin{enumerate}
\item We describe the string web lattice by parameterizing all sets $(\bde_1,\ldots,\bde_A)$ of $A$ integer pairs characterizing the fiber 1-homology cycles of webs ending at $A$ singularities in the CB.  This set has to satisfy the charge neutrality condition \eqref{webdef}, as well as the boundary conditions \eqref{InBC}, \eqref{In*BC}, \eqref{III*BC}, and \eqref{IV*BC} depending on the monodromy associated to each singularity.  This gives a finite-rank lattice with metric given by \eqref{webinner2} and \eqref{webinner3}.  
\item In order to determine the shape of this lattice, we then search for the shortest and next-to-shortest non-zero elements in this lattice, and use them to construct a lattice basis in which the metric has a recognizable form.  
\item Finally, we compare these lattice shapes to those expected to occur as root lattices of semisimple Lie algebras.   These latter are orthogonal sums of one of four exceptional lattices, $E_8$, $E_7$, $E_6$, and $F_4$, corresponding to the simple Lie algebras of the same names, and one of three infinite series which we name $\CUB_r$, $\FCC_r$, and $\HEX_r$.  Here the subscript refers to the rank of the lattice and the names stand for (higher-dimensional analogs of) cubic, face-centered cubic, and hexagonal lattices, respectively.  These are defined in appendix \ref{appA}, where the relations with simple Lie algebra root lattices,
\begin{align}\label{}
\L(A_r) &= \HEX_r, \qquad
\L(G_2) = \HEX_2,\nonumber\\
\L(B_r) &= \CUB_r, \qquad
\L(C_r) = \FCC_r, \qquad
\L(D_r) = \FCC_r,
\end{align}
and the low-rank equivalences,
\begin{align}\label{}
\CUB_1 &=\HEX_1=\FCC_1,\qquad
\CUB_2=\FCC_2,\qquad
\FCC_3=\HEX_3,
\end{align}
are explained.
\end{enumerate}
Of these steps, the second is the most difficult, and can be computationally intensive for high rank lattices.  Luckily, we will only have to compute at rank five or lower.

\subsection{CB geometries of $I_1$ type}

We start by first looking at the SCFTs appearing in the blue-shaded rows in table \ref{tablist} whose deformation patterns contain singularities only of $I_1$ type.  These are also called the maximal deformations, and are the ones which are realized as configurations of $(p,q)$-7branes in F-theory, as reviewed in section \ref{Fthry} above.  As we showed at length in the last sections, our middle-homology construction of the lattices coincides in these cases with the F-theory construction.  The neutral string web lattices in these cases have been studied in detail in \cite{Gaberdiel:1997ud, Gaberdiel:1998mv, DeWolfe:1998zf}, and give the results recorded in the corresponding entries in table \ref{tablist}.


\subsection{CB geometries of $I_{n>1}$ type}\label{sec5.2}

A more complicated set of CB geometries are those with deformation patterns all of $I_n$ type with at least one entry not $I_1$.  These are numbers 2, 3, 13, 20, 24, and 25 in table \ref{tablist}.  Thus the topology of the SK structure of the CBs in these cases is given by an ordered list $\{ M(I_{n_1},g_1), \ldots, M(I_{n_A},g_A) \}$ of monodromies around $A$ singularities.

Recall that a general $I_n$ type monodromy, $M(I_n,g)$ is characterized by $n$ and a pair, $(C,D)$, of coprime integers.  We will use the shorthand notation
\begin{align}\label{Insh}
\bpm C\\D \epm_{\!\!n} := M(I_n,g),
\end{align}
and drop the $n$ subscript when $n=1$.  Furthermore, if there are repeated adjacent entries in the list of monodromies, we will indicate them by a superscript.

The boundary condition \eqref{InBC} for the allowed fiber homology cycle for a web ending on an $M(I_n,g)$ singularity states that the corresponding $\bde = a n (C\ D)$ for some integer $a$.  Thus for this set of geometries we can specify a string web (homology 2-cycle) just by listing the integers $a_i$, that is, the web is given by $\cW(a_i) := \sum_i \bde_i S_i$ with $\bde_i = a_i n_i (C_i \ D_i)$.  Allowed string webs (closed homology 2-cycles) are those which satisfy the charge neutrality condition \eqref{webdef}.

We then construct the neutral string web lattice, find a convenient basis of short webs and compute the lattice metric to find the results listed in table \ref{typeIn}.
For instance, the first line in table \ref{typeIn} refers to CB geometry \#2 in table \ref{tablist} corresponding to the deformation pattern $II^* \to \{ I_4, I_1^6 \}$.   As constructed in \cite{Argyres:2015gha}, the CB topological data is realized by the ordered set of monodromies
\begin{align}
\left\{
\bpm1\\1\epm\;
\bpm-1\\2\epm\;
\bpm0\\1\epm^4\;
\bpm0\\1\epm_{\!\!4}\;
\right\} ,
\end{align}
in the notation defined in \eqref{Insh}.  
Numbering the integers $a_i$ describing the string web boundary conditions from left to right, 
the charge neutrality condition \eqref{webdef} becomes
\begin{align}
\bpms[l]
a_1-a_2 \\ 
a_1 + 2 a_2 + a_3 + a_4 + a_5 + a_6 +4 a_7
\epms = 0 ,
\end{align}
giving two independent conditions, and thus a rank-5 lattice.
By scanning over the webs in the lattice and computing their inner products from \eqref{webinner2}, we obtain the basis $\{\cW_1,\cW_2,\cW_3,\cW_4,\cW_5\}$ given in the second column of table \ref{typeIn}.   The lattice metric is given by the third column of table \ref{typeIn}.

\begin{table}[h]
\centering
$\begin{array}{lcc}
\text{CB SK topology}\, & \text{basis of webs} & \qquad\quad\text{metric}\ \ g=\qquad\quad\,\\[2mm]
\hline\\
\#2: \left\{
{(\bsm1\\1\esm)}\;
{(\bsm-1\\2\esm)}\;
{(\bsm0\\1\esm)}^4\;
{(\bsm0\\1\esm)}_{4}
\right\} \ \ \, &
{\small 
\begin{array}{c|rrrrrrr}
 & a_1 & a_2 & a_3 & a_4 & a_5 & a_6 & a_7 \\
\hline
\cW_1 & 0 & 0 & 0 & 0 & -1 & 1 & 0\\
\cW_2 & 0 & 0 & 0 & -1 & 1 & 0 & 0 \\
\cW_3 & -3 & -3 & 1 & 2 & 1 & 1 & 1 \\
\cW_4 & 3 & 3 & -2 & -1 & -1 & -1 & -1 \\
\cW_5 & 2 & 2 & 0 & 0 & -1 & -1 & -1
\end{array}
}
& {\small 
\bpms[r]
2 & -1 & 0 & 0 & 0 \\ 
-1 & 2 & -1 & 0 & -1\\
0 & -1 & 2 & -1 & 0\\
0 & 0 & -1 & 2 & 0 \\ 
0 & -1 & 0 & 0 & 2
\epms 
}
\\[16mm]
\#3: \left\{
{(\bsm1\\1\esm)}\;
{(\bsm-1\\2\esm)}\;
{(\bsm0\\1\esm)}_{4}^2
\right\} &
\begin{array}{c|rrrr}
& a_1 & a_2 & a_3 & a_4 \\
\hline
\cW_1 & -4 & -4 & 1 & 2\\
\cW_2 & -4 & -4 & 2 & 1
\end{array}
& \bpm4&0\\0&4\epm 
\\[9mm]
\#13: \left\{
{(\bsm1\\1\esm)}\;
{(\bsm-1\\2\esm)}\;
{(\bsm0\\1\esm)}^3\;
{(\bsm0 \\1 \esm)}_{4}
\right\} &
\begin{array}{c|rrrrrr}
& a_1 & a_2 & a_3 & a_4 & a_5 & a_6 \\
\hline
\cW_1 & 2 & 2 & 0 &-1 &-1 &-1  \\
\cW_2 & 0 & 0 &-1 & 0 & 1 & 0  \\
\cW_3 &-2 &-2 & 2 & 0 & 0 & 1  \\
\cW_4 &-1 &-1 & 1 & 1 & 1 & 0
\end{array}
& \bpms[r]
\ 2 & -1 & \ \,0 & \ 0 \\ 
-1 & \ \,2 & -2 & \ 0\\
\ 0 & -2 & \ \,4 & \ 0\\
\ 0 & \ \,0 & \ \,0 & \ 2
\epms \\[14mm]
\#20: \left\{
{(\bsm1\\1\esm)}\;
{(\bsm-1\\2\esm)}\;
{(\bsm0\\1\esm)}^2\;
{(\bsm0\\1\esm)}_4
\right\} &
\begin{array}{c|rrrrr}
& a_1 & a_2 & a_3 & a_4 & a_5 \\
\hline
\cW_1 &-2 &-2 & 1 & 1 & 1 \\
\cW_2 & 0 & 0 & -1 & 1 & 0  \\
\cW_3 &-1 &-1 & -1 & 0 & 1
\end{array}
& \bpm
2 & 0 & 1 \\ 0 & 2 & 1 \\ 1 & 1 & 4
\epm 
\\[10mm]
\#24: \left\{
{(\bsm1 \\0\esm)}_2\;
{(\bsm-1\\1\esm)}_2\;
{(\bsm0 \\1\esm)}_2
\right\} &
\begin{array}{c|rrr}
& a_1 & a_2 & a_3 \\
\hline
\cW & -1 & -1 & 1 
\end{array}
& \bpm2\epm 
\\[6mm]
\#25: \left\{
{(\bsm1 \\1\esm)}\;
{(\bsm-1\\1\esm)}\;
{(\bsm0 \\1\esm)}_4
\right\} &
\begin{array}{c|rrr}
& a_1 & a_2 & a_3 \\
\hline
\cW & -2 & -2 & 1 
\end{array}
& \bpm4\epm 
\end{array}$
\caption{Topologies of the special K\"ahler (SK) structure of the CB, string web lattice bases, and metrics for CB geometries with deformation patterns having only $I_n$ singularities with at least one $n>1$.  The numbers on the left refer to table \ref{tablist}.}
\label{typeIn}
\end{table}

From these lattice metrics we easily read off the lattice types for most of the entries: the metric for \#2 is the $D_5$ Cartan matrix, and so, as discussed in appendix \ref{appA}, the lattice is an $\FCC_5$ lattice; \#3 is a $\CUB_2$ lattice; \#13 is the Killing form of $C_3 \oplus A_1$ thus giving an $\FCC_3 \oplus \CUB_1$ lattice; \#24 and \#25 are both $\CUB_1$ lattices.  These lattice types are shown in the rightmost column of table \ref{tablist}.

The only entry for which the lattice is not semisimple is
\#20.  In this case $\{\cW_1,\cW_2\}$ span a $\CUB_2$ lattice, while $\{\cW_3\}$ spans (trivially) a $\CUB_1$ lattice which is not orthogonal to the $\CUB_2$ lattice.  We therefore characterize this lattice as the non-orthogonal sum $\CUB_2 \niplus \CUB_1$ in table \ref{tablist}.  The $\CUB_1$ sublattice is interpreted as the flavor charge lattice of a $U_1$ flavor factor under which the $C_2$ flavor root vectors (spanning the $\CUB_2$ sublattice) are charged.  This is an example of flavor charge mixing with $\U(1)$ factors mentioned in the introduction.  It will be explored in more detail in section \ref{sec6.1}.


\subsection{SCFTs of $III^*$ and $IV^*$ type}

CB geometries \# 9, 10, 11, and 18 in table \ref{tablist} have deformation patterns which include $III^*$ or $IV^*$ type singularities.  The monodromies associated with these deformation patterns, found in \cite{Argyres:2015gha}, can be conjugated to put the $III^*$ and $IV^*$ monodromies into the canonical forms \eqref{III*monod} and \eqref{IV*monod}, giving 
\begin{align}\label{3*4*monods}
&\text{CB \#9:} & 
II^* &\to \{ IV^*, I_2 \} \simeq \{ (ST)^2, T^2 \}, 
\nonumber\\
&\text{CB \#10:} & 
II^* &\to \{ IV^*, I_1^2 \} \simeq \{ (ST)^2, T, T \}, 
\\
&\text{CB \#11:} & 
II^* &\to \{ III^*, I_1 \} \simeq \{ S, T \}, 
\nonumber\\
&\text{CB \#18:} & 
III^* &\to \{ IV^*, I_1 \} \simeq \{ (ST)^2, T \}.
\nonumber
\end{align} 
Each string web is labeled by an ordered set of pairs of integers $\bde_i=(p_i, q_i)$, each denoting the fiber homology cycles of each segment of the web ending at each singularity as in \eqref{webdef}.  These $\bde$'s must satisfy the boundary conditions \eqref{InBC}, \eqref{III*BC}, and \eqref{IV*BC} appropriate to the monodromies given in \eqref{3*4*monods}, as well as the overall charge neutrality condition.  The resulting lattice metrics are shown in table \ref{tab4}.  The corresponding lattice types are listed in table \ref{tablist}.

\begin{table}[h]
\centering
$\begin{array}{lcc}
\qquad\text{CB SK topology}\qquad\, & \text{basis of webs} & \qquad\text{metric}\ \ g=\qquad\,\\[2mm]
\hline\\
\#9: \{ M(IV^*), (\bsm0\\1\esm)_2 \} &
\begin{array}{c|cc}
 & \bde_1 & \bde_2 \\
\hline
\cW & (0,6) & (0,-6)
\end{array}
& \bpm6 \epm \\[8mm]
\#10: \{ M(IV^*) , (\bsm0\\1\esm)^2 \} &
\begin{array}{c|ccc}
 & \bde_1 & \bde_2 & \bde_3 \\
\hline
\cW_1 & (0,-3) & (0,2) & (0,1) \\
\cW_2 & (0,3) & (0,-1) & (0,-2)
\end{array}
& \bpms[r] 2&-1\\-1&2\epms \\[11mm]
\#11: \{ M(III^*) , (\bsm0\\1\esm) \} &
\begin{array}{c|cc}
 & \bde_1 & \bde_2 \\
\hline
\cW & (0,2) & (0,-2)
\end{array}
& \bpm2 \epm \\[8mm]
\#18: \{ M(IV^*) , (\bsm0\\1\esm) \} &
\begin{array}{c|cc}
 & \bde_1 & \bde_2\\
\hline
\cW & (0,3) & (0,-3)
\end{array}
& \bpm6 \epm 
\end{array}$
\caption{Topologies, string web lattice bases, and metrics for CB geometries with deformation patterns containing a $III^*$ or $IV^*$ singularity.}
\label{tab4}
\end{table}

\subsection{SCFTs of $I_n^*$ type}

The remaining scale-invariant CB geometries listed in table \ref{tablist} all have deformation patterns involving $I_{n_i}$ singularities and a single $I_n^*$ singularity.  It was shown in \cite{Argyres:2015gha} that the monodromy data of of most of these CB geometries can be written uniformly as
\begin{align}\label{In*defos}
\left\{
\bpm1 \\2 \epm,\;
M(I^*_n),\;
\bpm0 \\1 \epm^{m-n}
\right\},  \quad 0 \leq n \leq m, \quad
\text{and}\quad m \in \{1,2,3\},
\end{align}
using the column vector notation for the $M(I_n,g)$ monodromies defined in \eqref{Insh}.  These correspond to deformations patterns $\{I_1^{m+1-n},I_n^*\}$ of the $II^*$, $III^*$, and $IV^*$ singularities for $n=3$, 2, and 1, respectively.  These are entries \# 4, 5, 7, 8, 14, 15, 17, 21, and 22 in table \ref{tablist}.  
There are two remaining CB geometries with $I_n^*$ singularities in their deformation patterns.  In these cases we calculate their monodromy data from their Seiberg-Witten curves given in the appendices to \cite{Argyres:2015gha}, and find
\begin{align}\label{In*defos2}
&\text{CB \#6}:& 
II^* &\to \{I_1^*,I_3\}&
&\Leftrightarrow&
&\left\{ M[I^*_1],\;
\bpm1\\0\epm_{\!\!3} \right\}
\nn\\
&\text{CB \#16}:&
III^* &\to \{I_1^*,I_2\}& 
&\Leftrightarrow& 
&\left\{ M[I^*_1],\;
\bpm1\\0\epm_{\!\!2} \right\} .
\end{align}
Using a similar notation as in table \ref{tab4}, namely, $\bde_i$ denote the fiber homology cycles corresponding to the $m-n+2$ monodromies in \eqref{In*defos} or the two monodromies in \eqref{In*defos2}, numbered from left to right, we find the lattice bases and their metrics shown in table \ref{tab4}.   The resulting lattice is easily recognized from the metric in most cases, and is shown in table \ref{tablist}.  The cases where it may not be obvious are:  \#7, where the metric is the Cartan matrix of a $C_3$ root lattice, and so is of type $\FCC_3$; \#8, where the metric is the Cartan matrix of an $F_4$ root lattice; and \#15, where the metric describes a $\CUB_1 \oplus\CUB_1$ lattice which is not a $\CUB_2$ lattice, since the lattice basis vectors are of different lengths in the two $\CUB_1$ sub-lattices.

\begin{table}[htbp]
\centering
\small
$\begin{array}{lcc}
\text{CB SK topology} & \text{basis of webs} & \text{metric} \ \ g=\\ \hline
\#4: \{ (\bsm1\\2\esm) , M(I_0^*) , (\bsm0\\1\esm)^3 \} &
\begin{array}{c|ccccc} 
&\bde_1&\bde_2&\bde_3&\bde_4&\bde_5\\ \hline
\cW_1&(0,0)&(0,0)&(0,0)&(0,-1)&(0,1)\\
\cW_2&(0,0)&(0,0)&(0,-1)&(0,1)&(0,0)\\
\cW_3&(0,0)&(0,-2)&(2,0)&(0,0)&(0,0)\\
\cW_4&(-2,-4)&(2,4)&(0,0)&(0,0)&(0,0)
\end{array}
& \bpms[r] 2&-1&0&0\\-1&2&-2&0\\0&-2&4&-2
\\0&0&-2&4\epms\\
\#5: \{ (\bsm1\\2\esm) , M(I_1^*) , (\bsm0\\1\esm)^2 \} &
\begin{array}{c|cccc}
& \bde_1 & \bde_2 & \bde_3 & \bde_4  \\ \hline
\cW_1 & (0,0) & (0,-2) & (0,1) & (0,1)\\
\cW_2 & (-2,-4) & (2,3) & (0,1)& (0,0)\\
\cW_3 & (0,0) & (0,0) & (0,-1) & (0,1)
\end{array}
&  \bpms[r] 2&-1&0\\-1&2&-1\\0&-1&2 \epms\\
\#6: \{ M(I_1^*) , (\bsm1\\0\esm)_3 \} &
\begin{array}{c|cc}
& \bde_1 & \bde_2 \\ \hline
\cW & (12,0) & (-12,0)
\end{array}
&  \bpm 12 \epm\\ 
\#7: \{ (\bsm1\\2\esm) , M(I_2^*) , (\bsm0\\1\esm) \} &
\begin{array}{c|ccc}
 & \bde_1 & \bde_2 & \bde_3 \\ \hline
\cW_1 & (-2,-4) & (2,2) & (0,2) \\
\cW_2 & (2,4) & (-2,-4) & (0,0) 
\end{array}
&  \bpm2&0\\0&2 \epm\\
\#8: \{ (\bsm1\\2\esm) , M(I_3^*) \} &
\begin{array}{c|cc}
& \bde_1 & \bde_2 \\ \hline
\cW & (4,8) & (-4,-8)
\end{array}
& \bpm4 \epm\\
\#14: \{ (\bsm1\\2\esm), M(I_0^*), (\bsm0\\1\esm)^2\} &
\begin{array}{c|cccc}
& \bde_1 & \bde_2 & \bde_3 & \bde_4 \\ \hline
\cW_1 &(0,0) &(0,2) &(0,-1) & (0,-1)\\
\cW_2 &(0,0) &(0,0) &(0,-1) & (0,1)\\
\cW_3 &(2,4) &(-2,-2) &(0,-1) & (0,-1)\\
\end{array}
&  \bpm 2 & 0& 0\\0&2&0\\0&0&2 \epm\\
\#15: \{ (\bsm1\\2\esm) , M(I_1^*) , (\bsm0\\1\esm) \} &
\begin{array}{c|ccc}
& \bde_1 & \bde_2 & \bde_3 \\ \hline
\cW_1 &  (2,4) & (-2,-3) & (0,-1)\\
\cW_2 &  (0,0) & (0,2) & (0,-2)
\end{array}
&  \bpm2&0\\0&4 \epm\\
\#16: \{ M(I_1^*) , (\bsm1\\0\esm)_2 \} &
\begin{array}{c|cc}
& \bde_1 & \bde_2 \\ \hline
\cW & (4,0) & (-4,0)
\end{array}
&  \bpm 4 \epm\\ 
\#17: \{ (\bsm1\\2\esm) , M(I_2^*) \} &
\begin{array}{c|cc}
& \bde_1 & \bde_2 \\ \hline
\cW & (2,4) & (-2,-4)
\end{array}
&  \bpm2 \epm\\ 
\#21: \{ (\bsm1\\2\esm) , M(I_0^*) , (\bsm0\\1\esm) \} &
\begin{array}{c|ccc}
& \bde_1 & \bde_2 & \bde_3 \\ \hline
\cW_1 & (0,0) & (0,-2) & (0,2)\\
\cW_2 & (2,4) & (-2,-2) & (0,-2)
\end{array}
&  \bpms[r] 4&-2\\-2&4 \epms\\
\#22: \{ (\bsm1\\2\esm) , M(I_1^*) \} &
\begin{array}{c|cc}
& \bde_1 & \bde_2 \\ \hline
\cW & (4,8) & (-4,-8)
\end{array}
&  \bpm12 \epm 
\end{array}$
\caption{Topologies, string web lattice bases, and metrics for CB geometries with deformation patterns containing an $I_n^*$ singularity.}
\label{tab5}
\end{table}


\section{Geometries with general IR free configurations}\label{sec6}

Coulomb branches with a single non-scale invariant singularity --- i.e., of $I_n$ or $I_n^*$ type in table \ref{kodaira} --- are not on the same footing as the scale invariant geometries.  These cases correspond to IR free theories with a lagrangian description.  At rank 1, we have two possibilities: a $\U(1)$ or an $\SU(2)$ gauge theory with sufficient massless charged matter.  These two families of theories are identified in the geometrical classification with deformations of the $I_n$ and $I_n^*$ geometries respectively.

Because these are weakly coupled lagrangian theories, and because there are (infinitely) many of them with distinct flavor symmetries, they are an excellent place to test and diagnose our homological (string web) approach to flavor symmetries.  

But these IR free gauge theories also present two unique challenges.  The first is that since these theories are not UV complete, they break down at some UV strong coupling scale (the scale of the ``Landau pole"), $\L$, where typically they are nonunitary unless additional degrees of freedom are included.  This is reflected in the low energy theory on the CB in the fact that without adding any additional structures into the scaling solution for the CB geometries of an $I_n$ or $I^*_n$ singularity, their CB metrics are no longer positive definite for $|u|^{1/\D} \ge \L$.  Because of the necessity for and ambiguity in UV completions of these theories, their IR free field theory descriptions are only effective descriptions valid for energy scales $\m\ll\L$.  Thus their CB geometries are only valid near the singularity, i.e., for $|u|^{1/\D} \ll \L$ and for mass deformation parameters $|m| \ll \L$.  (We give explicit constructions of these CB geometries in terms of Seiberg-Witten curves and one-forms in appendices \ref{appU1} and \ref{appSU2}.)  

The second challenge is presented by the IR free $\SU(2)$ gauge theories.  Because the gauge group is nonabelian, its charged fields come in representations of dimension greater than 1 (unlike in $\U(1)$ gauge theories).  With respect to the low energy unHiggsed $\U(1)$ gauge factor on the CB, a field in such a representation will generally have components with a spectrum of different $\U(1)$ charges.  Upon turning on a mass, $m$, for such a field, the singularity at the origin of the CB will split into a spectrum of correlated singularities on the CB with positions proportional to $m^2$ and inversely proportional to the square of the $\U(1)$ charges of its components: each hypermultiplet in an irreducible representation of the $\SU(2)$ gauge group generally contributes multiple singularities on the CB.

With this multiplication of the number of singularities, the rank of the neutral string web lattice generally far exceeds the rank of the flavor symmetry (which is the number of independent mass parameters).  For instance, a theory with a single hypermultiplet in an isospin-$j$ irrep of $\SU(2)$ has a rank-1 flavor symmetry but has neutral string webs stretching between the singularities associated to this single multiplet spanning a lattice of rank $\lceil j\rceil$.  Thus some additional rule is necessary to select a smaller set of representative string webs which do not overcount within $\SU(2)$ multiplets in this manner.  We present and test such a rule in section \ref{sec6.2}.  Note that in the case of hypermultiplets in $j=1/2$ and $j=1$ irreps of $\SU(2)$, which are the only ones which enter into asympotically free and scale invariant $\SU(2)$ gauge theories, this over-counting problem does not arise since in both cases $\lceil j\rceil =1$.

\subsection{Flavor lattices and deformations of IR free U(1) gauge theories}
\label{sec6.1}

A general IR free $\cN=2$ $\U(1)$ gauge theory is specified by its spectrum of massless charged hypermultiplets.  If there are $H$ such hypermultiplets of charges $\pm q_i$, $i=1,\ldots,H$, then the coefficient of the 1-loop $\b$ function is proportional to \begin{align}\label{u1}
N := \sum_{i=1}^H q_i^2,
\end{align}
and the Coulomb branch geometry has a singularity at the origin with $T^N$ monodromy, so an $I_N$-type singularity; see table \ref{kodaira}.  

The possible $\cN{=}2$-preserving mass deformations of this theory govern the possible deformations of the CB geometry.  The effect of mass deformations follows easily from the $\U(1)$ gauge theory lagrangian.  Write the hypermultiplets in $\cN=1$ superfield notation as a pair of chiral superfields $(Q_i, \tQ_i)$ with opposite $\U(1)$ gauge charges, $(q_i, -q_i)$, and the $\U(1)$ vector multiplets as an $\cN=1$ vector multiplet and neutral chiral superfield $\Phi$.  Partition the set of hypermultiplets into $T$ subsets $\{Q_i^I,\tQ_{iI}\}$ with $i=1,\ldots,T$, where the hypermultiplets in each subset have the same charge $\{q_i\}$, and where $I=1, \ldots, f_i$ labels the fields within that subset.  So the total number of hypermultiplets satisfies $H = \sum_{i=1}^T f_i$ and the $N$ in \eqref{u1} is now
\begin{align}\label{u2}
N = \sum_{i=1}^T f_i q_i^2.
\end{align}
The global flavor symmetry group of the massless theory is then
\begin{align}\label{u3}
F = U(f_1) \times \cdots \times U(f_T)
\end{align}

The $\cN=1$ superpotential for this theory is $\cW = q_i^2 Q_i^I \Phi \tQ_{iI} - M_{iI}^{\ J} Q_i^I \tQ_{iJ}$.  Here $M_i^{IJ}$ are complex mass matrices which satisfy the $\cN=2$ supersymmetry condition $[M_i , M_i^\dag]=0$ for each $i$.  Such $M_i$ can be diagonalized by flavor rotations so that the superpotential becomes
\begin{align}\label{u4}
\cW = Q_i^I (q_i^2 \Phi - m_i^I) \tQ_{iI} .
\end{align}
This implies that whenever the CB complex coordinate $u := \vev{\Phi} = q_i^{-2} m_i^I$, a single hypermultiplet of $\U(1)$ electric charge $q_i$ becomes massless.  The presence of a massless charged state results in a singularity (non-analyticity) in the effective metric on the CB.  Since a charge-$q_i$ massless hypermultiplet contributes a coefficient of $q_i^2$ to the 1-loop $\b$ function, the EM duality monodromy about this singularity will be $T^{n_i}$, with
\begin{align}\label{u5}
n_i := q_i^2,
\end{align}
and thus the singularity will be of $I_{n_i}$ type in the classification of possible rank-1 CB singularities given in table \ref{kodaira}.  Though this structure of the CB geometry was deduced from the classical lagrangian description, by keeping all the masses small compared to the strong coupling scale, $\L$, of the $\U(1)$ gauge theory, the theory is kept arbitrarily weakly coupled, and so is arbitrarily well described by its lagrangian.

The general deformation pattern of an $I_N$ singular CB geometry is thus
\begin{align}\label{u6}
I_N \to \left\{ (I_{n_1})^{f_1}, \ldots, (I_{n_T})^{f_T} \right\} .
\end{align}
Furthermore, since all the massless states at these singularities are electrically charged with respect to the $\U(1)$ gauge group, it follows that all the singularities will have $T^n$ monodromies (i.e., not relatively conjugated ones).  Thus the topological data describing the special K\"ahler (SK) structure of a CBs with deformation pattern \eqref{u6} is the ordered set of monodromies
\begin{align}\label{u7}
\left\{ {(\bsm0\\1\esm)_{n_1}}^{f_1} \ \cdots\ \ 
{(\bsm0\\1\esm)_{n_T}}^{f_T} \right\},
\end{align}
where we are using the notation introduced in \eqref{Insh} for $I_n$-type monodromies.  Because they are all mutually local their ordering does not matter.

Constructing the SK geometry of these CBs (not just the topology of their SK structure) is quite a bit harder.  Though it is not needed for our homological calculation of the string web lattice, we nevertheless present its solution in appendix \ref{appU1};  it has not appeared elsewhere in the literature.

Having constrained the possible deformations that can appear in \eqref{u7}, we only need to distinguish two cases which will have a different qualitative behavior.
The first is when all singularities on the right hand side are of the same $n_i$ type, so only the first entry in \eqref{u7} occurs.  The second is when we have multiple $n_i$ types, in which case we only need to look at the case with two entries in \eqref{u7}, since additional ones will follow the same pattern. These two families of examples are described in the following subsections.

\subsubsection{One family of $I_n$-type singularities}
\label{sec:In1type}

We start with the case of a simple configuration of $I_n$ singularities all having the same monodromy $T^n$. This has deformation pattern and monodromy data
\begin{align}\label{u8}
I_N \to &\left\{ {I_n}^f \right\},&
&\text{with monodromies}&
& \left\{ {(\bsm0\\1\esm)_n}^f \right\},&
&\text{where}&
N &= f n .
\end{align}
As explained above, this corresponds to a $\U(1)$ gauge theory with $f$ hypermultiplets all of electric charge $q=\sqrt n$, and thus has flavor symmetry $\U(f) = \U(1)\times\SU(f)$.  Let $I,J\in\{1,\ldots,f\}$ be flavor indices, which can also be taken to label the $f$ $I_n$ singularities.

The boundary conditions on string webs are that segments ending on the $J$th singularity can carry fiber 1-homology class $\d_J = a_J (0\ n)$ for any integer $a_J$.  The string web neutrailty condition is then just the condition that $\sum_{J=1}^f a_J = 0$.  A simple choice of basis string webs for the resulting rank-$(f-1)$ lattice is 
\begin{align}\label{u9}
\begin{array}{c|rrrrrr}
 & a_1 & a_2 & a_3 & \ldots & a_{f-1} & a_f \\
\hline
\cW_1 & 1 & -1 & 0 & \ldots & 0 & 0 \\
\cW_2 & 0 & \phantom{+}1 & -1 & \ldots & 0 & 0\\
\vdots & & & & \ddots & & \\
\cW_{f-1} & 0 & 0 & 0 & \ldots & 1 & -1
\end{array} .
\end{align}
The inner products of these webs, from \eqref{webinner2}, is simply $\cW\cdot\til\cW = \sum_{J=1}^f n a_J \til a_J$, giving the metric in the above basis
\begin{equation}\label{u10}
g = n 
\bpms[r]
2 & -1 & 0 & \ldots & 0 & 0 \\
-1 & 2 & -1 &  & 0 & 0 \\
0 & -1 & 2 &  & 0 & 0\\
\vdots & & & \ddots & & \vdots \\
0 & 0 & 0 & & 2 & -1 \\
0 & 0 & 0 & \ldots & -1 & 2 
\epms .
\end{equation}
Since this is proportional to the Cartan matrix for $A_{f-1}$, this is a $\HEX_{f-1}$ lattice.

We can compare this result to the predictions from the $\U(1)$ gauge field theory.  The lattice of neutral string webs should be generated by the gauge-neutral BPS states of the theory.  These correspond to the holomorphic ``meson'' operators $M^I_J := Q^I \tQ_J$.  They fill out the adjoint representation of the $\U(f) \simeq \U(1) \times \SU(f)$ flavor symmetry.  These span an $A_{f-1}$ root lattice, i.e., a $\HEX_{f-1}$ lattice.  Thus even though the flavor symmetry has rank $f$, the flavor root lattice has rank $f-1$ because none of the gauge-neutral states are charged under the overall $\U(1)$ flavor factor.  This matches with the calculation of the string web metric in \eqref{u10}.

The normalization of the metric in \eqref{u10}, though, is not a property of the representation of the flavor symmetry on neutral states.  The length-squared of the shortest (non-zero) lattice elements is $2n$ in \eqref{u10}, which is twice the electric charge squared of the hypermultiplet fields in the theory.  This charge is observable in the low energy theory on the CB as the smallest non-zero electric charge of BPS states in the theory.  

This suggests the conjecture that the normalization of the neutral string web metric is twice the square of the smallest invariant EM charge of BPS states on the CB.  In this form the statement can be tested in non-lagrangian theories:  the
normalizations of the lattice metrics computed in tables \ref{typeIn}, \ref{tab4}, and \ref{tab5} can be checked against the BPS spectra of the SCFTs in the cases where it has been worked out.  We postpone the discussion of this test to section \ref{sec6.3} where we present a more general form of the conjecture.

\subsubsection{Two or more families of $I_n$-type singularities}

Consider now the case of mutually local singularities divided into two sets.  This is a deformation of the form 
\begin{align}\label{u11}
I_N \to &\left\{ (I_n)^f, (I_{n'})^{f'} \right\},&
&\text{with monodromies}&
& \left\{ {(\bsm0\\1\esm)_n}^f\  
{(\bsm0\\1\esm)_{n'}}^{f'}\right\},
\end{align}
where 
\begin{align}\label{u11a}
N = f n + f' n'.
\end{align}
This corresponds to a $\U(1)$ gauge theory with $f$ hypermultiplets all of electric charge $q=\sqrt{n}$ and $f'$ hypermultiplets all of electric charge $q'=\sqrt{n'}$.  It thus has flavor symmetry $\U(f)\times\U(f') = \U(1)^2\times\SU(f)\times\SU(f')$.  Let $I,J\in\{1,\ldots,f\}$ label the $f$ $I_n$-type singularities and $I',J'\in\{1,\ldots,f'\}$ label the $f'$ $I_{n'}$-type singularities.  Thus the $I,J$ can be taken to be flavor indices for the $\U(f)$ factor, while the $I',J'$ indices are flavor indices for the $\U(f')$ factor.

Consistency with Dirac quantization requires that the hypermultiplet charges be commensurate.\footnote{Even though there are no magnetic charges in this IR free field theory, any UV completion of the theory will introduce them, and so requires the Dirac quantization condition.}  This means that the possible physically consistent $n$ and $n'$ must satisfy
\begin{align}\label{u11.1}
n &= P \hq^2, &
& \text{and} &
n' &= P (\hq')^2, &
& \text{where} & 
P, \hq, \hq' &\in \Z &
& \text{and} & 
\gcd(\hq, \hq') &=1.
\end{align}
Note that here
\begin{align}\label{u11.2}
P := \gcd(n,n') ,
\end{align}
and the extra condition \eqref{u11.1} on the singularity types is that $n/P$ and $n'/P$ are perfect squares.  Physically, we can take $\sqrt{P}$ to be the unit of charge quantization, and the hypermultiplet charges, $q=\sqrt{P} \hq$ and $q' = \sqrt{P} \hq'$, are commensurate.   

Each subset of $f_i$ singularities of $I_{n_i}$ type will make up its own rank-$(f_i-1)$ $\HEX_{f_i-1}$ sublattice, and together they span a rank-$(f+f'-2)$ lattice.  However, we have a total of $f+f'$ singularities and, since all the monodromies are mutually local, only one constraint equation coming from web neutrality.  Thus the total rank of the neutral string web lattice is $f + f' - 1$, indicating that besides the $\HEX_{f-1} \oplus \HEX_{f'-1}$ factors, there has to be an additional rank 1 contribution to the lattice.  This is generated by an additional neutral web connecting the two subsets.    

Parameterize the string webs as
\begin{align}\label{u12}
\cW = \sum_{J=1}^f a_J S_J + \sum_{J'=1}^{f'} a'_{J'} S_{J'},
\end{align}
where, as in \eqref{webdef}, $S_J$ and $S_{J'}$ are formal symbols denoting the $J$th or $J'$th singularity.  Take the first $f-1$ basis webs, $\cW_i$, $i=1,\ldots,f-1$, of the form expressed in \eqref{u9}, connecting the $I_n$-type singularities, and the final $f'-1$ basis webs, $\cW'_j$, $j=1,\ldots,f'-1$, to be of the same form, but now connecting the $I_{n'}$-type singularities.  The remaining basis web, $\cW_*$, comes from connecting the set of $I_n$-type to the set of $I_{n'}$-type singularities.  The web neutrality condition for $\cW_*$ is
\begin{align}\label{u13}
0 = n \sum_{J=1}^f a_J + n' \sum_{J'=1}^{f'} a'_{J'}.
\end{align}
It is not hard to see that any choice of $a_J$s and $a'_{J'}$s satisfying this condition which minimize $|\sum_J a_J|$ and $|\sum_{J'} a'_{J'}|$ while keeping them positive is a basis web.  A basis web can thus be taken to have
\begin{align}\label{u14}
\sum_{J=1}^f a_J &= n'/P,&
&\text{and}&
\sum_{J'=1}^{f'} a'_{J'} &= -n/P .
\end{align}
It is convenient to choose the shortest such basis web for $\cW_*$, so that the shape of the resulting lattice is clear from the metric.  The length-squared of $\cW_*$ is, from \eqref{webinner2},
\begin{align}\label{u15}
\cW_* \cdot \cW_* = 
n \sum_{J=1}^f (a_J)^2 +
n' \sum_{J'=1}^{f'} (a'_{J'})^2 .
\end{align}
This is minimized by choosing the integers $|a_J|$ and $|a'_{J'}|$ to be as small as possible, subject to \eqref{u14}.
If $n'/P \gtrsim f$ or $n/P \gtrsim f'$, then the solutions minimizing \eqref{u15} subject to \eqref{u14} depend in a complicated way on $n$, $n'$, $f$, and $f'$.  But in two limiting cases  these solutions are easy to describe.

The first limiting case to consider is when 
\begin{align}\label{u16}
n'/P < f \qquad \text{and} \qquad n/P < f' .
\end{align}
Then the solution is to set $n'/P$ of the $a_J=1$ and the rest to $0$, and set $n/P$ of the $a'_{J'}=-1$ and the rest to $0$, giving the basis
\begin{align}\label{u17}
\begin{array}{l|rrrrr:rrrrr}
& a_1 & a_2 & \ldots & a_{f-1} & a_f \ \, 
& a'_1 & a'_2 & \ldots & a'_{f'-1} & a'_{f'} \\
\hline
\cW_1 & 1 & -1 & & & \\
\ \vdots & & & \ddots & & \\
\cW_{f-1} & & & & 1 & -1\ \,  \\
\hdashline
\cW_* & \multicolumn{5}{c:}{\longleftarrow {\footnotesize n'/P\ \text{``$+1$''s}}\longrightarrow} 
& \multicolumn{5}{c}{\longleftarrow {\footnotesize n/P\ \text{``$-1$''s}} \longrightarrow} \\
\hdashline
\cW'_1 & & & & & & 1 & -1 \\
\ \vdots & & & & & & & & \ddots \\
\cW'_{f'-1} & & & & & & & & & 1 & -1
\end{array}
\end{align}
(where all the empty entries are zero).  Furthermore, by placing the $+1$'s and $-1$'s in $\cW_*$ strategically --- e.g., by setting $a_1 = \cdots = a_{n'/P}=1$ and $a'_1 = \cdots = a'_{n/P} = -1$ --- we find that $\cW_*$ has a non-vanishing inner product with only the $\cW_{n'/P}$ and $\cW'_{n/P}$ webs, giving the metric
\begin{align}\label{u18}
g = 
\left(\begin{array}{rrrr:c:rrrr}
2n & -n & & &  & & & & \\
-n & 2n & \ddots & & n & & & & \\
& \ddots & \ddots & -n & & & & & \\
& & -n & 2n & & & & & \\[0.5mm]
\hdashline
& n & & & \,2 n n'/P \,\, & & -n' & &\\
\hdashline
& & & & & 2n' & -n' & & \\
& & & & -n' & -n' & \ddots & \ddots& \\
& & & & & & \ddots & 2n' & -n' \\
& & & & & & & -n' & 2n' 
\end{array}\right) ,
\end{align}
where the ``$n$'' entry in the $\cW_*$ row and column are in the $a=n'/P$ column and row, respectively, and the ``$-n'$'' entry in the $\cW_*$ row and column are in the $a'=n/P$ column and row, respectively.  (If $n'/P=f$ or if $n/P=f'$, then the corresponding $\cW_*$ row and column entries vanish.)

A second limiting case is when
\begin{align}\label{u19}
n'/P \gg f \qquad \text{and} \qquad n/P \gg f' .
\end{align}
Then the solution is to set all the $a_J \approx n'/(fP)$ and all the $a'_{J'} \approx -n/(f'P)$.  Here the ``$\approx$''  means that the $a_J$ are either $\lfloor n'/(fP) \rfloor$ or $\lceil n'/(fP) \rceil$ such that \eqref{u14} is satisfied, and similarly for the $a'_{J'}$'s.\footnote{$\lfloor x \rfloor$ is the greatest integer $\leq x$ and $\lceil x \rceil$ is the least integer $\geq x$.}  We then have a basis of string webs
\begin{align}\label{u20}
\begin{array}{l|rrrrr:rrrrr}
& a_1 & a_2 & \ldots & a_{f-1} & a_f\ \, 
& a'_1 & a'_2 & \ldots & a'_{f'-1} & a'_{f'} \\
\hline
\cW_1 & 1 & -1 & & & \\
\ \vdots & & & \ddots & & \\
\cW_{f-1} & & & & 1 & -1\ \, \\
\hdashline
\cW_* 
& \lceil \frac{n'}{fP}\rceil & \lceil \frac{n'}{fP}\rceil & \cdots 
& \lfloor\frac{n'}{fP}\rfloor & \lfloor\frac{n'}{fP}\rfloor \ \,
& -\lceil\frac{n}{f'P}\rceil & -\lceil\frac{n}{f'P}\rceil & \cdots 
& -\lfloor\frac{n}{f'P}\rfloor & -\lfloor\frac{n}{f'P}\rfloor \\[1.5mm]
\hdashline
\cW'_1 & & & & & & 1 & -1 \\
\ \vdots & & & & & & & & \ddots \\
\cW'_{f'-1} & & & & & & & & & 1 & -1
\end{array}
\end{align}
giving the metric
\begin{align}\label{u21}
g=
\left(\begin{array}{rrrr:c:rrrr}
2n & -n & & &  & & & & \\
-n & 2n & \ddots & & n & & & & \\
& \ddots & \ddots & -n & & & & & \\
& & -n & 2n & & & & & \\
\hdashline
& n & & & \phantom{\biggl(}\left[\frac{n n'}{f f' P^2} + \cO(1) \right] N \ \,
& & -n' & &\\[1mm]
\hdashline
& & & & & 2n' & -n' & & \\
& & & & -n' & -n' & \ddots & \ddots& \\
& & & & & & \ddots & 2n' & -n' \\
& & & & & & & -n' & 2n' 
\end{array}\right) ,
\end{align}
where $N=fn+f'n'$, as in \eqref{u11a}.

In both cases, the lattices \eqref{u18} and \eqref{u21} have the shape $\HEX_{f-1} \oplus \HEX_{f'-1} \niplus \CUB_1$, with the ``additional'' rank-1 $\CUB_1$ sublattice becoming more accurately orthogonal to the $\HEX$ sublattices as $n$ and $n'$ become large.  Also, in the special cases where $fP | n'$ and $f'P |n$, then the $\CUB_1$ sublattice is exactly orthogonal to both $\HEX$ sublattices.

It should be clear that these qualitative features also hold for the lattice of string webs of the general deformation pattern \eqref{u6} when the number of singularity types, $T$, is greater than 2.  The lattice shape is $\oplus_{i=1}^T \HEX_{f_i-1} \niplus_{j=1}^{T-1} \CUB_{1_j}$, with the $\CUB_{1_j}$ sublattices becoming more accurately orthogonal to each other and to all the $\HEX_{f_i-1}$ sublattices in the limit where all the $n_i$ satisfy $n_i \gg 1$.

We can, again, compare this result to the predictions from the $\U(1)$ gauge field theory.  Recall that the field theory had flavor symmetry $\U(f)\times \U(f') \simeq \U(1) \times \U(1)' \times \SU(f)\times \SU(f')$.  An algebraic basis of gauge-neutral holomorphic meson operators is  
\begin{align}\label{u22}
M^I_J &= Q^I \tQ_J,  \nn\\
(M')^{I'}_{J'} &= {Q'}^{I'} {\tQ'}_{J'}, & 
I,J &\in\{1,\ldots,f\} \nn\\
(M_*)^{I_1 \cdots I_{\hq'}}_{J'_1 \cdots J'_\hq} 
&= Q^{I_1} \cdots Q^{I_{\hq'}} 
{\tQ'}_{J'_1} \cdots {\tQ'}_{J'_\hq} , &
I', J' &\in\{1,\ldots,f'\} \nn \\
(\til M_*)^{I'_1 \cdots I'_\hq}_{J_1 \cdots J_{\hq'}} 
&= {Q'}^{I'_1} \cdots {Q'}^{I'_\hq} 
\tQ_{J_1} \cdots \tQ_{J_{\hq'}} ,
\end{align}
where $\{Q^I, \tQ_I\}$ are the $f$ charge $q= \sqrt P \, \hq$ hypermultiplets, and $\{ {Q'}^{I'}, {\tQ'}_{I'}\}$ are the $f$ charge $q= \sqrt P \, \hq$ hypermultiplets, where $P$, $\hq$ and $\hq'$ are defined in \eqref{u11.1} and \eqref{u11.2}.  The $M$ and $M'$ mesons fill out the flavor root lattices of the $\SU(f)$ and $\SU(f')$ simple flavor factors, just as in the simpler example in section \ref{sec:In1type}.  In the basis for the $\U(1)\times\U(1)'$ flavor factors in which the hypermultiplets have unit charges, the $M_*$ and $\til M_*$ operators have charges
\begin{align}\label{u23}
\begin{array}{l|rcl c|c ccc}
& \U(1) \!\!\! &\times &\!\! \U(1)' 
& \multicolumn{2}{c}{\simeq} 
& \U(1)_0\!\! &\times & \!\!\U(1)_1\\
\hline
Q^I & +1 && \ph{+}0 &&& \ph{+}\hq && \ph{+}\hq\\
\tQ_I & -1 && \ph{+}0 &&& -\hq && -\hq\\
{Q'}^{I'} & \ph{+}0 && +1 &&& \ph{+}\hq' && -\hq' \\
{\tQ'}_{I'} & \ph{+}0 && -1 &&& -\hq' && \ph{+}\hq' \\
M_* & \hq' && -\hq &&& \ph{+}0 && \ph{+}2\hq\hq'\\
\til M_* & -\hq' && \ph{+}\hq &&& \ph{+}0 && -2\hq\hq'\\
\end{array}
\end{align}
Because the $\U(1)^2$ charge vectors for $M_*$ and $\til M_*$ are parallel, there are bases of the $\U(1)^2$ flavor factor for which they are both neutral under one of the $\U(1)$'s.  One such basis is shown in \eqref{u23}, where $\U(1)_0$ is the neutral factor, and $\U(1)_1$ is an arbitrary choice for the other factor.  Since the $M$ and $M'$ operators are neutral under the $\U(1)^2$, it follows that no states are charged under the $\U(1)_0$ factor.  On the other hand, the $M_*$ and $\til M_*$ operators do carry charges under the $\SU(f)\times\SU(f')$ flavor factors.

Since $M$ and $M'$ fill out the adjoint representation of the $\SU(f) \times \SU(f')$ flavor symmetry, states built from powers of them will span an $A_{f-1} \oplus A_{f'-1}$ root lattice, i.e., an orthogonal sum $\HEX_{f-1} \oplus \HEX_{f'-1}$ sublattice.  The flavor charges of the $M_*$ and $\til M_*$ operators and their powers span an additional rank-1 sublattice but which is not orthogonal to the $\HEX_{f-1} \oplus \HEX_{f'-1}$ sublattice.  This matches with the above calculations \eqref{u18} and \eqref{u21} of the string web lattice metric, giving the $\HEX_{f-1} \oplus \HEX_{f'-1} \niplus \CUB_1$ structure found there.

The normalization of the metric for the simple $\HEX_{f-1} \oplus \HEX_{f'-1}$ sublattice factors follows the rule proposed in the last subsection:  the length-squared of the flavor charge lattice basis of simple roots is twice the electric charge squared of the hypermultiplet fields in the theory.

But the normalization of the $\CUB_1$ sublattice and the angle it makes with the $\HEX_{f-1} \oplus \HEX_{f'-1}$ sublattice does not seem to be related in an obvious way to the flavor $\U(1)_1$ charges of the generating $M_*$ and $\til M_*$ operators.  For instance, the $\U(1)_1$ charge-squared of the generating $M_*$ and $\til M_*$ fields is $4 nn'/P^2$, and twice the electric charges squared of the fundamental hypermultiplets are $2n/P$ and $2n'/P$.  None of these match the lengths-squared of the shortest generating $\cW_*$ web found in \eqref{u18} and \eqref{u22}; in particular, they do not show the interesting and intricate dependence on $f$ and $f'$ that $\cW_*\cdot\cW_*$ does.  So, beyond the fact that they have something to do with $\U(1)$ flavor symmetry factors, it is unclear to us what gauge-invariant and convention-independent property of the $\cN=2$ field theory is captured by the normalization of the non-orthogonal $\CUB_1$ factors in the string web lattice.


\subsection{Flavor lattices and deformations of IR free SU(2) gauge theories}
\label{sec6.2}

The SK structure topologies for CBs of IR free $\SU(2)$ gauge theories is quite a bit more complicated than what we found in the last subsection for IR free $\U(1)$ gauge theories.  The result, which we will derive in the next few pages, is that there are two classes of deformation patterns and associated ordered set of monodromies,
\begin{align}\label{su0}
(i)& & \text{deform.\ pattern:}& &
I_N^* \to & \ \left\{ (I_1)^2, (I_{n_1})^{f_1}, \ldots, (I_{n_X})^{f_X} \right\}, \nn\\
& & \text{monodromies:}& &
& \ \left\{ (\bsm1\\1\esm) \ \ (\bsm-1\\ \ph{-}1\esm)\ \ 
{(\bsm0\\1\esm)_{n_1}}^{f_1} \ \cdots\ \ 
{(\bsm0\\1\esm)_{n_X}}^{f_X} \right\}, \nn\\
& & \text{constraint:}& &
N = & - 4 + \textstyle{\sum_{a=1}^X} n_a f_a ,\nn \\[2mm]
(ii)& & \text{deform.\ pattern:}& &
I_N^* \to & \ \left\{ I^*_{n_0}, (I_{n_1})^{f_1}, \ldots, (I_{n_X})^{f_X} \right\}, \\
& & \text{monodromies:}& &
& \ \left\{ M(I^*_{n_0}) \ \
{(\bsm0\\1\esm)_{n_1}}^{f_1} \ \cdots\ \ 
{(\bsm0\\1\esm)_{n_X}}^{f_X} \right\}, \nn\\
& & \text{constraint:}& &
N = & \ n_0 + \textstyle{\sum_{a=1}^X} n_a f_a ,\nn
\end{align}
where we are using the notation introduced in \eqref{Insh} for $I_n$-type monodromies.  Furthermore, there is a somewhat complicated relationship between the integers $n_a$ and $f_a$ that can appear in \eqref{su0}, analogous to the constraints \eqref{u11.1} and \eqref{u11.2} coming from the Dirac quantization condition in the $\U(1)$ gauge case, but which also reflects the $\SU(2)$ irreducible representations in which the hypermultiplets reside.  These constraints are described in detail in section \ref{sec6.2.1} below.

In section \ref{sec6.2.2} we motivate an additional rule on allowed string webs so that they are not only neutral under the unconfined low energy $\U(1)$ on the CB, but also under the microscopic $\SU(2)$ gauge group as well.  Then in section \ref{sec6.2.3} we use this prescription to calculate the neutral string web lattice geometries of general IR free $\SU(2)$ gauge theory CBs, and compare them to the field theory expectations in section \ref{sec6.2.4}.


\subsubsection{SK structure topology of the CB of IR free SU(2) gauge theories}
\label{sec6.2.1}

Hypermultiplets of $\cN=2$ supersymmetric $\SU(2)$ gauge theories are in isospin-$j$ irreducible representations of the gauge group.  The integer $j$ irreps are orthogonal (real) and odd-dimensional, so we must have an even number, $2O_j$, of half-hypermultiplets.  Half-odd-integer $j$ irreps are symplectic (pseudo-real) and even-dimensional, so we can have any number, $S_j$, of half-hypermultiplets. We will call the resulting theory the $\SU(2)$ gauge theory with matter representation
\begin{align}\label{su0.5}
{\bf R}=\bigoplus_{j\in\N} (2O_j \cdot {\bf j}) \ \oplus \bigoplus_{j\in\N+\frac12} (S_j \cdot {\bf j}) .
\end{align}
where we denote the isospin-$j$ irrep as $\bf j$, i.e., by its highest weight (and not by its dimension, which is the more common convention).
With massless hypermultiplets, the flavor symmetry is then
\begin{align}\label{su1}
F = \bigoplus_{j\in\N} \Sp(2O_j) \ \oplus \bigoplus_{j\in\N+\frac12} \SO(S_j) .
\end{align}
The $\Z_2$ global anomaly \cite{Witten:1982fp} for gauge theories with Weyl fermions in symplectic representations imposes a further constraint that
\begin{align}\label{su2}
\sum_{j\in2\N+\frac12} S_j = \text{even}.
\end{align}

On the CB where the $\SU(2)$ gauge group is Higgsed to a $\U(1)$, the $2j+1$ components of a field in the $j$-irrep will carry $\U(1)$ charges $m \in \{-j, -j+1, \ldots, j\}$, which are the weights of the $j$-irrep.
The $\SU(2)$ vector multiplet together with the hypermultiplets contribute a beta function whose one-loop coefficient, $N$, is
\begin{align}\label{su2.5}
N := -4 + \sum_{j\in\N} O_j T(j) 
+ \frac12 \sum_{j\in\N+\frac12} S_j T(j) ,
\end{align}
where the quadratic index for an $\SU(2)$ irrep is 
\begin{align}\label{su3}
T(j) := \sum_{m=-j}^{j} 2 m^2 
= {2j+2\choose3}.
\end{align}
For massless hypermultiplets, the CB geometry of such a theory has a singularity at the origin with $-T^N$ monodromy, so is an $I_N^*$-type singularity; see table \ref{kodaira}.
Note that the global anomaly constraint \eqref{su2} precisely ensures that $N$ is integral.

The possible $\cN{=}2$-preserving mass deformations of this theory govern the possible deformations of the CB geometry.  Write the $2O_j$ or $S_j$ half-hypermultiplets in the isospin-$j$ $\SU(2)$ irrep in $\cN=1$ notation as chiral superfields $Q^J_j$ with $J=1,\ldots, 2O_j$ or $S_j$.  The $\cN=1$ superpotential has the general form $\cW = \Tr [Q_j^J \Phi Q_j^J - M_{JK} Q_j^J Q_j^K]$ where $\Phi$ is the $\cN=1$ chiral multiplet part of the $\cN=2$ vector multiplet, the trace is in the isospin-$j$ irrep, and $M$ is a complex $\cN=2$ supersymmetric mass matrix.  $\Phi$ is valued in the gauge Lie algebra, and on the CB its components in a complexified Cartan subalgebra get vevs, $\vev{\Phi}:=a t$, where $t$ is the Cartan generator.  In a basis of the isospin-$j$ irrep which diagonalizes $t$, the components of the hypermultiplets, $Q^J_{j,m}$, will be labeled by their eigenvalues under the $t$ action, i.e., by their weights, $m$.  The superpotential thus has the form $\cW = \sum_{J,m} (a m - M_J) Q^J_{j,-m} Q^J_{j,m}$ where $M_J$ are the (skew) eigenvalues of the mass matrix.  Thus one finds a massless hypermultiplet with electric charges $\pm m$ coming from the $J$th flavor whenever $a = M_J/m$.  The CB vev $a$ is gauge equivalent to $-a$, so in terms of the gauge-invariant CB vev $u := a^2$, there is a massless hypermultiplet of charge proportional to $m$ whenever $u = M_J^2/m^2$ for $m\in\{-j,\ldots,j\}$.  Note that the component with weight $m=0$ does not contribute a singularity on the CB.\footnote{Such components instead contribute massless neutral hypermultiplets everywhere on the CB, turning the CB into an ``enhanced CB'', discussed extensively in \cite{Argyres:2016xmc}.}

So turning on a mass $M_J$ for a hypermultiplet in an isospin-$j$ irrep of the gauge group results in massless hypermultiplets of electric charges $q \propto m$ with $m\in \{ -j, -j+1, \ldots, j\}$ at points $u = M_J^2/m^2$ on the CB.   The CB geometry will have metric singularities at these points.  Since the $\Z_2$ Weyl group orbit of a weight $m$ is $\{m,-m\}$, there are always two electric-charge-$q$ hypermultiplets at each of these singularities.  But the $\Z_2$ Weyl group is part of the gauge group, so the weight $m$ and $-m$ hypermultiplets are gauge-identified.  In particular, there is no gauge-invariant mass deformation that can separate a weight-$m$ from a weight-$(-m)$ singularity on the CB.  Therefore they should not really be thought of as contributing two electric-charge-$q$ hypermultiplets in the low energy theory, but rather as a single hypermultiplet of $\SU(2)$-invariant effective charge $q \propto |m|$ for each pair of weights $\{\pm m\}$. 

We can fix the proportionality factor between the hypermultiplet effective $\U(1)$ electric charge $q$ and the $\SU(2)$ Weyl orbit of weights $\{\pm m\}$ to be
\begin{align}\label{su4}
q = 2 \, |m|
\end{align}
as follows.  By the effective electric charge $q$, we mean the charge in the normalization where such a hypermultiplet contributes $q^2$ to the coefficient of the one-loop beta function for the low energy $\U(1)$ coupling, or, equivalently, gives rise to a singularity on the CB with EM duality monodromy $T^{q^2}$.  Follow the standard charge normalization conventions\footnote{\label{ft11}
Dirac quantization \cite{Dirac:1931kp, Schwinger:1969ib, Zwanziger:1968rs} says that the $(p,q)$ (magnetic, electric) charges of dyons obey $p q'-q p'\in\Z$, which implies that $p = cm$ and $q = de$ for integers $(c, d)$ (with both $c = 1$ and $d = 1$ realized in the spectrum) and some fixed positive real $(m,e)$ satisfying $me:=P\in\N$.  $m$ and $e$ are the electric and magnetic unit charges which can be rescaled by inverse factors by a change in the definition of the $\U(1)$ gauge coupling.  The invariant $\sqrt{P}$ of a given theory is the \emph{charge quantization unit} defined in section \ref{rk1SK}.  The EM duality group --- the subgroup of $\GL(2,\R)$ preserving the EM charge lattice and Dirac inner product --- is the group of matrices of the form $(\bsm \a & \b m/e \\ \g e/m & \d \esm)\in\SL(2,\R)$ with $\a,\b,\g,\d \in \Z$.  By choosing conventions in which we rescale the $\U(1)$ coupling so that $m = e = \sqrt P$, the EM duality group becomes $\SL(2,\Z)$.  These are the standard conventions, used in, e.g., \cite{Seiberg:1994aj}.
Furthermore, there are inequivalent possible choices of normalization of the $\SU(2)$ gauge generators consistent with the Dirac quantization condition on the CB, and they are closely related to the choice of the charge quantization unit $\sqrt{P}$.  This is the $a$ parameter introduced and discussed in detail in section 4.2 of  \cite{Argyres:2015ffa}.  In this section we will assume for simplicity that $a=1$; this implies that an isospin $j=\frac12$ hypermultiplet contributes a single hypermultiplet of charge 1 with respect to the $\U(1)$ gauge field on the CB.} 
where a massless $\SU(2)$ isospin-$\frac12$ hypermultiplet gives a singularity on the CB with an EM monodromy in the conjugacy class of $T\in\SL(2,\Z)$.  $T^{q^2}=T$ implies $q=1 = 2 \, |m|$, since the $j=\frac12$ irrep only has weights $|m|=\frac12$.  Note that this coincides with the normalization in which a root (a highest weight of the $j=1$ irrep) has length-squared $q^2 = 4 m^2 = 4$.  This is implicitly the normalization we used in defining the quadratic index $T(j)$ in \eqref{su3} which can be rewritten as
\begin{align}\label{su4.5}
T(j) := \sum_{m>0}^{j} 4 m^2 = \sum_{m>0}^j q^2.
\end{align}

Thus the EM duality monodromy of the singularity at $u=M_J^2/m^2$ due to the components of the isospin-$j$ hypermultiplet with weights $\pm m$ is $T^{q^2}=T^{4m^2}$, and so the singularity is of $I_{4m^2}$-type.  Recalling that the weight $m=0$ component gives no singularity, we have found the rule: a massive hypermultiplet in an isospin-$j$ irrep of the $\SU(2)$ gauge group contributes $\lfloor j + \frac12 \rfloor$ singularities of types
\begin{align}\label{su5}
\text{$\SU(2)$ isospin-$j$ hyper} & \to 
\begin{cases}
\left\{ I_1, I_{3^2}, I_{5^2}, \ldots, I_{(2j-2)^2}, I_{(2j)^2} \right\} 
& \text{if}\ j\in \N + \frac12 ,\\
\left\{ I_{2^2}, I_{4^2}, I_{6^2}, \ldots, I_{(2j-2)^2}, I_{(2j)^2} \right\} 
& \text{if}\ j\in \N .
\end{cases}
\end{align}

Recall that the number, $S_j$, of isospin-$j$ \emph{half}-hypermultiplets for $j\in \N + \frac12$ can be odd.  $\cN=2$ supersymmetric masses can occur only for full hypermultiplets (pairs of half-hypermultiplets), so a general mass deformation gives masses to all $S_j$ half-hypers if $S_j$ is even, and to all but one if $S_j$ is odd.  It will be useful to define 
\begin{align}\label{su5.5}
p(S_j) := S_j \ (\text{mod}\ 2) =
\begin{cases}
0 & \text{if $S_j$ is even,}\\
1 & \text{if $S_j$ is odd.}
\end{cases}
\end{align}
For the integer-isospin irreps, the number, $2O_j$, of half-hypers is even, so all gain masses under a generic mass deformation.  Thus even after a generic mass deformation of an IR free $\SU(2)$ gauge theory \eqref{su0.5}, there will be left at the origin a singularity on the CB corresponding to an IR free $\SU(2)$ gauge theory with massless matter in the representation
\begin{align}\label{su6}
{\bf R}=\bigoplus_{j\in\N+\frac12} 
p(S_j) \cdot {\bf j}  .
\end{align}
where we are using the notation of \eqref{su0.5}.  This is a ``frozen'' theory, since it follows from \eqref{su1} that it has no flavor symmetry and hence no mass deformation.  From \eqref{su2.5} it follows that it gives an $I_{n_0}^*$ singularity on the CB with $-T^{n_0}$ monodromy and
\begin{align}\label{su7}
n_0 = -4 + \frac12 \sum_{j\in\N+\frac12} p(S_j) \cdot T(j) .
\end{align}
Note that the global anomaly condition \eqref{su2} together with the properties of the quadratic index \eqref{su3} ensure that $n_0$ is a positive integer for all such theories for which at least one $p(S_j)=1$.  For instance, the smallest $n_0$ occurs when $p(S_j)=0$ for all $j$ except for $p(S_{\frac32})=1$, giving $n_0=1$; the next-smallest has
$p(S_{\frac12})=p(S_{\frac52})=1$ and the rest 0, giving $n_0=14$; and so on. 

This $I_{n_0}^*$ singularity together with the $I_n$-type singularities contributed by all the massive hypermultiplets as in \eqref{su6} then gives the deformation pattern of the type shown in case $(ii)$ of \eqref{su0}.  Note that the multiplicities, $f_a$, of the $I_{n_a}$-type singularities are related to the representation content \eqref{su0.5} of the theory through the deformation patterns typical of each $\SU(2)$ irrep, shown in \eqref{su6}.

This structure of the CB geometry was deduced from the classical lagrangian description, which is valid if there is a regime with all masses turned on where the $\U(1)$ gauge theory is weakly coupled at all scales (well below its UV strong coupling scale $\L$).  But this is not always possible.  In particular, if $p(S_j)=0$ for all $j\in\N+\frac12$, then all the half-hypermultiplets will be massive, and the effective theory at energy scales much less than those masses is pure $\cN=2$ $\SU(2)$ superYang-Mills (sYM).  This theory is asymptotically free, and so flows to strong coupling, and its CB geometry cannot be deduced from the classical lagrangian.  But, famously, this CB geometry was found in \cite{Seiberg:1994rs} to have two $I_1$-type singularities separated by the strong coupling scale of the sYM theory, and correspond to the occurrence of a massless dyonic hypermultiplet in the low energy $\U(1)$ theory at each singularity.  Furthermore, in the duality frame in which the total monodromy around these dyonic singularities is $-T^{-4}$ --- which is what follows from \eqref{su7} when all the $p(S_j)=0$ --- the ordered monodromies of the dyonic singularities are 
\begin{align}\label{su8}
\left\{ (\bsm1\\1\esm)\ \ (\bsm-1\\ \ph{-}1\esm) \right\}.
\end{align}
In a regime where the masses of the original IR free $\SU(2)$ theory are all small compared to its UV strong coupling scale, and also all of the same order of magnitude, then the strong coupling scale of the effective sYM theory is parametrically small compared both to the masses and UV strong scale.  Thus in this regime where there is a large separation of scales, the sYM solution for the CB geometry is reliable for CB vevs much smaller than the masses, while the mass-induced singularities found in \eqref{su5} are reliable for CB vevs on the order of the masses.  This then shows that the topological structure of the CB geometry in the cases where $p(S_j)=0$ for all $j\in\N+\frac12$ is given by case $(i)$ in \eqref{su0}.  Thus case $(i)$ is just like case $(ii)$ but with the naive (semi-classical) $I^*_{-4}$ singularity at the origin replaced by an $(I_1)^2$ pair of singularities with the dyonic  monodromies of \eqref{su8} reflecting the strong coupling effects of the effective sYM description at small vevs.

The explicit CB geometry --- not just the topology of its SK structure which we have determined above --- can be determined from its Seiberg-Witten curve and one-form.  Even though we do not need them for the homological calculations we do here, we describe these curves and one-forms in appendix \ref{appSU2}, since they have not appeared elsewhere in the literature.


\subsubsection{SU(2) gauge-neutrality conditions on string webs}\label{sec6.2.2}

One of the main results of the above analysis is that a hypermultiplet in a single isospin-$j$ $\SU(2)$ gauge irrep gives rise to $\lfloor j+\frac12\rfloor$ $I_n$-type singularities on the CB, as shown in \eqref{su5}.  With $f$ such isospin-$j$ multiplets, the flavor symmetry has a rank-$f$ simple factor, but $f\cdot \lfloor j+\frac12\rfloor$ mutually local $I_n$-type singularities.  This gives a neutral string web lattice of rank $\sim f\cdot \lfloor j+\frac12\rfloor$, greater than that expected for $j>1$.

From the field theory perspective it is clear that we are over-counting neutral string webs.  The reason is that the $\SU(2)$ gauge invariance requires fields to occur in irreps of the gauge group, and so upon Higgsing to a $\U(1)\subset \SU(2)$ subalgebra on the CB, each isospin-$j$ field, $\Phi_j$, contributes precisely one component for each weight, $m \in \{-j,\ldots,j\}$, of the irrep where each component has $\U(1)$ charge proportional to that weight as found in \eqref{su4}.  $\SU(2)$-invariant states, e.g., one created by an operator such as $\Tr(\Phi_j \Phi_j)$, will thus be formed only from a specific sum of $\U(1)$-neutral combinations of components.

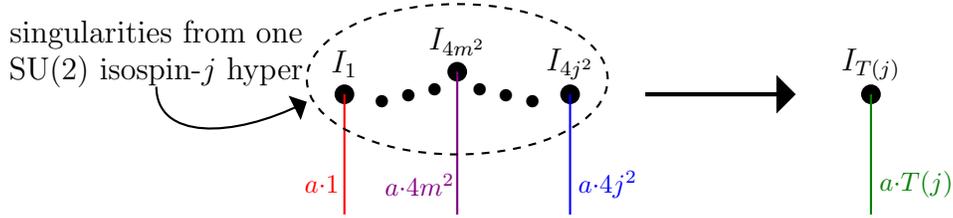
\begin{figure}[htbp]\centering
\begin{tikzpicture}[thick]
\draw (-4,0) node {$I_1$};
\node[bl] at (-4,-.4) {};
\draw[red,thick] (-4,-.4) -- (-4,-2);
\draw (-3.5,-.5) node {$\bullet$};
\draw (-3.15,-.425) node {$\bullet$};
\draw (-2.8,-.35) node {$\bullet$};
\draw (-2.5,0.3) node {$I_{4m^2}$};
\node[bl] at (-2.5,-.1) {};
\draw[red!50!blue,thick] (-2.5,-.1) -- (-2.5,-2);
\draw (-2.2,-.35) node {$\bullet$};
\draw (-1.85,-.425) node {$\bullet$};
\draw (-1.5,-.5) node {$\bullet$};
\draw (-1,0) node {$I_{4j^2}$};
\node[bl] at (-1,-.4) {};
\draw[blue,thick] (-1,-.4) --  (-1,-2);
\draw[black] (3,0) node {$I_{T(j)}$};
\node[bl] at (3,-.4) {};
\draw[green!50!black,thick] (3,-.4) -- (3,-2);
\draw[ultra thick,->] (0,-.4) -- (2,-.4);
\draw[dashed] (-2.5,-.2) ellipse (2cm and 1cm);
\draw[black,->] (-6.5,-.3) .. controls (-6.5,-1) and (-5.5,-1) .. (-4.5,-.5);
\node[black] at (-6.5,0.4) {singularities from one};
\node[black] at (-6.5,-0.1) {SU(2) isospin-$j$ hyper};
\draw[red,font=\footnotesize] (-4.3,-1.6) node {$a{\cdot} 1$};
\draw[red!50!blue,font=\footnotesize] (-3.0,-1.6) node {$a{\cdot}4m^2$};
\draw[blue,font=\footnotesize] (-0.5,-1.6) node {$a{\cdot}4j^2$};
\draw[green!50!black,font=\footnotesize] (3.6,-1.6) node {$a{\cdot}T(j)$};
\end{tikzpicture}
\caption{The selection rule from demanding gauge invariance of string webs requires them to end with equal multiplicity, $a$, on each $I_{4m^2}$ singularity, $m\in\{-j,\ldots,j\}$, coming from a single $\SU(2)$ isospin-$j$ hypermultiplet.  For the purpose of computing the neutral string web lattice, this is equivalent to replacing the singularities from the isospin-$j$ hypermultplet with a single $I_n$-type monodromy with $n=T(j)$, the quadratic index of the isospin-$j$ irrep.}
\label{fig8}
\end{figure}

This is a selection rule on the kinds of states that will appear in the low energy theory on the CB.  In terms of string webs, this selection rule is equivalent to requiring that only string webs which end with equal multiplicity on all the $I_n$-type singularities coming from a single $\SU(2)$ isospin-$j$ hypermultiplet are allowed.  This is illustrated in figure \ref{fig8}.  Effectively, the requirement of $\SU(2)$ gauge invariance imposes that the $\lfloor j+\frac12\rfloor$ $I_n$-type singularities on the CB coming from a single $\SU(2)$ isospin-$j$ hypermultiplet be treated as a single ``bound''  singularity.  Since each bound group consists of an $I_{4m^2}$-type singularity for each positive weight $m$, and all have mutually local monodromies, the effective type of the bound singularity is $I_{T(j)}$, since $T(j) = \sum_{m>0}^j 4m^2$, by \eqref{su3}.

This selection rule leads to a great simplification of the string web lattice calculation.  Using it, we can effectively replace the CB monodromy data of \eqref{su0} with its complicated dependence of the $n_a$'s and $f_a$'s on the field content of the IR free $\SU(2)$ theory, with
\begin{align}\label{su9}
(i)& & \text{deform.\ pattern:}& &
I_N^* \to & \ \left\{ (I_1)^2, (I_{T(j_1)})^{f_1}, \ldots, (I_{T(j_X)})^{f_X} \right\}, \nn\\
& & \text{monodromies:}& &
& \ \left\{ (\bsm1\\1\esm)\ \ (\bsm-1\\ \ph{-}1\esm) \ \
{(\bsm0\\1\esm)_{T(j_1)}}^{f_1} \ \cdots\ \ 
{(\bsm0\\1\esm)_{T(j_X)}}^{f_X} \right\}, \nn\\[2mm]
(ii)& & \text{deform.\ pattern:}& &
I_N^* \to & \ \left\{ I^*_{n_0}, (I_{T(j_1)})^{f_1}, \ldots, (I_{T(j_X)})^{f_X} \right\}, \\
& & \text{monodromies:}& &
& \ \left\{ M(I^*_{n_0}) \ \
{(\bsm0\\1\esm)_{T(j_1)}}^{f_1} \ \cdots\ \ 
{(\bsm0\\1\esm)_{T(j_X)}}^{f_X} \right\}, \nn
\end{align}
where
\begin{align}\label{su10}
N = & \ -4 + \sum_{j_a \in \N+\frac12} T(j_a) p(S_{j_a}) 
+ \textstyle{\sum_{a=1}^X} T(j_a) f_a ,
\end{align}
and recall that $p(S_j)=1$ if $S_j$ is odd, and zero if it is even.  These monodromy patterns are simply related to the flavor data \eqref{su0.5} and \eqref{su1} of the theory by
\begin{align}\label{su11}
f_a &= 
\begin{cases}
O_{j_a} & \text{if}\ j_a \in \N, \\
\frac12 \left[ S_{j_a} - p(S_{j_a}) \right] & 
\text{if}\ j_a \in \N +\frac12, \\
\end{cases}
\end{align}
Case $(i)$ applies when $p(S_{j_a})=0$ for all $j_a \in \N + \frac12$; otherwise case $(ii)$ holds with $n_0$ given by \eqref{su7}.

This selection rule on the string webs depends on grouping the singularities on the CB into subsets associated to $\SU(2)$ isospin-$j$ irreps.  It might seem that extra, non-topological, data about the singularities --- namely, their associated $\SU(2)$ $(j,m)$ weights --- is necessary in order to implement the selection rule.  But, in fact, a purely topological prescription exists:
\begin{description}
\item[(1)] Locate the electric $I_n$ singularity with the largest value of $n = 4j^2$.
\item[(2)] Choose $\lfloor j -\frac12\rfloor$ of the remaining electric $I_n$-type singularities with $n=4m^2$ for $m = j{-}1, j{-}2, \ldots, (1 \ \text{or}\ \frac12)$, and group these with the singularity located in step (1).
\item[(3)] Repeat the process starting from step (1) on the remaining un-grouped singularities.
\end{description}
This terminates with a set of grouped electric $I_{T(j)}$ effective singularities and either two dyonic $I_1$ singularities or an $I_{n_0}^*$ singularity, as in \eqref{su9}.  

This topological prescription, however, depends on knowing ``in advance'' that the singularities on the CB are organized by $\SU(2)$ representation theory, which itself is not a strictly topological piece of information.  One can dispense with this advance knowledge of the IR free gauge group by using some analytic information about the CB geometry.  In particular, the sets of CB singularities which should be grouped together are those which have mutually local $I_n$-type singularities and whose CB positions are not independent under general variation of the (mass) deformation parameters.

Note that the need for this additional rule grouping singularities on the CB into gauge-invariant combinations is only necessary when there is matter in gauge representations whose weights fall into multiple distinct non-zero Weyl orbits.  In the case of $\SU(2)$, that means that the rule is needed only if there are fields in irreps with isospin $j>1$.  $\SU(2)$ gauge theories with $j>1$ matter are all IR free theories.  But this fact is a coincidental property of $\SU(2)$:  higher-rank gauge groups can have irreps with multiple distinct non-zero Weyl orbits and not lead to IR free gauge theories.  Thus this kind of additional rule picking out gauge-neutral string webs is likely to be required on higher-rank CBs.


\subsubsection{String web lattices for IR free SU(2) theories}\label{sec6.2.3}

Given the CB monodromy data for IR free $\SU(2)$ gauge theories summarized in \eqref{su9}--\eqref{su11} above, it is a simple matter to compute the resulting string web lattice geometries.  In fact, the calculation can be made even simpler by noting that in the case of deformation patterns of type $(i)$ in \eqref{su9}, since the total monodromy around the initial ${1\choose1}$ and ${-1\choose\ 1}$ singularities is $-T^{-4}$, so of $I^*_{-4}$-type, any neutral string can be replaced by an equivalent neutral string web of type $(ii)$ with initial singularity monodromy $M(I^*_{-4})$.  Thus, for the purposes of computing string web lattices, we can treat type $(i)$ geometries as special cases of type $(ii)$ geometries with $n_0 = -4$.

Since the monodromies are therefore of the form 
\begin{align}\label{su13}
\left\{ M(I^*_{n_0}) \ \
{(\bsm0\\1\esm)_{T(j_1)}}^{f_1} \ \cdots\ \ 
{(\bsm0\\1\esm)_{T(j_X)}}^{f_X} \right\},
\end{align}
any neutral string web, by \eqref{In*H1classes}, must end on the $I^*_{n_0}$ singularity with charge (i.e., fiber cycle) $\bde_* = (0\ 2a_*)$ for some integer $a_*$.  Consider just the subset of string webs ending on, say, the $I^*_{n_0}$ singularity and one set of $f$ $I_{T(j)}$ singularities.   The neutrality condition for these webs is 
\begin{align}\label{su14}
0 = 2a_* + T(j) \sum_{i=1}^{f} a_i ,
\end{align}
where we are using a notation for the string webs of $I_n$-type singularities as in \eqref{u12}, and a convenient basis is
\begin{align}\label{su15}
\begin{array}{c|c:rrr}
& a_* & a_1 & \ldots & a_f \\
\hline
\cW_1 & -\frac12 T(j)& 1\ & &  \\
\vdots & \vdots & &\ \ddots\ & \\
\cW_f & -\frac12 T(j) & & &\ 1
\end{array}
\qquad
\mbox{if $T(j)$ is even,}
\end{align}
or
\begin{align}\label{su16}
\begin{array}{l|c:rrrr}
                &a_*&a_1&   a_2&\ldots &a_f\\
\hline
\cW_1       &     &   1&    -1 &           &  \\
\ \ \vdots   &     &    &\ddots&\ddots &   \\
\cW_{f-1} &     &    &          &       1 &-1\\
\hdashline
\cW^*_f        &-T(j)&    &          &         & 2
\end{array}
\qquad
\mbox{if $T(j)$ is odd.}
\end{align}
These sublattices have then have metrics
\begin{align}\label{su17}
g = T(j) \bpm1\;\ & & \\ &\ \ddots\ & \\ & &\;\ 1\epm 
\qquad
\mbox{if $T(j)$ is even,}
\end{align}
and
\begin{align}\label{su18}
g = T(j)
\left(\begin{array}{rrrr:c}
2 & -1 & & & \\
-1 & 2 & \ddots & & \\
& \ddots & \ddots & -1 & \\
& & -1 & 2 & -2\\[0.5mm]
\hdashline
& & & -2 & \ph{+}4
\end{array}\right) 
\qquad
\mbox{if $T(j)$ is odd.}
\end{align}
Recognizing the first as a cubic lattice and the second as the Killing form the $C_f$ Lie algebra, we learn that
\begin{align}\label{su19}
\text{an isospin-$j$ sublattice is }
\begin{cases}
\CUB_{f} & \text{if $T(j)$ is even,}\\
\FCC_{f} & \text{if $T(j)$ is odd.}\\
\end{cases}
\end{align}

Furthermore, using similar bases for each isospin-$j_k$ block, it is easy to see that we span a lattice of neutral string webs in which each isospin-$j_k$ sublattice is orthogonal to every other one.  The reason for this is that the only basis webs from different sublattices which end on the same singularity and so could potentially have a non-zero inner product are the ones which all have a common end on the $I_{n_0}^*$ singularity.   But since their fiber cycle at this singularity is purely electric, i.e., of the form $\bde_* = (0\ 2a_*)$, their web inner product vanishes due to the projector onto magnetic charges appearing in $G(\bde_*,\til\bde_*)$ for $I^*_n$-type singularities in \eqref{webinner3}. 

However, in the case where there are two or more isospin-$j_k$ blocks with $T(j_k)$ odd, the choice of basis webs as in \eqref{su18} for each block does not provide a basis for the whole lattice.  To see this, consider a case with two such blocks:
\begin{align}\label{su20}
& \left\{ M(I^*_{n_0}) \ \
{(\bsm0\\1\esm)_{T}}^{f} \ \cdots\ \ 
{(\bsm0\\1\esm)_{T'}}^{f'} \right\}, &
& \text{with $T:=T(j)$ and $T':=T(j')$ both odd.}
\end{align}
Then a lattice basis is
\begin{align}\label{su21}
\begin{array}{l|c:rrrr:rrrr}
              &a_*&a_1&   a_2&\ldots &a_f
                     &a'_1& a'_2&\ldots &a'_{f'}\\ \hline
\cW_1     &    &    1&    -1 &         &  &&&&\\
\ \vdots   &    &    &\ddots&\ddots&   &&&&\\
\cW_{f-1}&   &    &          &       1&-1&&&&\\ \hdashline
\cW_f     & -\frac12(T{+}T')&     &&& 1&1&&&\\ 
\cW'_{f'}& -\frac12(T{-}T')&     &&& 1&-1&&&\\ \hdashline
\cW'_{f'-1}& &&&&& 1& -1&   &   \\
\ \vdots  & &&&&& &\ddots&\ddots& \\
\cW'_1   & &&&&& &      & 1      & -1 
\end{array}
\end{align}
with resulting metric
\begin{align}\label{su22}
g = \left(\begin{array}{rrrr:rr:rrrr}
2T&-T&    &   &       &           &     &   &     &\\
-T&2T&\ds&   &      &            &    &    &     &\\
  &\ds&\ds&-T&      &            &    &    &     &\\
  &    & -T&2T&    -T&         -T&    &    &     &\\[0.5mm]
\hdashline
  &    &   &-T&T{+}T'&T{-}T'&  T'&     &    &\\
  &    &   &-T&T{-}T'&T{+}T'& -T'&     &    &\\[0.5mm] 
\hdashline
  &    &    &   &      T'&       -T'&2T'& -T'&    &\\
  &    &    &   &        &           & -T'&2T'&\ds&\\
  &    &    &   &        &           &    &\ds&\ds&-T'\\
  &    &    &   &        &           &    &     & -T'&2T'
\end{array}\right) .
\end{align}
This does not decompose as the orthogonal sum of simpler sublattices, though it contains the orthogonal sum $\FCC_f \oplus \FCC_{f'}$ as a sublattice, since the $\cW^*_f$ webs of \eqref{su16} are $\cW_f \pm \cW'_{f'}$ in terms of the \eqref{su21} basis.  Note that the lengths-squared of the new basis webs in \eqref{su22}, $T+T'$, are shorter than one or both of the length-squared $4T$ or $4T'$  webs they replace in an orthogonal sum $\FCC_f \oplus \FCC_{f'}$ of $T(j)$-odd lattices of the form \eqref{su18}.   However they are longer than the shortest basis webs of an orthogonal sum $\CUB_f \oplus \CUB_{f'}$ lattice normalized as in \eqref{su17} to have lengths-squared $T$ and $T'$.

We will denote the lattices \eqref{su18}, \eqref{su22} and their generalization to more than two factors with $T(j_a)$ odd as the non-orthogonal lattice sums $\niplus_a\!\!  \til\FCC_{f_a}$.   With the $\CUB$ and $\FCC$ lattice metric normalizations as in \eqref{su17}, \eqref{su18}, we have the lattice inclusions
\begin{align}\label{su22.5}
\oplus_a \FCC_{f_a} \subset
\ \niplus_a\!\! \til\FCC_{f_a} \subset
\oplus_a \CUB_{f_a}.
\end{align}

Since $T(j)$ is odd when $j \in 2\N+\frac12$, and is even otherwise, it follows that the final result for the string web lattice for an IR free $\SU(2)$ gauge theory with hypermultiplets in representations \eqref{su0.5} is the lattice
\begin{align}\label{su22.6}
\L_{F\text{-string web}} = 
\bigoplus 
\begin{cases}
\niplus_a\!\! \til\FCC_{f_a} & \text{for $j_a\in2\N+\frac12$,}\\
\oplus_a \CUB_{f_a} & \text{for $j_a\in2\N+\frac32$,}\\
\oplus_a \CUB_{f_a} & \text{for $j_a\in\N$,}\\
\end{cases}
\end{align}
where $f_a$ is given by \eqref{su11}.   


\subsubsection{Comparison to flavor-charge lattice of IR free SU(2) gauge theory}\label{sec6.2.4}

The $\SU(2)$ gauge theory with the same hypermultiplet representation content that gives rise to the string web lattice found in \eqref{su22.6} has flavor symmetry \eqref{su1}, and so its flavor root lattice is 
\begin{align}\label{su22.7}
\L_{F\text{-root}} = 
\bigoplus 
\begin{cases}
\oplus_a \CUB_{f_a} & \text{for $j_a\in\N+\frac12$ and $S_{j_a}$ odd,}\\
\oplus_a \FCC_{f_a} & \text{for $j_a\in\N+\frac12$ and $S_{j_a}$ even,}\\
\oplus_a \FCC_{f_a} & \text{for $j_a\in\N$.}
\end{cases}
\end{align}
The results for the neutral string web lattice and the flavor root lattice clearly do not agree.

But --- as we also learned in the IR free $\U(1)$ gauge theory case --- the flavor charge lattice, $\L_{F}$, does not coincide with the flavor root lattice:  $\L_{F}\supset \L_{F\text{-root}}$.  We can see this by looking at gauge-neutral BPS states in the gauge theory.   Denote the half-hypermultiplets ($\cN=1$ chiral multiplets) in the isospin-$j$ $\SU(2)$ irrep by $Q^{(j)}_J$ for $J=1,\ldots, H_j$ where $H_j=2O_j$ if $j\in\N$ and $H_j=S_j$ if $j\in\N+\frac12$, and denote the $\SU(2)$-adjoint $\cN=1$ chiral multiplet in the $\cN=2$ vector multiplet by $\Phi$.  Then gauge neutral BPS states are created by holomorphic gauge-singlet combinations of the $Q^{(j)}_J$ and $\Phi$ fields. 

For $j\in\N$, an algebraic basis of such holomorphic monomials are $\Tr(\Phi^j Q^{(j)}_J)$ for $J=1,\ldots,2O_j$.  These transform in the fundamental $2O_j$-dimensional representation of the $\Sp(2O_j)$ flavor symmetry factor, and so their flavor weights span a simple cubic lattice
\begin{align}\label{su23}
\L_{F\text{-BPS}} = \oplus_a \CUB_{f_a} \quad
\text{for $j_a \in \N$,}
\end{align}
which matches the string web computation \eqref{su22.6}.  Furthermore, note that in the normalization where the shortest $\Sp(2O_j)$ root has length-squared $2 T(j)$, as in the $\FCC$ lattice \eqref{su18}, the $\Sp(2O_j)$ $\CUB$ weight lattice will have shortest element of length-squared $T(j)$, as in the $\CUB$ lattice \eqref{su17} computed from the string web picture.\footnote{This statement is true for $O_j>1$, but fails when $O_j=1$, where the flavor symmetry is $\Sp(2) \simeq \SU(2)$.  In that case one has instead that if the root has length-squared $2 T(j)$ then the weight lattice has shortest element of length-squared $T(j)/2$.}

For $j\in\N+\frac12$, an algebraic basis of holomorphic monomials are quadratic in the $Q^{(j)}_J$'s, since gauge-invariant monomials which are linear do not exist.  These are all of the form $\Tr(Q^{(j)}_J \Phi^k Q^{(j')}_{L} )$ for some integer $k$.  For $j=j'$, these then transform in the tensor product of two vector representations of the $\SO(S_j)$ flavor symmetry factor, and so span the root lattice of this factor.  For $j \neq j'$ (and appropriate $k \ge |j-j'|$), they are in the tensor product of vector representations of two different flavor factors, $\SO(S_j)\times\SO(S_{j'})$.  The $\CUB$ lattice generated by the $\SO(S_j)$ vector weights will have shortest element of length-squared $T(j)$ in the normalization where the shortest root has length-squared $2 T(j)$.   We thus may expect that the flavor charge lattice generated by the $\Tr(Q^{(j)}_J \Phi^k Q^{(j')}_{L} )$ BPS states will be a sublattice of a $\CUB_f \oplus \CUB_{f'}$ lattice with factors normalized as in \eqref{su17}, and will be generated by shortest elements of length-squared $T(j)+T(j')$.  These are properties of the $\til\FCC_{f} \niplus\!\! \til\FCC_{f'}$ string web lattice found in \eqref{su22}.  Combining this with the conclusion \eqref{su23} in the $j\in\N$ case, it is thus plausible to conclude that the flavor charge lattice generated by the gauge neutral BPS states of an IR free $\SU(2)$ gauge theory is 
\begin{align}\label{su24}
\L_{F\text{-BPS}} = 
\bigoplus 
\begin{cases}
\niplus_a\!\! \til\FCC_{f_a} 
& \text{for $j_a\in\N+\frac12$,}\\
\oplus_a \CUB_{f_a} 
& \text{for $j_a\in\N$,}
\end{cases}
\end{align}
and with the same normalizations of the sublattices as in the string web lattice \eqref{su21}.

While \eqref{su24} is close to the string web lattice \eqref{su22.6}, it is not the same for flavor factors coming from hypermultiplets with $j \in 2\N+\tfrac32$.  For those factors the string web lattice is larger (finer) than the flavor charge lattice generated by gauge neutral BPS states, since $\oplus_a \CUB_{f_a} \supset \ \niplus_a\!\! \til\FCC_{f_a}$.  It is a logical possibility that the string web lattice is a flavor charge lattice which is at least partially generated by non-BPS states, and so the BPS-generated flavor charge lattice \eqref{su24} is a sub-lattice of the string web lattice.


\subsection{On the normalization of the string web lattice metric}
\label{sec6.3}

We emphasize that the comparison of the string web lattice to the flavor charge lattice computed from the field theory in the last section suffers from the fact that we have no clear prescription for how to compute a metric on the flavor charge lattice from the field theory.  Instead, the field theory gives us only the flavor weights of gauge neutral BPS fields.  Each simple factor of the flavor symmetry has a Killing form, providing a natural metric up to normalization, but the overall and relative normalizations of the factors is undetermined.  Furthermore, for sublattices charged under two or more simple flavor factors (as we encountered in the last section), even their metrics up to normalization are undetermined.

The unambiguous metric determined by the intersection form on the string web lattice thus reflects some property of $\cN=2$ SCFTs which goes beyond just the flavor symmetry.  We do not have a clear conjecture for what property of the SCFT the string web lattice metric measures, but we can make a partial conjecture based on the experience gained from the IR free field theories examined in the last section.  

In the case of IR free $\U(1)$ gauge theories we saw that the lengths-squared of the string webs corresponding to roots of the simple $A_f \simeq \SU(f+1)$ flavor sublattices is $2q^2$ where $\pm q$ is the $\U(1)$ electric charge of the hypermultiplets transforming under that simple flavor factor.

In the case of IR free $\SU(2)$ gauge theories we saw that the lengths-squared of the string webs corresponding to roots of the simple flavor sublattices is either $T(j)$, $2T(j)$, or $4T(j)$, where $T(j)$ is quadratic index for the $\SU(2)$ isospin-$j$ irrep of the hypermultiplets transforming under that simple flavor factor.  Recall \eqref{su4.5} that $T(j)$ is the sum of the squares of the $\U(1)$ electric charges of the (effective) hypermultiplets coming from components of an isospin-$j$ $\SU(2)$ hypermultiplet.  For the simply-laced $D_f \simeq \SO(2f)$ flavor factors the roots have length-squared $2T(j)$; for the non-simply-laced $C_f \simeq \Sp(2f)$ flavor factors the short roots have lengths-squared $2T(j)$ while the long roots are $4T(j)$; and for the non-simply-laced $B_f \simeq \SO(2f+1)$ flavor factors the short roots have lengths-squared $T(j)$ while the long roots are $2T(j)$.  And for the low-rank cases where there are equivalences $B_1 \simeq C_1$ and $B_2 \simeq C_2$, which root lengths are realized depends on the $\SU(2)$ gauge isospin $j$ of the hypermutiplets which are involved.

Furthermore, the sets of $\lfloor j + \tfrac12 \rfloor$ singularities on the CB coming from the components of a single isospin-$j$ hypermultiplet and which we argued should be grouped together in computing the string web lattice are those which have mutually local $I_n$-type singularities and whose CB positions are not independent under general variation of the (mass) deformation parameters.  The $T(j)$ factor in the string web metric normalization effectively comes from summing over the contributions of webs ending on these grouped singularities.

In both the $\U(1)$ and $\SU(2)$ gauge cases the gauge-charged hypermultiplets create BPS states which transform in a representation of the flavor group, and a single hypermultiplet component --- i.e., the hypermultiplet field labelled by a single weight of its flavor representation --- contributes a set of BPS states all electrically charged with respect to the low energy $\U(1)$ gauge field on the CB, and the sum of the squares of their $\U(1)$ charges is $T(j)$.  

These observations can be summarized in the unfortunately somewhat imprecise and unwieldly statement:
\begin{quote}
The length-squared of an element of the string web lattice corresponding to a root of a simple flavor symmetry factor is proportional to the sum of squares of the smallest positive invariant EM charges of BPS states on the CB which: (1) carry a single weight of the simple flavor symmetry factor in question, (2) whose EM charge vectors are parallel, and (3)  whose masses are not independent under general variation of the deformation parameters.
\end{quote}
This statement relates the normalizations of the lattice metrics computed in tables \ref{typeIn}, \ref{tab4}, and \ref{tab5} to the BPS spectra of these theories.   
In the case of the SCFTs in table \ref{tablist} corresponding to maximal deformation patterns (the blue-shaded deformations in the table) the conjecture is satisfied with the factor of proportionality being 2.  This follows since it is known \cite{Minahan:1996fg, Minahan:1996cj, Shapere:1999xr} that the smallest-charge BPS states which carry the flavor symmetry have invariant EM charge 1 on the CB, and the string web metric \cite{Gaberdiel:1997ud} is the flavor root lattice with shortest lengths-squared 2.  

It would be interesting to compare the string web lattices computed in this paper for the rank 1 SCFTs in table 1 to the BPS spectra of these theories, perhaps all computable using a combination of $\cS$-class superconformal index calculations \cite{Beem:2014rza} and BPS quiver calculations \cite{Shapere:1999xr, Cecotti:2010fi, Cecotti:2011rv, Alim:2011ae, Alim:2011kw}.

\section{Generalization to higher ranks}
\label{sec7}

Our string web construction is naturally extended to higher-rank CB geometries by considering the middle homology of the total space of the CB.  

A rank-$r$ CB, $\cB$, is a connected $r$-complex-dimensional special K\"ahler (SK) space with metric non-analyticities (as well as complex structure singularities \cite{Argyres:2017tmj, Bourget:2018ond, Argyres:2018wxu}) in a complex codimension 1 variety $\cS$.  The SK structure of $\cB$ is captured by a holomorphic algebraically completely integrable Hamiltonian system \cite{Donagi:1994, Donagi:1995cf, Freed:1997dp} whose complex phase space we call the \emph{total space}, $X$, of the CB.  This means that $X$ is a $2r$-complex-dimensional space holomorphically fibered over the CB, $\pi: X \to \cB$, whose fibers $X_p = \pi^{-1}(p)$ are $r$-complex dimensional abelian varieties for $p\in\cB\setminus\cS$ a regular point of the CB base.  

Denote by $\tX = X \setminus \pi^{-1}(\cS)$ the total space  of the regular points, $\cB\setminus\cS$, of the CB.  Then $\tX$ is a connected but not simply connected non-compact $2r$-complex-dimensional manifold.  $\tX$ is endowed with a complex symplectic structure, a holomorphic (2,0)-form $\Om$, with respect to which the fibers are lagrangian.  Furthermore, the fibers are endowed with a choice of polarization, a non-degenerate integral (1,1)-form $t$, which is extended to $X$ by continuity.  $t$ provides an integral skew pairing of 1-homology cycles of the fibers $X_p$ which encodes the Dirac pairing on the low energy $\U(1)^r$ EM charge lattice in the vacuum corresponding to $p\in\cB$.   

The analog of the central charge map \eqref{ZOm} on $\tX$ is the map $Z: C_2(\tX,\Z)\times \cB \to \C$ from compactly supported 2-chains on $\tX$ with boundary in a fiber given by
\begin{align}\label{r-Zmap}
Z(p) = \int_\a \Om, \qquad
\del\a \subset X_p.
\end{align}
If $\del\a$ is the 1-cycle $\bla\in H_1(X_p)$, then $Z(p)$ is the central charge in the $p$ vacuum and in the $\bla$ charge sector.  EM-neutral states then correspond to 2-cycles on $\tX$ via this map.  But, unlike the rank-1 case, 2-cycles are no longer in the middle homology, and do not have an interesting intersection form.

We propose instead to generalize the central charge map \eqref{Zmap} used in the rank-1 case to the following ``string web map'', $W$, on the compact middle homology,
\begin{align}\label{r-sw-map}
W: H_r(\tX,\Z) \to \C , \quad
\a \mapsto \int_\a \Om \wedge t^{r-1} .
\end{align}
Then the string web lattice is defined by
\begin{align}\label{r-sw-lattice}
\L_\text{string web} = H_{2r}(\tX,\Z)/\text{ker}W ,
\end{align}
in analogy to the rank-1 case given in \eqref{flavlatt} or \eqref{LFH3}.  Besides being a natural generalization of the string web construction in rank-1, there are further reasons why this is a physically reasonable proposal for characterizing the lattice of EM charge-neutral states in the theory.   First, the middle homology cycles which are not in the kernel of the $W$ map are $r$-cycles in $\tX$ with one leg in the CB base over which a $(2r{-}1)$-cycle in the fiber, $\l\in H_{2r-1}(X_p)$, is suspended.   This agrees with association \cite{Argyres:2001pv} on general grounds (i.e., that applies to all $\cN=2$ theories whether or not they have F-theory realizations) between particle states in the effective theory on the CB and string webs on the CB.  Second, the suspended fiber $(2r{-}1)$-cycle $\l$ which is not in the kernel of $W$ will be dual in the fiber relative to the fiber fundamental class $t^r$ to a non-zero fiber 1-cycle $\l^*$.   This gives an isomorphism of the string web lattice \eqref{r-sw-lattice} with the more familiar picture derived from \eqref{r-Zmap} of EM charge-neutral states as 2-cycles in the total space filling out the lattice $H_2(\tX,\Z)/\text{ker}Z$.

Finally, recall that in our exploration of IR free $\SU(2)$ gauge theories in the last section we uncovered the occurrence of distinct singularities on the CB which are $\SU(2)$ gauge-related.  This implied that gauge-neutral states in these theories were represented by string webs with special rules for their boundary conditions on these gauge-related singularities, as discussed in section \ref{sec6.2.2}.  We point out not only that this phenomenon persists at higher ranks, but also that it is no longer associated only to IR free gauge theories.

To see this, consider a theory with a general Lie algebras, $\gf$.  Say there are $f_j$ (full) hypermultiplets, $H^J_j$ with $J=1,\ldots,f_j$, in an irrep $j$ of $\gf$.  A component of $H^J_j$ labeled by a weight $m\in\tf^*$ in the dual of a Cartan subalgebra of $\gf$ becomes massless along hypersurfaces in the CB parameterized by elements $a\in\tf_\C$ of the complexified Cartan subalgebra satisfying $m(a)=M_J$ for $M_J$ a mass eigenvalue.  The gauge-inequivalent points on the CB are parameterized by $\tf_\C/\text{Weyl}(\gf)$, and the weights, $m$, of $j$ are partitioned into a disjoint set of Weyl$(\gf)$ orbits.  So the $m(a)=M_J$ hyperplanes in the Cartan subalgebra descend to a set of distinct hypersurfaces, $\cH^J_{j,o}$, in the CB corresponding to the different Weyl orbits, $o$, of weights in $j$ and to the flavor index, $J$.  If $r=\text{rank}(\gf)$, then $m$ are the low energy $\U(1)^r$ charges of the hypermultiplet components on the CB, and $H^J_{j,m}$ becomes massless at $\cH^J_{j,o}$ if $m\in o$.  

Thus, for each flavor index there will be a set of distinct hypersurfaces in the CB corresponding to the different (non-zero) Weyl orbits of weights in the irrep of the gauge group under which the hypermultiplets reside.  In the case of gauge group $\SU(2)$ the only irreps for hypermultiplets that do not make the theory IR free happened to be the isospin $j=\frac12$ and $j=1$ irreps, which each have only a single non-zero Weyl orbit of weights.  But at higher rank this is no longer true.  For example, already at rank 2, the adjoint irreps of the non-simply laced gauge groups $\gf = \mathfrak{sp}(4) \simeq \mathfrak{so}(5)$ and $\gf = G_2$ each have two distinct non-zero Weyl orbits of roots, and so the corresponding $\cN{=}2^*$ theories (which have a hypermultiplet in the adjoint irrep and are $\cN{=}4$ SCFTs in the massless limit) will have multiple component hypersurface singularities which are tied together by the microscopic gauge invariance.  It is therefore natural to suppose that such a phenomenon will also occur in non-lagrangian SCFTs at higher rank.



\acknowledgments

It is a pleasure to thank S. Cecotti, M. Caorsi, J. Distler, S. Gukov, C. Long, Y. L\"u, and especially M. Martone for helpful comments and discussions.  This work was supported in part by DOE grant DE-SC0011784.  Also, while part of this work was being done, PA was hosted by the California Institute of Technology and partially supported by Simons Foundation Fellowship 506770
.


\appendix


\section{Root systems and their lattices}\label{appA}

We give a brief overview of the lattices that appear as root lattices of compact Lie algebras.

Let $\left\{e_i\right\}$ be an orthonormal basis of $\R^\infty$. Then the roots of the classical simple Lie algebras are
\begin{align}\label{}
A_N &: & &\{\pm (e_i-e_j), \  1\le i,j \le N+1\} ,
\nonumber\\
B_N &: & &\{\pm e_i \pm e_j, \ 1\le i<j \le N \ 
\text{and}\  \pm e_i,\  1\le i\le N\} ,
\nonumber\\
C_N &: & &\{\pm e_i\pm e_j, \ 1\le i<j\le N \ 
\text{and}\  \pm 2e_i,\ 1\le i\le N\} ,
\\
D_N &: & &\{\pm e_i \pm e_j, \ 1\le i<j\le N\} .
\nonumber
\end{align}
So their root lattices are the sets of vectors $v=\sum _i v_ie_i$ with $(v_i)\in \Z^N$ or $\Z^{N{+}1}$ such that
\begin{align}\label{}
\L_{A_N} &: & & (v_i)\in \Z^{N+1},\ 
\sum _i v_i = 0
& &:= \HEX_N \ \text{lattice},
\nonumber\\
\L_{B_N} &: & & (v_i)\in \Z^N
& &:= \CUB_N \ \text{lattice},
\nonumber\\
\L_{C_N} &: & & (v_i)\in \Z^N,\ 
\sum _i v_i=0 \ (\text{mod}\ 2)
& &:= \FCC_N \ \text{lattice},
\\
\L_{D_N} &: & & (v_i)\in \Z^N,\ 
\sum _i v_i=0 \ (\text{mod}\ 2)
& &:= \FCC_N \ \text{lattice}.
\nonumber
\end{align}
The names on the right stand for hexagonal, cubic, and face-centered cubic, in analogy to their low-dimensional cases.

There are some equivalences for low N, namely:
$\HEX_1 = \CUB_1 = \FCC_1$ (since all rank-1 lattices are equivalent), $\CUB_2 = \FCC_2$, and $\HEX_3 = \FCC_3$.  

Here is the argument for the $\CUB_2= \FCC_2$ equivalence up to rescaling and rotation.  A basis of $\CUB_2$ is $\{ e_1, e_2 \}$ which is orthonormal, while one for $\FCC_2$ is $\{ e_1{+}e_2, e_1{-}e_2 \}$
which are orthogonal but each has length $\sqrt{2}$.  So they describe the same square lattice, just rotated and rescaled relative to one another.

Similarly, $\HEX_3=\FCC_3$ because a basis of $\FCC_3$ is $\{\a_1 := -e_1{-}e_2,\ \a_2:=e_1{-}e_2,\ \a_3:=e_2{-}e_3 \}$ which satisfies $\a_j\cdot \a_j=2$ for $j=1,2,3$, $\a_1\cdot \a_2 = \a_2\cdot \a_3=-1$, and $\a_1\cdot \a_3=0$.  But this is exactly the Killing form\footnote{For the simply laced $A_n$, $D_n$, $E_n$ algebras, the Killing form and the Cartan matrix coincide, though not for the non-simply-laced cases.} of $A_3$, i.e., the matrix of inner products of a basis of simple roots of the
$A_3$ root lattice (e.g., the basis $\{e_1{-}e_2,\ e_2{-}e_3,\  e_3{-}e_4 \}$).  Another (quicker) way of seeing this equivalence is to recall that $A_3 = \SU(4)$ and $D_3=\SO(6)$ describe the same Lie algebra, so share the same root system and therefore root lattice.

Another, obvious, equivalence is $\CUB_N=(\CUB_1)^N$, where the $N$th power on the right denotes the $N$-fold orthogonal sum.  But note that this is only an equivalence if all the $\CUB_1$ lattices in the sum on the right have generating vectors of the same length.

Similarly, the exceptional root lattices are easily described from the usual descriptions of their root systems:
\begin{align}\label{}
\L_{E_8} &: & &\sum_i v_i=0 \ (\text{mod}\ 4) \ \text{and all $v_i$ even or all $v_i$ odd,}\ (v_i)\in\Z^8,
\nonumber\\
\L_{E_7} &: & & \text{the $E_8$ lattice with the extra constraint that} \ \sum_i v_i=0,
\nonumber\\
\L_{E_6} &: & & \text{the $E_7$ lattice with the extra constraint that} \ v_7+v_8=0,
\\
\L_{F_4} &: & & \text{all $v_i$ even or all $v_i$ odd,} \ (v_i)\in \Z^4 ,
\nonumber\\
\L_{G_2} &: & &\HEX_2\ \text{lattice}.
\nonumber
\end{align}

Finally, the metric for these lattices in a basis corresponding to the simple roots of the corresponding Lie algebra are simply the appropriately symmetrized Cartan matrix (the Killing form) of the Lie algebra.  In particular,
\begin{align}\label{}
\HEX_N \ \text{lattice} &: & 
g & \propto  
\bpms[r]
2 & -1       &           &    \\ 
-1& \ddots & \ddots &    \\
   & \ddots &        2 & -1 \\ 
   &           &       -1 &  2
\epms ,
\nn\\
\CUB_N \ \text{lattice} &: & 
g & \propto 
\bpms[r]
1\ &          & \\ 
  & \ \ddots\ & \\
  &           & \ 1
\epms
\ \text{or}\ 
\bpms[r]
2 & -1       &           &    & \\ 
-1& \ddots & \ddots &    & \\
   & \ddots &        2 & -1 & \\ 
   &           &       -1 &  2 &-1\\
   &           &           & -1 & 1
\epms
,
\nn\\
\FCC_N \ \text{lattice} &: & 
g & \propto 
\bpms[r]
2 & -1       &           &    & \\ 
-1& \ddots & \ddots &    & \\
   & \ddots &        2 & -1 & \\ 
   &           &       -1 &  2 &-2\\
   &           &           & -2 & 4
\epms
\ \text{or}\ 
\bpms[r]
2 & -1     &         &   &  & \\ 
-1&\ddots&\ddots&  &   & \\
   &\ddots& 2      &-1&  &-1\\
   &         & -1     & 2&-1& \\
   &         &         &-1& 2& \\ 
   &         & -1     &   &   &2
\epms ,
\\
F_4 \ \text{lattice} &: & 
g & \propto
\bpms[r] 
2&-1&&\\-1&2&-2&\\&-2&4&-2\\&&-2&4
\epms ,
\nn
\end{align}
where the empty entries are zero.


\section{Curves and 1-forms of IR free U(1) gauge theories}\label{appU1}


We outline the construction of the SW curve and 1-form for all IR free $\U(1)$ gauge theories.  In the massless limit all these theories have an $I_n$-type singularity, with a CB parameter $u$ of dimension $[u]=1$.  We will start with deriving the curve and 1-form for the maximally deformed $I_n$ singularity.  This corresponds to the $\U(1)$ theory with $n$ charge-1 hypermultiplets, and has a $\U(n)$ flavor symmetry.  We then outline how to generalize this construction to arbitrary sub-maximal deformations of the singularity, corresponding to $\U(1)$ gauge theories with hypermultiplets of arbitrary charges.

\subsection{Maximally deformed $I_n$ singularity.}

\paragraph*{The curve.}

From \cite{Argyres:2010py} the SW curve describing an $I_n$ singularity with its general (maximal) complex deformation is
\begin{align}\label{curve}
y^2&= (x-1) (x^2+\L ^{-n} \D),\\
\text{with}\qquad
\D &:= u^n+\sum _{j=1}^n u^{-j+n} M_j ,\nonumber
\end{align}
describing a family of Riemann surfaces $X\to$ CB varying holomorphically over the CB.  We denote the specific Riemann surface in $X$ at the point $u\in$CB as the fiber $X_u$.
Here $\L$ is the strong coupling scale of this IR free theory, the theory with gauge group $\U(1)$ and light charged matter with flavor mass invariants $M_j$.   Here light means light with respect to $\L$.  Though the $M_1$ mass invariant can be absorbed by a shift of $u$, it will prove convenient to keep it explicit.  

In general, the low energy theory on the CB of such an IR free theory only has meaning at scales $\ll \L$.   For masses and vevs $\ll \L$, we see that the curve has a singularity at $x=0$ of Kodaira type $I_n$, while the branch point at $x=1$ is far away and plays no role.  So, to isolate the IR theory for the free CFT, we should scale in to $x=0$ while keeping $u$ and the $M_j$ finite as $\L\to\infty$.  Note that the discriminant of the curve is $\propto \D \L^{-3n} (\D+\L^n)^2 \sim \D$ in the $\L\to\infty$ limit: the $(\D+\L^n)^2$ factor only vanishes when $(u,M_j)\sim\L$ which is outside the range of validity of the effective theory.

With the curve presented as in (\ref{curve}), the usual scaling arguments \cite{Argyres:1995xn} then give the mass scaling dimensions $[x]=[y]=0$, $[u]=[\L]=1$, and $[M_j]=j$.  Actually, to get these scaling dimensions we made the choice that the holomorphic 2-form on $X$ be normalized as $\Om \sim du\wedge dx/y$. This is a choice since we have an extra parameter $\L$ in our curve, and could have chosen $\Om \sim \L^\a du \wedge dx/y$ for some $\a\neq0$.  We chose $\a=0$ precisely to get $[u]=1$ which comes from the physics input that the theory in question is an IR free $\U(1)$ gauge theory.

\paragraph*{The massless 1-form.}

We now show that the SW 1-form of the massless $I_n$ curve is
\begin{align}\label{*2}
\l_0=u \frac{dx}y .
\end{align}
However, because of the presence of the scale $\L$, the argument is a bit subtle.  

Recall \cite{Seiberg:1994aj, Donagi:1995cf} that the SW 1-form is a meromorphic $(1,0)$-form, $\l$, on the fibers $X_u$ such that the holomorphic $(2,0)$-form on the total space $X$ is $\Om = du\wedge (\del_u \l - dg)$ where $dg$ is a closed $(1,0)$-form on the fiber.  In addition, the residues of the poles of $\l$ on $X_u$ are required to depend linearly on the dimension-1 mass parameters, $m_i$.  So $\l$ satisfies 
\begin{align}\label{*1}
\del_u\l  = \frac{dx}y + d_x g ,
\end{align}
on the fiber.  By writing the 1-form as $\l=\ell\, dx$, this becomes a differential equation for the meromorphic function $\ell$ on $X_u$.   Since we are only in one variable, it can always be integrated locally, but the answer need not be a globally defined function on $X_u$ for a given $u$. 

For example, directly integrating (\ref{*1}) as a function of $u$, $\ell = \int^u y^{-1} du$, gives the hypergeometric function
\begin{align}
\ell = \frac2{(2-n)} 
\frac{(\L/u)^{(n-2)/2}}{\sqrt{x-1}}\, 
{}_2F_1\! 
\left( \tfrac12, \tfrac12-\tfrac1n ;
\tfrac32-\tfrac1n; -\tfrac{x^2}{(u/\L)^n} \right),
\nn
\end{align}
which, as a function of $x$, has square-root branch points at $x\in\{1$, $\pm i (u/\L)^{n/2}, \infty\}$, as well as a $2/n$ branch point at $\infty$.  The square-root branch points are those needed to make $\ell$ meromorphic on $X_u$, but the $2/n$ branch point is more problematic: it indicates that $\ell$ is not a function on $X_u$, but only on some $n$-fold cover, $\til X_u \to X_u$.   
The projection $\til X_u\to X_u$ pulls back the Jacobian torus of $X_u$ to an $n$-fold cover.  Thus the charge lattice realized in this way will be an index-$n$ sublattice of the one implied by a principal polarization of $X_u$, meaning the theory can only realize a subset of all possible charges.  There is a physical consistency condition that (massless) charged states exist to explain all singularities in the effective theory, so such $n$-isogenous solutions are typically not physical.  


The fact that we cannot find an unramified SW 1-form for the $I_n$ singularity just reflects the fact that physically our theory is UV incomplete.  It is just a low energy effective theory, good for energies below $\L$, so there is no need for a 1-form to exist for the whole curve (\ref{curve}) including the structure at $x=1$ (which governs the singularities on the Coulomb branch at $u,M_j \sim \L$).  
By keeping the $(x-1)$ factor in the curve (\ref{curve}), we are keeping too much information since the singularity should not care in the $\L\to\infty$ limit about what is happening out at $x\sim 1$.  

So we should take the $\L \to \infty $ limit first in such a way as to send off to infinity all the strong coupling poles while retaining the interesting features of the singularities at small values of $u$ in the CB.  Geometrically, this limit is equivalent to degenerating the torus fiber to become a twice-punctured sphere with the punctures identified.  We can then throw away the information about the identification of the punctures ``at infinity" since they encode UV physics which we are not interested in.  Thus the small-$u$ behavior of SK deformations of an $I_n$ CB geometry will effectively be described by a family of $S^2$-fibrations over the CB.

To start, rescale and redefine
$$
x = \L^{-n/2} \til x, \qquad 
y = i \L^{-n/2} \til y .
$$
By doing this, we re-express the massless version of the curve \eqref{curve} as
$$
\til y^2 = \left(1- \L^{-n/2} \til x\right)
\left(\til x^2+ u^n\right) ,
$$
and dropping the subleading pieces and the tildes, gives
$$
y^2 = x^2  + u^n .
$$
This describes a Riemann surface which is a two-sheeted cover of the $x$-plane branched at two points, which is topologically a 2-sphere.
The $u^n$ term is extended in the general mass-deformed case to the discriminant $\D = \D(u,m_i)$ function \eqref{curve} of the $u$ coordinate and of the mass deformation parameters,
\begin{align}\label{Indegen}
y^2 = x^2  + \D(u,m_i) .
\end{align}

Next we consider the Seiberg-Witten one-form, $\l$, and how it gets modified in the genus-0 limit.  The one form satisfies the SK condition \eqref{*1} for some meromorphic function $g$ on the fiber such that $\l$ has poles with residues which are integer linear combinations of the linear mass deformation parameters.

The 1-form $\l_0$ of \eqref{*2} satisfies \eqref{*1} in the degeneration limit \eqref{Indegen}.  Indeed, plugging this ansatz into the SK condition and using the Euler homogeneity condition for $y^{-1}$ gives after some algebra
\begin{align}
\partial_u \left( u \frac{dx}{y} \right) = 
\frac{dx}{y} -
\frac{dx}{2y^3}\left( n \D -m_i \partial_{m_i} \D \right) .
\end{align}
But since $d_x (x/y) = \D \cdot (dx/y^3)$, we see that indeed \eqref{*2} satisfies \eqref{*1} with
\begin{align}\label{}
g = \frac12 \left(\frac{x}{y}\right) \left( n - \sum_i \frac{\del\log\D}{\del\log m_i} \right) .
\end{align}

Note that the fact that $\l_0$ satisfies \eqref{*1} does not make it the 1-form for the massive curve: it has to be modified to have poles with residues proportional to the linear mass parameters.

\paragraph{The massive 1-form.}

We follow \cite{Minahan:1996fg, Minahan:1996cj} to find the mass-deformed curve and 1-form in 3 steps:
\begin{description}
\item{\bf (i)} Write the invariant mass parameters $M_j$ of dimension $j$ in the curve \eqref{curve} as polynomials in linear masses $m_a$ such that there is an enhanced singularity whenever the $\U(n)$ flavor symmetry is not maximally broken.
\item{\bf (ii)} Find polynomial sections (which we label by ``$\a$"), $x = x_\a\left(u,m_a\right)$ and $y = y_\a\left(u,m_a\right)$, of the curve \eqref{curve}.
\item{\bf (iii)} Test that
\begin{align}\label{*4}
\l =\l_0 + \sum _{\a\in W_F\ \text{orbits}}
\frac{\r_\a(m) y_\a}{x-x_\a}
\frac{dx}{y}
\end{align}
satisfies \eqref{*1}.  
\end{description}
The form of $\l$ in \eqref{*4} is dictated \cite{Minahan:1996fg, Minahan:1996cj} by its having poles with constant residues proportional to integral linear combinations of the masses.  This implies we should take the residues, $\r_\a(m)$, to be integral linear combinations of the linear mass parameters $m_a$.  These masses break the flavor symmetry to its Cartan subalgebra, $\tf_F$, times the discrete flavor Weyl symmetry group, $W_F$.  Thus the linear masses $m_a$ can be though of as coordinates of a point $m\in\tf_F$, and so $\r_\a\in\tf^*_F$ are flavor weight vectors.  Invariance under the residual flavor Weyl group symmetry implies that the sum over $\a$ in \eqref{*4} must be over whole $W_F$ orbits of the weights $\r_\a$.  

\paragraph*{Step (i).}

Minimal flavor $\U(n)$ breaking is when all but one $m_a$ vanish.  
When all $m_a=0$, $\D =u^n$, so for minimal breaking, we need $\D \sim u^{n-1}$, since there are still $n-1$ more sub-breakings to go.  The only way for this to happen is that
\begin{align}\label{Un}
\D =\prod _a\left(u-m_a\right),
\end{align}
so the $(-)^j M_j$ mass invarants in the curve are just the $n$ independent symmetric polynomials in the $m_a$.
More simply, $\U(n)$ flavor symmetry requires the $m_a$ to enter Weyl-symmetrically.  But the Weyl group is just the symmetric group permuting the $m_a$, so only (\ref{Un}) (up to shift by $M_1$) is allowed.

\paragraph*{Step (ii).}

Rewrite the curve \eqref{curve} as
\begin{align}\label{pinch2}
(y-x)(y+x)=\D := \prod_{a=1}^n(u-m_a) ,
\end{align}
making it obvious that possible factorizations are
\begin{align}
y+x = \D_\a
\quad\text{and}\quad
y-x = \D_{\a^c},
\end{align}
where we have defined the order-$\n$ polynomial in $u$, $\D_\a$, and its  complementary order-($n-\n$) polynomial, $\D_{\a^c}$, by
\begin{align}
\D_\a := \prod _{a\in \a }\left(u-m_a\right), 
\quad\text{and}\quad
\D_{\a^c} := \D /\D_\a, 
\end{align}
where $\a \subset \{1,2,...,n\}$ has $\n$ elements.  Then a solution to the factorization equation is 
\begin{align}
x_\a=\frac{1}{2}\left(\D_\a - \D_{\a^c} \right)
\quad\text{and}\quad 
y_\a=\frac{1}{2}\left(\D_\a + \D_{\a^c} \right).
\end{align}
Note that we have a different set of factorization solutions for each $\n \in\{0,\ldots, \lfloor n/2 \rfloor \}$.  

\paragraph*{Step (iii).}

Fix a choice of $\n$.  To write the 1-form, we need to choose the residues $m_\a$ of the poles.  They should be some integer linear combination of the $m_a$ and should treat all the $a\in\a$  the same (this is enforced by the sum over all $\a$).  So we must have $m_\a=p \sum _{a\in\a}m_a + q \sum_a m_a$, where $p$, $q$ are arbitrary integers.  The $q$-term can always be absorbed in a shift in $u$, but it will be convenient below to choose $p$ and $q$ to give the symmetrical combination
\begin{align}\label{*5}
m_\a=(n-\n)\sum _{a\in\a}m_a 
- \n\sum_{a\not\in\a}m_a.
\end{align}
(This is the ``traceless" combination: it vanishes if all $m_a$ are equal.  It can thus be interpreted as a flavor $\SU(n)$ weight.)

All that remains is to check that \eqref{*1} is satisfied.  This turns out to be simple to show using the following identity.  Suppose that $\D=P\cdot P^c$ is a factorization into polynomials in $u$, and define
$\vf := (x+y-P)/(x-y-P)$.
Then direct computation shows
\begin{align}\label{ident}
d\ln\vf = \left[1
+\frac{\tfrac12(P+P^c)}{x-\tfrac12(P-P^c)}\right] \frac{dx}{y}.
\end{align}
If $P=\D_\a$ then $\tfrac12(P-P^c)=x_\a$, and $\tfrac12(P+P^c)=y_\a$, giving
\begin{align}
\frac{y_\a}{x-x_\a}\frac{dx}{y} = d \ln \vf
- \frac{dx}{y}.
\end{align}
Comparing to (\ref{*4}) gives
\begin{align}\label{Piform}
\l  
= \left(u - \sum_\a m_\a\right) \frac{dx}{y} + d \ln \vf
= u \frac{dx}{y} + d \ln \vf ,
\end{align}
where the second equality follows from the identities 
$\sum_\a \sum_{a\in\a}m_a = {n-1\choose\n-1}\sum_a m_a$ and
$\sum_\a \sum_{a\not\in\a}m_a = {n-1\choose\n}\sum_a m_a$, which, because of the way we chose the factors in (\ref{*5}), imply $\sum_\a m_\a = 0$.  So
\begin{align}
\del _u\l  = \del _u\left(u \frac{dx}{y}\right) 
+ d_x \left(\del _u\ln\vf \right) .
\end{align}
This then implies \eqref{*1} using the fact that $\del _u\ln \vf$ is a single-valued meromorphic function, and that $\l_0$ satisfies \eqref{*1} as shown previously.

This construction has actually given us a different 1-form, $\l_\n$, for each value of $1\le\n\le n$.  The weights (linear combinations of the masses) entering show that $\l_\n$ is associated to the $\n$-fold antisymmetric representation of $\U(n)$.  (The symmetry under interchanging $\D_\a\leftrightarrow\D_{\a^c}$ shows there is no difference between $\n$ and $n-\n$, reflecting the fact that only real representations enter into $\cN=2$ gauge theories.)   Arguments along the lines of \cite{Noguchi:1999xq} show that all these $\l_\n$'s are equivalent as far as the low energy CB physics is concerned.

\subsection{Submaximal deformations of $I_n$ singularities}

The SW curves and 1-forms for the IR free $\U(1)$ gauge theories corresponding to submaximal deformations of an $I_n$ singularity are somewhat more intricate than in the maximal deformation case, but follow the same pattern.

The beta-function and singularity structure on the CB of the IR free $\U(1)$ gauge theories with different hypermultiplet charges is described in section \ref{sec6.1}.  It is easy to write the curves reproducing these beta-functions and singularities:
\begin{align}\label{curve-sub}
y^2&= (x-1) (x^2+\prod_{q=1}^\infty[\L ^{-n_q} \D_q]^{q^2}),\\
\text{with}\qquad
\D_q &:= \prod_{J=1}^{n_q} (u-m_{q,J}),\nonumber
\end{align}
describes the curve of a $\U(1)$ gauge theory with $n_q$ hypermultiplets $H_{q,J}$, $J=1,\ldots,n_q$, of charge $\pm q$.  This theory has flavor symmetry $F = \prod_q \U(n_q)$.

Just as in the maximal deformation case, the $\L\to\infty$ degeneration limit reduces the curve to the form \eqref{Indegen},
\begin{align}\label{genpinch1}
y^2&=x^2+\prod_q \D_q^{q^2}
\end{align}
and the 1-form for the massless curve is still given by \eqref{*2}.  We then construct massive SW 1-form(s) for these curves by following the procedure of the last subsection.  We have already done step (i) in \eqref{curve-sub}.  

For step (ii), note that polynomial solutions of  \eqref{genpinch1} are given by
\begin{align}
y+x = \D_\a
\quad\text{and}\quad
y-x = \D_{\a^c},
\end{align}
for any polynomial factorization $\D_\a \cdot \D_{\a^c} =  \prod_q \D_q^{q^2}$.
Thus $\a$ is now labels all subsets of the set of all $(q,J)$ pairs each counted with multiplicity $q^2$, and $\a^c$ are the complementary subsets.  Each such factorization will be associated to a pole of the 1-form \eqref{*4}.

Finally, to construct the SW 1-form in step (iii), we need to specify the residues, $\r_\a(m)$, for each choice of pole, labelled by $\a$.  As in the maximal deformation case, the possible inequivalent choices of $\r_\a$ are limited by the sum over orbits of the flavor Weyl group. 

Denote the space $\simeq \C^{\sum_q n_q}$ of $m_{q,j}$'s by $\tf_F$ (the flavor symmetry complexified Cartan subalgebra).  This can be thought of as a vector space with basis $\{e^{q,j}\}$ such that a general point $m\in\tf_F$ is given by $m = \sum_{q,j} m_{q,j} e^{q,j}$.  Then denote by $\{e_{q,j}\}$ a basis of the dual space $\tf^*_F$ dual to the $e^{q,j}$.
The integral span of $\{e_{q,j}\}$ generates a lattice $\L^* \subset \tf^*_F$, the flavor weight lattice.  The flavor Weyl group, $W_F$, acts as $W_F \simeq \prod_q \cS_{n_q}$ permuting the weights for each charge $q$ among themselves.  

We specify the pole subset $\a$ by giving a set, $\{\n_{q,j}\}$, of integral multiplicities $0\le \n_{q,j} \le q^2$ for each $(u-m_{q,j})$ factor in \eqref{curve-sub}.  
We associate to this set a $\n\in\tf_F$ by
\begin{align}\label{Fco-weight}
\n := \sum_{q,j} \n_{q,j} e^{q,j}\in \L ,
\qquad 0\le \n_{q,j} \le q^2,
\end{align}
which thus specifies the pole position of the SW 1-form.  So we will now name poles, $\a$, by elements, $\n$, of $\L$, the co-weight lattice of $F$.  Furthermore, to each such pole position, $\n$, we would like to assign a residue of the SW 1-form, $\r(m)$, where $\r\in\L^* \subset\tf^*_F$.  Thus poles and their residues are labeled by pairs $(\r,\n)\in\L^*\times\L$.

The flavor Weyl group, $W_F$, acts on both $\L^*\times\L$ by permutations of the $e^{q,j}$ for each value of $q$ and similarly for their duals.  A typical term in the 1-form is
\begin{align}
\l \supset \r(m) \frac{dx}{x-x_\n} \frac{y_\n}{y}.
\end{align}
Invariance of $\l$ under the Weyl group implies we should average this term over the Weyl group action on the pair $(\r,\n)\in\L^*\times\L$.  Now say $\n$ is fixed by the subgroup $W_\n \subset W_F$, then upon averaging over $W_F$, $\r$ will be replaced by $\bar\r = \sum_{w\in W_\n} w\cdot \r$ which will therefore automatically also be fixed by $W_\n$.  Thus we need only consider pairs $(\r,\n)$ such that the stabilizer of $\r$ in $W_F$ contains that of $\n$: $W_\r \supset W_\n$.  But if $W_\r$ is strictly greater than $W_\n$, then the multiplicity of images of $\n$ with the same residue $\r$ gives, once the sum over the Weyl group action on the $\a$ in \eqref{*4} is performed, the same 1-form as if we had chosen to amalgamate those $\n$'s by choosing a factorization with with $W_\r=W_\n$.  This implies that the flavor weight $\r$ specifying the residue and the flavor co-weight $\n$ specifying its associated pole position can be taken to be integrally dual to one another: $\r=\n^*$.  Thus, the general form of the SW 1-forms satisfying the ansatz \eqref{*4} are
\begin{align}
\l_\n = \l_0 + \sum _{W_F(\n)} \n^*(m)
\frac{y_\n}{x-x_\n}
\frac{dx}{y},
\end{align}
where $\n$ is any non-zero flavor co-weight satisfying \eqref{Fco-weight}.


\section{Curves and 1-forms of IR free SU(2) gauge theories}\label{appSU2}

We briefly outline the construction of the SW curves and 1-forms for IR free $\SU(2)$ gauge theories.   The steps parallel those described in appendix \ref{appU1} for the IR free $\U(1)$ gauge theories, though both the physical interpretation and the required algebraic manipulations are more intricate.

\paragraph{The curve.}

From \cite{Argyres:2010py} the curve for the $I^*_n$ singularity with its general complex deformation can be presented as
\begin{align}\label{Dncurve}
y^2&= x^3 + u \,  x^2+2 B \, x \, \L^{-n}
- C \, \L^{-2n},\\
\text{with}\qquad
B & := \til M_{n+4}, \qquad
C := u^{n+3}+\sum _{j=1}^{n+3} u^{n+3-j} M_{2j} .\nn
\end{align}
Here we have used the freedom to rescale and shift $x$, $y$, and $u$ to fix the form of the terms that survive when $M_{2k}=\til M_{n+4}=0$.  The complex deformation parameterized by the $M_{2k}$ and $\til M_{n+4}$ is then the most general one possible that does not change the asymptotics of the $I^*_n$ singularity.

This curve describes the CB geometry of an $\SU(2)$ gauge theory with beta function $\propto n$ and strong coupling mass scale $\L$.  It thus describes an IR free $\SU(2)$ gauge theory for $n>0$.  With the curve presented as in \eqref{Dncurve} and with the choice of holomorphic 2-form $\Om \sim du\wedge dx/y$, the usual scaling arguments \cite{Argyres:1995xn} then give the mass scaling dimensions $[y]=3$, $[u]=[x]=2$, $[\L]=1$, and $[M_j]=j$.  Such a low energy theory only has meaning for masses and vevs $\ll \L$.   

For $u=M_i=0$ the curve has an $I^*_n$ singularity at $x=0$ where 3 branch points collide.  So the $I^*_n$ singularity describes a torus fiber degeneration where both cycles pinch, unlike the $I_n$ case where only a single fiber pinched.  This reflects the fact that an IR free $\SU(2)$ theory with UV strong coupling scale $\L$ generically deformed by mass terms which give masses $\sim m$ to all the hypermultiplets, has an effective asymptotically free $\SU(2)$ pure superYang-Mills theory at scales below $m$.   This sYM theory then runs to strong coupling at $\L_{IR} \sim m (m/\L)^\a$ for some positive $\a$ (determined, e.g., by matching the 1-loop RG running), where dyonically charged states become light at a pair of singularities in the $u$ plane separated by $|u|\sim \L_{IR}$ \cite{Seiberg:1994rs}.  In the cases where there are an odd number of half-hypermultiplets in symplectic representations of $\SU(2)$, then --- as discussed in detail in section \ref{sec6.2} --- there is an $I_{n'}^*$ singularity at $u=0$ for some $n'<n$ for generic masses, describing a ``frozen'' IR free $\SU(2)$ gauge theory.

To isolate the curve for the IR free theory we keep $u$ and the $M_j$ finite as $\L\to\infty$.  But this limit has the effect of sending $\L_{IR}\to 0$ as well.  So, even for generic masses, the pair of dyonic singularities from the effective sYM theory will effectively coalesce to form a single new singularity.  We can see this explicitly from the curve by noting that the discriminant of the right side of the curve with respect to $x$ is proportional to $u^2 (u C + B^2) - B (9 uC + 8 B^2) \L^{-n} - \tfrac{27}4 C^2 \L^{-2n}$, which has $n{+}6$ roots at $|u| \ll \L^2$.
In the $\L\to\infty$ limit the first term dominates, and splits $n{+}4$ of the roots generically, but fails to split two roots at $u=0$.  As we showed in section \ref{sec6.2}, the new singularity at $u=0$ can be thought of as an $I_{-4}^*$ singularity.  

So, the leading form of the curve in the $\L\to\infty$ limit, whatever the hypermultiplet representation content, for generic masses will describe an $I_{n'}^*$-type singularity at $u=0$.  As we saw in the IR free $\U(1)$ case in appendix \ref{appU1}, taking the $\L\to \infty$ limit in the SW curve collides two of the $x$-plane branch points at infinity, describing a degeneration of the torus fiber to a 2-sphere with two marked (and identified) points.  The same happens in this case.
Explicitly, rescale $x\to\L^{-n} x$ and $y\to \L^{-n} y$ so as to preserve the 2-form, giving in the $\L\to\infty$ limit
\begin{align}\label{x0splitcrv}
y^2 &= u x^2 + 2 B x - C .
\end{align}
Here $[y]=n-1$, $[x]=n-2$, $[\D]=2n-2$, $[u]=2$, and $\Om\sim du\wedge dx/y$.  This describes a 2-sphere as a double cover of the $x$ plane branched at two points.
The identification of the two points at $x=\infty$ is lost in this description (just as it was in the $\U(1)$ case), but should be retained as additional information about the curve.  In particular, the $I^*_{n'}$ singularity at $u=0$ is encoded now in the fact that one branch point of \eqref{x0splitcrv} goes to $x=\infty$ as $u\to0$.

\paragraph{Linear mass dependence of the curve.}

The discriminant (divided by 4) of the right side of \eqref{x0splitcrv} is  
\begin{align}\label{SU2disc}
\D :=  u C + B^2 .
\end{align}
Recall from \eqref{Dncurve} that $C = u^{n+3} + \cdots$ is a polynomial in $u$ and the masses, while $B$ only depends on the masses, so $\D$ is an order-$(n{+}4)$ polynomial in $u$, and can be written as
\begin{align}\label{SU2disc2}
\D = \prod_{a=1}^{n+4} (u + m_a^2)
\end{align}
for some $m_j$ with dimensions of mass, and implies
\begin{align}\label{SU2disc3}
B = \prod_{a=1}^{n+4} m_a .
\end{align}

The maximal mass deformation is when all the $m_a$ are independent, and corresponds to the $\SU(2)$ theory with $n+4$ hypermultiplets in the isospin $j=\frac12$ irrep of $\SU(2)$.  This is easily seen to match the CB topology dervied in section \ref{sec6.2}:  the roots of $\D=0$ give the expected $n+4$ mutually-local $I_1$-type singularities on the CB at $u=-m_a^2$.  Furthermore, the coefficients of $u$ in $B$ and $C$ are a basis of invariant polynomials in the $m_a$ under the action of the Weyl group of the $F=\SO(2n+8)$ symmetry.

Sub-maximal mass deformations, occuring when (half-)hypermultplets are put in different $\SU(2)$ irreps, correspond to specializations where the $n+4$ $m_a$ parameters are not all independent.   The specific dependences between them for a given choice of $\SU(2)$ hypermultiplet representation content can be easily read off in any particular case from the CB topology described in detail in section \ref{sec6.2.1}.  For instance, a hypermultiplet in the isospin-$j$ irrep with mass $M$ will contribute a factor of
\begin{align}\label{SU2disc4}
\text{isospin-$j$ hyper} & \to \prod_m 
\left(u + \frac{M^2}{m^2} \right)^{4m^2} , &
\text{where } m= &
\begin{cases}
\frac12, \frac32, \ldots, j 
& \text{if}\ j\in \N + \frac12 ,\\
1, 2, \ldots, j & \text{if}\ j\in \N 
\end{cases}
\end{align}
in $\D$; compare to \eqref{su5}.

\paragraph{The 1-form.}

Just as in the $\U(1)$ case, $\l_0 = 2 u dx/y$ solves the 1-form differential equation \eqref{*1}.  Indeed, straight forward manipulations show that
\begin{align}\label{lam0eq}
\del_u \l_0 
& = \frac{dx}{y} -
(n{+}2) \frac{dx}{y^3} (Bx - C)  + 
\frac{dx}{2y^3} m_a \del_{m_a} (2 B x - C) 
 = 
\frac{dx}{y} + d_x \left(\frac{Fx+G}{y} \right),
\end{align}
for 
\begin{align}\label{FGeqns2}
F &= \D^{-1} \left(uC-u^2\del_u C-2B^2\right),&
G &= \D^{-1} B \left(3C - u\del_u C\right) .
\end{align}
But $\l_0$ is not the SW 1-form, $\l$, in the massive case since it does not have poles with residues linear in the $m_a$ mass parameters.  

Suitable SW 1-forms can be found, again as in the $\U(1)$ case, using the Minahan-Nemeschansky ansatz \eqref{*4}, but with one modification:  instead of $(x_\a,y_\a)$ being a section of the curve over the $u$-plane, it is allowed to be a 2-section.  In particular, we take $x_\a$ to be rational in $u$ and the $m_a$, and $y_\a$ to be $u^{-1/2}$ times a rational function of $u$ and the $m_a$.  This order-2 ramification at $u=0$ reflects the fact that we have ``hidden'' an $I^*_{n'}$-type singularity there by taking the $\L\to\infty$ degeneration limit.  Indeed, one can check that the SW 1-forms for the asymptotically free or scale-invariant $\SU(2)$ gauge theories found in \cite{Seiberg:1994aj} have this form in the degeneration limit.

Such 2-sections are easy to find because the curve \eqref{x0splitcrv} can be written as
\begin{align}\label{SU2factor}
(ux+B - \sqrt u y) (ux+B + \sqrt u y) = \D.
\end{align}
Then picking a factorization
\begin{align}\label{SU2factor1}
\D = \D_\a \cdot \D_{\a^c}
\end{align}
of the discriminant polynomial \eqref{SU2disc2}, 2-sections are given by
\begin{align}\label{SU2factor2}
x_\a &= \frac1{2u} (\D_\a + \D_{\a^c} - 2B) ,&
y_\a &= \frac1{2\sqrt u} (\D_\a - \D_{\a^c}) .
\end{align}
Then, much as in the $\U(1)$ case described in appendix \ref{appU1}, the different choices of factorization can be associated to weight vectors of flavor symmetry representations, and the sum over the flavor Weyl orbits of these weights enforce that the residues, $\r_\a(m)$, are corresponding co-weights.

\bibliographystyle{JHEP}
\bibliography{stringweb}

\providecommand{\href}[2]{#2}\begingroup\raggedright\begin{thebibliography}{10}

\bibitem{Sen:1996vd}
A.~Sen, {\it {F theory and orientifolds}},  {\em Nucl.Phys.} {\bf B475} (1996)
  562--578, [\href{http://arxiv.org/abs/hep-th/9605150}{{\tt hep-th/9605150}}].

\bibitem{Banks:1996nj}
T.~Banks, M.~R. Douglas, and N.~Seiberg, {\it {Probing F theory with branes}},
  {\em Phys. Lett.} {\bf B387} (1996) 278--281,
  [\href{http://arxiv.org/abs/hep-th/9605199}{{\tt hep-th/9605199}}].

\bibitem{Aharony:1996xr}
O.~Aharony, J.~Sonnenschein, and S.~Yankielowicz, {\it {Interactions of strings
  and D-branes from M theory}},  {\em Nucl. Phys.} {\bf B474} (1996) 309--322,
  [\href{http://arxiv.org/abs/hep-th/9603009}{{\tt hep-th/9603009}}].

\bibitem{Sen:1996sk}
A.~Sen, {\it {BPS states on a three-brane probe}},  {\em Phys. Rev.} {\bf D55}
  (1997) 2501--2503, [\href{http://arxiv.org/abs/hep-th/9608005}{{\tt
  hep-th/9608005}}].

\bibitem{Gaberdiel:1997ud}
M.~R. Gaberdiel and B.~Zwiebach, {\it {Exceptional groups from open strings}},
  {\em Nucl.Phys.} {\bf B518} (1998) 151--172,
  [\href{http://arxiv.org/abs/hep-th/9709013}{{\tt hep-th/9709013}}].

\bibitem{Bergman:1997yw}
O.~Bergman, {\it {Three pronged strings and 1/4 BPS states in N=4
  superYang-Mills theory}},  {\em Nucl. Phys.} {\bf B525} (1998) 104--116,
  [\href{http://arxiv.org/abs/hep-th/9712211}{{\tt hep-th/9712211}}].

\bibitem{Gaberdiel:1998mv}
M.~R. Gaberdiel, T.~Hauer, and B.~Zwiebach, {\it {Open string-string junction
  transitions}},  {\em Nucl.Phys.} {\bf B525} (1998) 117--145,
  [\href{http://arxiv.org/abs/hep-th/9801205}{{\tt hep-th/9801205}}].

\bibitem{Mikhailov:1998bx}
A.~Mikhailov, N.~Nekrasov, and S.~Sethi, {\it {Geometric realizations of BPS
  states in N=2 theories}},  {\em Nucl. Phys.} {\bf B531} (1998) 345--362,
  [\href{http://arxiv.org/abs/hep-th/9803142}{{\tt hep-th/9803142}}].

\bibitem{DeWolfe:1998zf}
O.~DeWolfe and B.~Zwiebach, {\it {String junctions for arbitrary Lie algebra
  representations}},  {\em Nucl.Phys.} {\bf B541} (1999) 509--565,
  [\href{http://arxiv.org/abs/hep-th/9804210}{{\tt hep-th/9804210}}].

\bibitem{DeWolfe:1998eu}
O.~DeWolfe, T.~Hauer, A.~Iqbal, and B.~Zwiebach, {\it {Uncovering the
  symmetries on [p,q] seven-branes: Beyond the Kodaira classification}},  {\em
  Adv.Theor.Math.Phys.} {\bf 3} (1999) 1785--1833,
  [\href{http://arxiv.org/abs/hep-th/9812028}{{\tt hep-th/9812028}}].

\bibitem{Hauer:2000xy}
T.~Hauer, A.~Iqbal, and B.~Zwiebach, {\it {Duality and Weyl symmetry of 7-brane
  configurations}},  {\em JHEP} {\bf 0009} (2000) 042,
  [\href{http://arxiv.org/abs/hep-th/0002127}{{\tt hep-th/0002127}}].

\bibitem{Argyres:2015ffa}
P.~Argyres, M.~Lotito, Y.~L{\"u}, and M.~Martone, {\it {Geometric constraints
  on the space of N=2 SCFTs. Part I: physical constraints on relevant
  deformations}},  {\em JHEP} {\bf 02} (2018) 001,
  [\href{http://arxiv.org/abs/1505.04814}{{\tt arXiv:1505.04814}}].

\bibitem{Argyres:2015gha}
P.~C. Argyres, M.~Lotito, Y.~L{\"u}, and M.~Martone, {\it {Geometric
  constraints on the space of N=2 SCFTs. Part II: construction of special
  KŠhler geometries and RG flows}},  {\em JHEP} {\bf 02} (2018) 002,
  [\href{http://arxiv.org/abs/1601.00011}{{\tt arXiv:1601.00011}}].

\bibitem{Argyres:2016xua}
P.~C. Argyres, M.~Lotito, Y.~L{\"u}, and M.~Martone, {\it {Expanding the
  landscape of N=2 rank 1 SCFTs}},  {\em JHEP} {\bf 05} (2016) 088,
  [\href{http://arxiv.org/abs/1602.02764}{{\tt arXiv:1602.02764}}].

\bibitem{Argyres:2016xmc}
P.~Argyres, M.~Lotito, Y.~L{\"u}, and M.~Martone, {\it {Geometric constraints
  on the space of N=2 SCFTs. Part III: enhanced Coulomb branches and central
  charges}},  {\em JHEP} {\bf 02} (2018) 003,
  [\href{http://arxiv.org/abs/1609.04404}{{\tt arXiv:1609.04404}}].

\bibitem{Minahan:1996fg}
J.~A. Minahan and D.~Nemeschansky, {\it {An N=2 superconformal fixed point with
  E(6) global symmetry}},  {\em Nucl.Phys.} {\bf B482} (1996) 142--152,
  [\href{http://arxiv.org/abs/hep-th/9608047}{{\tt hep-th/9608047}}].

\bibitem{Minahan:1996cj}
J.~A. Minahan and D.~Nemeschansky, {\it {Superconformal fixed points with E(n)
  global symmetry}},  {\em Nucl.Phys.} {\bf B489} (1997) 24--46,
  [\href{http://arxiv.org/abs/hep-th/9610076}{{\tt hep-th/9610076}}].

\bibitem{Seiberg:1994aj}
N.~Seiberg and E.~Witten, {\it {Monopoles, duality and chiral symmetry breaking
  in N=2 supersymmetric QCD}},  {\em Nucl.Phys.} {\bf B431} (1994) 484--550,
  [\href{http://arxiv.org/abs/hep-th/9408099}{{\tt hep-th/9408099}}].

\bibitem{Noguchi:1999xq}
M.~Noguchi, S.~Terashima, and S.-K. Yang, {\it {N=2 superconformal field theory
  with ADE global symmetry on a D3-brane probe}},  {\em Nucl.Phys.} {\bf B556}
  (1999) 115--151, [\href{http://arxiv.org/abs/hep-th/9903215}{{\tt
  hep-th/9903215}}].

\bibitem{Gaiotto:2009we}
D.~Gaiotto, {\it {N=2 dualities}},  {\em JHEP} {\bf 1208} (2012) 034,
  [\href{http://arxiv.org/abs/0904.2715}{{\tt arXiv:0904.2715}}].

\bibitem{Gaiotto:2009hg}
D.~Gaiotto, G.~W. Moore, and A.~Neitzke, {\it {Wall-crossing, Hitchin Systems,
  and the WKB Approximation}},  \href{http://arxiv.org/abs/0907.3987}{{\tt
  arXiv:0907.3987}}.

\bibitem{Xie:2015rpa}
D.~Xie and S.-T. Yau, {\it {4d N=2 SCFT and singularity theory Part I:
  Classification}},  \href{http://arxiv.org/abs/1510.01324}{{\tt
  arXiv:1510.01324}}.

\bibitem{Shapere:1999xr}
A.~D. Shapere and C.~Vafa, {\it {BPS structure of Argyres-Douglas
  superconformal theories}},  \href{http://arxiv.org/abs/hep-th/9910182}{{\tt
  hep-th/9910182}}.

\bibitem{Cecotti:2010fi}
S.~Cecotti, A.~Neitzke, and C.~Vafa, {\it {R-Twisting and 4d/2d
  Correspondences}},  \href{http://arxiv.org/abs/1006.3435}{{\tt
  arXiv:1006.3435}}.

\bibitem{Cecotti:2011rv}
S.~Cecotti and C.~Vafa, {\it {Classification of complete N=2 supersymmetric
  theories in 4 dimensions}},  {\em Surveys in differential geometry} {\bf 18}
  (2013) [\href{http://arxiv.org/abs/1103.5832}{{\tt arXiv:1103.5832}}].

\bibitem{Alim:2011ae}
M.~Alim, S.~Cecotti, C.~Cordova, S.~Espahbodi, A.~Rastogi, and C.~Vafa, {\it
  {BPS Quivers and Spectra of Complete N=2 Quantum Field Theories}},  {\em
  Commun. Math. Phys.} {\bf 323} (2013) 1185--1227,
  [\href{http://arxiv.org/abs/1109.4941}{{\tt arXiv:1109.4941}}].

\bibitem{Alim:2011kw}
M.~Alim, S.~Cecotti, C.~Cordova, S.~Espahbodi, A.~Rastogi, and C.~Vafa, {\it
  {N=2 quantum field theories and their BPS quivers}},  {\em Adv. Theor. Math.
  Phys.} {\bf 18} (2014), no.~1 27--127,
  [\href{http://arxiv.org/abs/1112.3984}{{\tt arXiv:1112.3984}}].

\bibitem{Beem:2013sza}
C.~Beem, M.~Lemos, P.~Liendo, W.~Peelaers, L.~Rastelli, et~al., {\it {Infinite
  chiral symmetry in four dimensions}},  {\em Commun.Math.Phys.} {\bf 336}
  (2015), no.~3 1359--1433, [\href{http://arxiv.org/abs/1312.5344}{{\tt
  arXiv:1312.5344}}].

\bibitem{Beem:2014rza}
C.~Beem, W.~Peelaers, L.~Rastelli, and B.~C. van Rees, {\it {Chiral algebras of
  class S}},  \href{http://arxiv.org/abs/1408.6522}{{\tt arXiv:1408.6522}}.

\bibitem{Donagi:1994}
R.~Donagi and E.~Markman, {\it {Cubics, integrable Systems, and Calabi-Yau
  threefolds}},  \href{http://arxiv.org/abs/alg-geom/9408004}{{\tt
  alg-geom/9408004}}.

\bibitem{Donagi:1995cf}
R.~Donagi and E.~Witten, {\it {Supersymmetric Yang-Mills theory and integrable
  systems}},  {\em Nucl.Phys.} {\bf B460} (1996) 299--334,
  [\href{http://arxiv.org/abs/hep-th/9510101}{{\tt hep-th/9510101}}].

\bibitem{Caorsi:2018ahl}
M.~Caorsi and S.~Cecotti, {\it {Special Arithmetic of Flavor}},
  \href{http://arxiv.org/abs/1803.00531}{{\tt arXiv:1803.00531}}.

\bibitem{Argyres:2017tmj}
P.~C. Argyres, Y.~L{\"u}, and M.~Martone, {\it {Seiberg-Witten geometries for
  Coulomb branch chiral rings which are not freely generated}},  {\em JHEP}
  {\bf 06} (2017) 144, [\href{http://arxiv.org/abs/1704.05110}{{\tt
  arXiv:1704.05110}}].

\bibitem{Argyres:2001pv}
P.~C. Argyres and K.~Narayan, {\it {String webs from field theory}},  {\em
  JHEP} {\bf 03} (2001) 047, [\href{http://arxiv.org/abs/hep-th/0101114}{{\tt
  hep-th/0101114}}].

\bibitem{Freed:1997dp}
D.~S. Freed, {\it {Special Kahler manifolds}},  {\em Commun.Math.Phys.} {\bf
  203} (1999) 31--52, [\href{http://arxiv.org/abs/hep-th/9712042}{{\tt
  hep-th/9712042}}].

\bibitem{Kodaira:1964}
K.~{Kodaira}, {\it {On the structure of compact complex analytic surfaces.
  I.}},  {\em {Am. J. Math.}} {\bf 86} (1964) 751--798.

\bibitem{Kodaira:1966}
K.~{Kodaira}, {\it {On the structure of compact complex analytic surfaces. II,
  III.}},  {\em {Am. J. Math.}} {\bf 88} (1966) 682--721.

\bibitem{Witten:1982fp}
E.~Witten, {\it {An SU(2) anomaly}},  {\em Phys.Lett.} {\bf B117} (1982)
  324--328.

\bibitem{Dirac:1931kp}
P.~A. Dirac, {\it {Quantized singularities in the electromagnetic field}},
  {\em Proc.Roy.Soc.Lond.} {\bf A133} (1931) 60--72.

\bibitem{Schwinger:1969ib}
J.~S. Schwinger, {\it {A magnetic model of matter}},  {\em Science} {\bf 165}
  (1969) 757--761.

\bibitem{Zwanziger:1968rs}
D.~Zwanziger, {\it {Quantum field theory of particles with both electric and
  magnetic charges}},  {\em Phys.Rev.} {\bf 176} (1968) 1489--1495.

\bibitem{Seiberg:1994rs}
N.~Seiberg and E.~Witten, {\it {Electric-magnetic duality, monopole
  condensation, and confinement in N=2 supersymmetric Yang-Mills theory}},
  {\em Nucl.Phys.} {\bf B426} (1994) 19--52,
  [\href{http://arxiv.org/abs/hep-th/9407087}{{\tt hep-th/9407087}}].

\bibitem{Bourget:2018ond}
A.~Bourget, A.~Pini, and D.~Rodr'guez-G—mez, {\it {The Importance of Being
  Disconnected, A Principal Extension for Serious Groups}},
  \href{http://arxiv.org/abs/1804.01108}{{\tt arXiv:1804.01108}}.

\bibitem{Argyres:2018wxu}
P.~C. Argyres and M.~Martone, {\it {Coulomb branches with complex
  singularities}},  {\em JHEP} {\bf 06} (2018) 045,
  [\href{http://arxiv.org/abs/1804.03152}{{\tt arXiv:1804.03152}}].

\bibitem{Argyres:2010py}
P.~C. Argyres and J.~Wittig, {\it {Mass deformations of four-dimensional, rank
  1, N=2 superconformal field theories}},  {\em J.Phys.Conf.Ser.} {\bf 462}
  (2013), no.~1 012001, [\href{http://arxiv.org/abs/1007.5026}{{\tt
  arXiv:1007.5026}}].

\bibitem{Argyres:1995xn}
P.~C. Argyres, M.~R. Plesser, N.~Seiberg, and E.~Witten, {\it {New N=2
  superconformal field theories in four-dimensions}},  {\em Nucl.Phys.} {\bf
  B461} (1996) 71--84, [\href{http://arxiv.org/abs/hep-th/9511154}{{\tt
  hep-th/9511154}}].

\end{thebibliography}\endgroup

\end{document}